\documentclass[twocolumn]{revtex4-2}
\usepackage{amsfonts}
\usepackage[dvipsnames]{xcolor}
\usepackage{amsmath}
\usepackage{amssymb}
\usepackage{amsthm}
\usepackage{mathtools}
\usepackage{graphicx}
\usepackage{enumitem}
\usepackage[colorlinks=true,
pdfpagelabels=true,hypertexnames=true,
plainpages=false,naturalnames=false]{hyperref}
\usepackage{graphicx}

\begin{document}

\title{Obstruction to ergodicity in nonlinear Schr\"{o}dinger equations with
resonant potentials}
\author{Anxo Biasi$^{1}$, Oleg Evnin$^{2,3}$, Boris A. Malomed$^{4,5}$
\vspace{2mm}}
\affiliation{ $^{1}$Laboratoire de Physique de l'Ecole Normale Sup\'erieure ENS Universit\'e PSL,
	CNRS, Sorbonne Universit\'e, Universit\'e de Paris, F-75005 Paris, France\\
	$^{2}$Department~of~Physics,~Faculty~of~Science,~Chulalongkorn~University,~Bangkok 10330,~Thailand\\
	$^{3}$Theoretische Natuurkunde, Vrije Universiteit Brussel and International Solvay Institutes, Brussels 1050, Belgium\\
	$^{4}$Department of Physical Electronics, School of Electrical Engineering,
	Tel Aviv University, Tel Aviv 69978, Israel\\
	$^{5}$Instituto de Alta Investigaci\'{o}n, Universidad de Tarapac\'{a}, Casilla 7D, Arica, Chile}

\begin{abstract}
\noindent We identify a class of trapping potentials in cubic nonlinear Schr%
\"{o}dinger equations (NLSEs) that make them non-integrable, but prevent the
emergence of power spectra associated with ergodicity. The potentials are
characterized by equidistant energy spectra (e.g., the harmonic-oscillator
trap), which give rise to a large number of resonances enhancing the
nonlinearity. In a broad range of dynamical solutions, spanning the regimes in which the nonlinearity may be either weak or strong in comparison with
the linear part of the NLSE, the power spectra are shaped as narrow
(quasi-discrete) evenly spaced spikes, unlike generic truly continuous
(ergodic) spectra. We develop an analytical explanation for the emergence of
these spectral features in the case of weak nonlinearity. In the strongly
nonlinear regime, the presence of such structures is tracked numerically by
performing simulations with random initial conditions. Some potentials that
prevent ergodicity in this manner are of direct relevance to Bose-Einstein
condensates: they naturally appear in 1D, 2D and 3D Gross-Pitaevskii
equations (GPEs), the quintic version of these equations, and a
two-component GPE system.
\end{abstract}

\maketitle


\section{Introduction}

The clash between integrability and ergodic behavior is a well-known basic
phenomenon in the dynamics of nonlinear systems \cite%
{BookNearIntegrability,BookStochasticity}. While the evolution of generic
systems with many degrees of freedom typically exhibits thermalization,
chaotization and stochasticity, dynamics of integrable systems are tightly
constrained by a large (or infinite) number of conservation laws. A conflict
between these scenarios arises when the system is ``close" to integrability
\cite{BookNearIntegrability}. In that case, a natural question is to what
extent the dynamics displays ergodic features. Such questions were
suggested, in particular, two decades ago by experiments with nearly-1D cold
atomic gases \cite{2004,2006} because the underlying basic model may be the
integrable Lieb-Liniger one \cite{Lieb-Liniger}, but integrability-breaking
effects cannot be completely eliminated from the real-world setup \cite%
{Olshanii}. The problem of the competition between the integrability and
ergodicity motivated studies of deviations from the standard framework of
non-equilibrium dynamics \cite{Langen}, bringing along intriguing ideas such
as generalized hydrodynamics \cite{GHD1,GHD2,GHD3,GHD4,GHD5,GHD6,GHD7},
prethermalization \cite{PreTherm0, PreTherm1,PreTherm2,PreTherm3},
generalized Gibbs ensembles \cite{Gibbs1,Gibbs2,Gibbs3}, etc.

\begin{figure}[t]
\centering	
\includegraphics[width=\columnwidth]{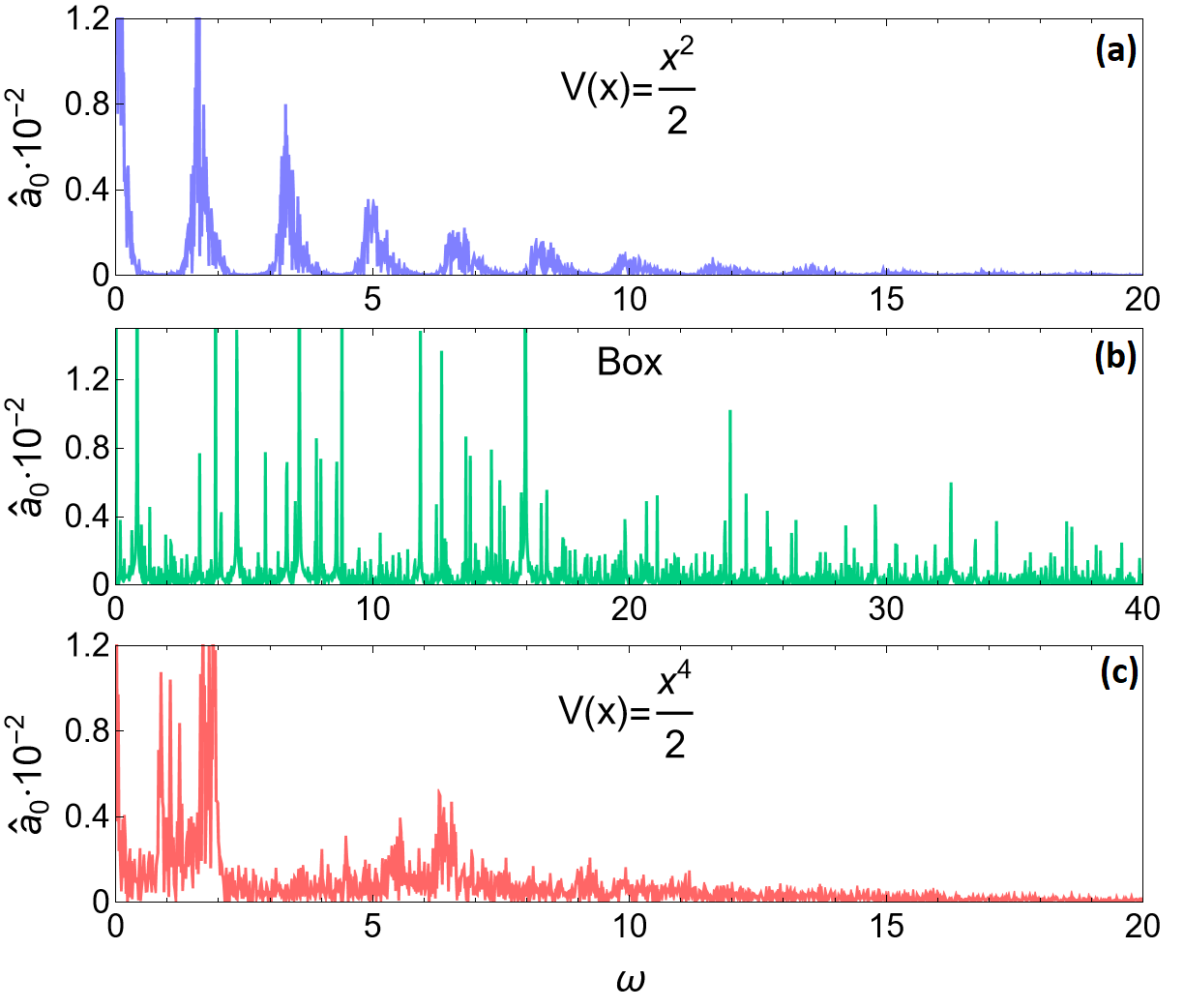}
\par
\vspace{-0.5cm}
\caption{The contrast between power spectra of the first-mode amplitude $%
\hat{a}_{0}$, defined according to Eq. (\protect\ref{eq:Fourier_transform}),
as produced by the numerical solution of the one-dimensional NLSE with the
HO potential (a), infinitely deep square potential well (b), and quartic
potential (c), initialized by the input with\ a random phase and amplitude.
Amplitudes of higher modes produce similar plots.}
\label{fig:harmonic_vs_box}
\end{figure}

A common approach to analytical and numerical studies of these problems
relies on perturbing an integrable equation by extra terms -- typically,
this is an external trap added to the nonlinear Schr\"{o}dinger equation
(NLSE) \cite{ErgodicBreakingTrap,GHD4}. Then, one explores consequences of
the integrability breaking in the perturbed model \cite%
{GHD4,GHD5,PreTherm0,ErgodicBreakingTrap,AtomicLosses,TDK}. One may,
however, wonder whether\emph{\ a mechanism other than integrability }exists
to produce essential deviations from ergodic signatures of non-integrable
dynamics. This question underlies the present work, leading to a class of
NLSEs including \textit{highly resonant} \textit{potentials} (HRP), namely,
ones that, for the linear Schr\"{o}dinger equation, yield equidistant
spectra of energy eigenvalues $E_{n}$:
\begin{equation}
E_{n}=an+b,  \label{eq:equidistant_linear_spectrum}
\end{equation}%
with integer $n$ and real constants $a$ and $b$. A commonly known example is
the harmonic-oscillator (HO) potential, whose equidistant spectrum is a
consequence of the hidden symmetry of the respective quantum Hamiltonian
\cite{Niederer}. Similarly, the equidistant structure of spectra of other
potentials is related to their symmetries \cite{Turbiner}.

The term \textit{highly resonant} reflects extreme abundance of resonances
in these systems. Indeed, the equidistant positioning of eigenvalues in Eq. (%
\ref{eq:equidistant_linear_spectrum}) ensures that the four-wave constraint,
$E_{n}+E_{m}-E_{l}-E_{j}=0$ with integers $n,m,l,j$, which is the resonance
condition for the cubic nonlinearity, reduces to a simple relation between
the integer numbers, $n+m-i-j=0$. It implies an infinite number of
resonances for any mode ($n=i+j-m$). It is shown below that the special
structure of energy eigenvalues (\ref{eq:equidistant_linear_spectrum}) has a
strong impact on the dynamics, producing a regime of non-ergodic evolution,
in contrast with the generic (non-equidistant) energy spectra. This
phenomenon is demonstrated, in particular, by the power spectra for the
cubic NLSE with the HO potential displayed in Fig.~\ref{fig:harmonic_vs_box}%
. In the case of generic trapping potentials, the system indiscriminately
excites a large range of frequencies, leading to ergodic (continuous and
unstructured) power spectra \cite{StochasticNLS}, as shown in Figs.~\ref%
{fig:harmonic_vs_box}(b) and (c), which correspond, respectively, to the
infinitely deep square well and an anharmonic potential. By contrast, HRPs,
in a parameter range spanning regimes in which the cubic nonlinearity may be
weak or strong, in comparison to the linear part of the NLSE, give rise to
unusually depopulated power spectra, in which the excited frequencies reside
in a \textit{\textquotedblleft comb-like"} arrangement of spikes, as shown
in Fig.~\ref{fig:harmonic_vs_box}(a). The comb-like spectra induced by HRPs
reveal an \emph{obstruction to ergodicity}, being drastically different from
the continuously distributed spectra created by the generic traps. This
conclusion is upheld by the similarity of the comb-like power spectra in
HRPs to the discrete power spectra which are a characteristic feature of the
integrable dynamics. The truly discrete spectra are associated with periodic
and quasi-periodic trajectories that the integrable dynamics track on the
surface of invariant tori in the phase space (with a very small share of the
invariant tori being destroyed by integrability-breaking perturbations,
according to the KAM theorem \cite{KAM}).

Our motivation to search for alternatives to exact integrability in
explaining non-ergodic behavior came from specific results for the 1D
Gross-Pitaevskii equation (GPE), which is a well-established model for the
dynamics of atomic Bose-Einstein condensates, based on the NLSE for the
mean-field wave function of the condensate \cite{GPE1,GPE2,GPE3}. It is
commonly known that the NLSE is integrable in the free 1D space \cite%
{NLS_Integrability_1,NLS_Integrability_2}, thus providing a good starting
point for the study of the integrability-ergodicity clash. The dynamical
behavior in the presence of an external trap, which breaks integrability
\cite{Yu_Boris}, has been addressed for non-equilibrium configurations \cite%
{StochasticNLS,OutEqui1,OutEqui2,OutEqui3,OutEqui4,OutEqui5,OutEqui6,OutEqui7}%
, coherent states in time-dependent traps \cite{TimeDepTraps0,TimeDepTraps1}%
, and propagation of a small number of solitons \cite%
{Soliton2,Soliton3,Soliton4,BorisQuasiIntegrability,Soliton1,Soliton5,Soliton6,Soliton7,Soliton8,Soliton9}
(see also Refs. \cite%
{OutEquiVVVV-3,OutEquiVVVV-2,OutEquiVVVV-1,OutEquiVVVV0,OutEquiVVVV1,OutEquiVVVV2,OutEquiVVVV3,OutEquiVVVV4,OutEquiVVVV5,Paredes,Paredes2}
for related models). Numerical works \cite%
{BorisQuasiIntegrability,Soliton2,Soliton3,Soliton4} suggested remarkable
contrast between the GPE with the HO potential, and the equation including
either anharmonic potentials or the infinitely deep potential box, which is
represented by zero boundary conditions at the box edges. In particular, a
single dark soliton trapped in the box potential displays a continuous power
spectrum, in consonance with ergodicity and indicating the emission of
radiation \cite{BorisQuasiIntegrability}. On the other hand, the evolution
of the dark soliton governed by the GPE with the HO potential gives rise to
a quasi-discrete power spectrum, reminiscent of discrete spectra associated
with the quasiperiodic dynamics of integrable systems \cite%
{BorisQuasiIntegrability}. The non-ergodic behavior of the 1D GPE with the
HO potential, as opposed to the apparent ergodicity maintained by other
potentials, is not restricted to the soliton motion, but also happens for
more generic initial conditions, such as random waves. As shown in Fig.~\ref%
{fig:harmonic_vs_box}, the evolution initialized by these configurations in
the case of the HO potential displays comb-like power spectra, while ergodic
ones (truly continuous and unstructured) are seen in case of the box and
quartic potentials. The specific shape of the power spectra supported by the
HO potential suggests the presence of an underlying mechanism constraining
the dynamics to a non-ergodic form. It was referred to as
``quasi-integrability" in Ref. \cite{BorisQuasiIntegrability}, because, as
said above, discrete spectra are a hallmark of integrable systems.

NLSEs with the HO potential display peculiar behavior which is not
restricted to 1D. In particular, in 2D there are analytical solutions
describing periodically modulated motion of a single-vortex \cite{BBCE1} and
multi-vortex configurations \cite{BBCE2,GGT,BEM}, as well as dark rings \cite%
{BBCE2}, as well as analytical and numerical manifestations of
Fermi-Pasta-Ulam recurrences \cite{BEM}. In Ref. \cite{BBE1}, the rich
structure exhibited by weakly nonlinear dynamics of the 2D GPE with the HO
potential was extended to a large family of related systems with similar
behaviors, and in Ref. \cite{Evnin}, it was connected to the presence of
breathing modes \cite{Pitaevskii_V1,Pitaevskii_V2}. Another setup where the
HO potential has shown quasi-periodic motions is the 1D quintic NLSE \cite%
{BBE2}.

The connection between the 2D GPE with the HO potential and other systems
with equidistant linear spectra subject to condition (\ref%
{eq:equidistant_linear_spectrum}), which were considered in Refs. \cite%
{BBE1,BBE2,Evnin}, is an incentive to find out whether the
quasi-integrability of the 1D GPE with the HO potential, established in Ref.
\cite{BorisQuasiIntegrability}, is an exceptional feature, or, on the
contrary, it is shared by a large class of NLSEs. To this end, we here
examine the role played by the potential and conclude that comb-like power
spectra similar to the one plotted in Fig.~\ref{fig:harmonic_vs_box}(a) are
displayed by NLSEs with HRPs, whose linear spectra of energy eigenvalues
take the form of Eq. (\ref{eq:equidistant_linear_spectrum}). On the other
hand, NLSEs with potentials that do not obey definition (\ref%
{eq:equidistant_linear_spectrum}) do not display comb-like spectra either,
even if the spectra admit resonances between some modes.

Our results suggest three essential implications. First, NLSEs including
HRPs constitute a broad class of models ranging from some of the most common
and physically relevant ones, such as the GPE with the HO potential in any
number of spatial dimensions, to more sophisticated potentials [e.g., the
one accounting for the ``superselection", see Eq. (\ref{eq:V1}) below] and
nonlinear terms. The availability of 2D and 3D models of this type is
particularly interesting for the experiment because they overcome
fundamental limitations inherent to studies of weakly broken integrable
dynamics. First, the perturbation theory applies, in the traditional form,
solely to 1D models \cite{Yu_Boris}. The second lifted limitation, which is
related to the first one, is that our models are not necessarily produced by
deformations of integrable equations. An example is the 1D quintic NLSE with
the HO potential, which features non-ergodic power spectra without proximity
to an exactly integrable equation (see details below). Finally, it is
relevant to stress that our results offer an example of how a linear
property, \textit{viz}., the equidistant linear energy spectrum (\ref%
{eq:equidistant_linear_spectrum}), may impose a fundamental constraint on
the full nonlinear dynamics, preventing the onset of ergodicity. For our
exposition of the results we mostly refer to two models, the 1D GPEs with
the HO and box potentials, which represent the HRPs and non-HRPs{},
respectively. Then, we explain how similar results are produced by other
potentials.

The rest of the paper is organized as follows. First, we introduce the setup
and make a direct comparison between the dynamics under the action of the HO
and box potential in Section II. Then, in Section III we develop an
analytical approximation for the power spectrum in the case of weak
nonlinearity, which makes it possible to explain differences between the
respective power spectra. Afterwards, in Section IV we show numerically how
the comb-like power spectrum depends on the magnitude and sign
(defocusing/focusing) of the nonlinear terms. This is followed in Section V
by the presentation of comb-like power spectra produced by \emph{eleven}
other HRP models, which provide a robust confirmation of the genericity of
our results. The paper is concluded, in Section VI, by a discussion of
prospects and implications of our findings. Some technical aspects of
numerical methods employed in this work are presented in Appendix.


\section{1D Gross-Pitaevskii equations with the harmonic-potential and box
potentials}

Throughout this paper, we use the 1D GPE with the cubic nonlinearity as the
main setup to illustrate the methods and results. In section~\ref%
{sec:Other_highly_resonant_equations}, we describe several other models,
related to the ones addressed here. The scaled form of the GPE, with time $t$
and coordinate $x$, is
\begin{equation}
i\partial _{t}\psi =-\frac{1}{2}\partial _{x}^{2}\psi +V(x)\psi +g|\psi
|^{2}\psi ,  \label{eq:1D-GPE}
\end{equation}%
where $V(x)$ is the potential, and $g$ the nonlinearity coefficient, with $%
g>0$ and $g<0$ representing the repulsive and attractive self-interactions,
respectively. This equation conserves the norm
\begin{equation}
M=\int_{-\infty }^{+\infty }|\psi |^{2}dx,  \label{eq:M}
\end{equation}%
and energy (Hamiltonian)
\begin{equation}
H=\int_{-\infty }^{+\infty }\left( \frac{1}{2}|\nabla \psi |^{2}+V(x)|\psi
|^{2}+\frac{g}{2}|\psi |^{4}\right) dx,  \label{eq:energy}
\end{equation}%
which includes the quadratic and quartic parts, associated with the linear
and nonlinear terms in Eq. (\ref{eq:1D-GPE}), respectively:
\begin{align}
& H_{2}=\int_{-\infty }^{+\infty }\left( \frac{1}{2}|\nabla \psi
|^{2}+V(x)|\psi |^{2}\right) dx,  \label{eq:E_L} \\
& H_{4}=\frac{g}{2}\int_{-\infty }^{+\infty }|\psi |^{4}dx.  \label{eq:E_NL}
\end{align}%
We fix the normalization by setting $M=1$ in Eq. (\ref{eq:M}). The equation will be studied in the full range from the weakly nonlinear
regime ($|g|\ll1$) to the strongly nonlinear one ($|g|\gg1$). As said above, the
latter case represents the situation in which the cubic term is large in
comparison with the linear ones, but higher-order nonlinear terms are still negligible.
Normally, such terms do not appear in the GPE, except for the specially designed
configuration, in which the cubic cross-attraction between two components of a binary
BEC is nearly compensated by the self-repulsion in each component, making it necessary
to consider the quartic self-repulsion, that represents effects of quantum fluctuations
around the respective mean-field states, thus giving rise to the \textit{quantum droplets}
\cite{Petrov,droplets-review}.

The HO and box potentials are our representative examples, chosen to
illustrate the differences between HRP{} and non-HRP{} cases, respectively:
\begin{equation}
\text{\textbf{HO}}:\ V(x)=\frac{1}{2}x^{2},\qquad \text{\textbf{box}}:\
\begin{cases}
0, & \text{for}\ x\in (0,L), \\
\infty , & \text{elsewhere,}%
\end{cases}
\label{eq:Potentials_HP_Box_SHAPE}
\end{equation}%
where the coefficient of the HO potential is fixed by scaling to be $1$, $L$
is the size of the box, and the Dirichlet boundary conditions $\psi
(t,0)=\psi (t,L)=0$ are implied in the latter case.

The linearized version of Eq. (\ref{eq:1D-GPE}) ($g=0$) gives rise to the
commonly known eigenvalues $E_{n}$ and eigenfunctions $f_{n}(x)$:
\begin{align}
\text{\textbf{HO:}}\ E_{n}=n+\frac{1}{2},& \ f_{n}(x)=\frac{H_{n}(x)}{\pi
^{1/4}\sqrt{2^{n}n!}}e^{-{x^{2}}/{2}},  \label{eq:eigensystem_HP} \\
\hspace{-3mm}\text{\textbf{box:}}\ E_{n}=\frac{\pi ^{2}(n+1)^{2}}{2L^{2}},&
\ \ f_{n}(x)=\sqrt{\frac{2}{L}}\sin \frac{\pi (n+1)x}{L},
\label{eq:eigensystem_Box}
\end{align}%
where $n\geq 0$ is the number of the bound state, and $H_{n}(x)$ are Hermite
polynomials. We fix $L={\pi }/\sqrt{2}$ for the box, to facilitate the
comparison of power spectra produced by the two models. The fact that the HO
potential belongs to the class of HRP{}s is determined by its commonly known
equidistant energy spectrum (\ref{eq:eigensystem_HP}), while the quadratic
spectrum (\ref{eq:eigensystem_Box}) clearly indicates that the box potential
belongs to the non-HRP{} class. It admits some resonances among its modes,
but much fewer than enabled by the equidistant spectrum.

In both cases, the sets of eigenstates $f_{n}(x)$ are used to rewrite the
solution to Eq. (\ref{eq:1D-GPE}) in terms of complex mode amplitudes $%
\alpha _{n}(t)$, defined so that
\begin{equation}
\psi (t,x)=\sum_{n=0}^{\infty }\alpha _{n}(t)f_{n}(x)e^{-iE_{n}t}.
\label{eq:mode_expansion_NLS_general}
\end{equation}%
The power spectrum of each amplitude was computed as
\begin{equation}
	\hat{a}_{n}(\omega )\equiv \mathcal{F}\left[ |\alpha _{n}(t)|^{2}\right]
	\label{eq:Fourier_transform}
\end{equation}where $\mathcal{F}$ stands for the Fourier transform. The
spectra are the main targets that we follow to observe the effect of the potential in the
underlying GPE, as motivated by Fig.~\ref{fig:harmonic_vs_box}. To produce $%
\hat{a}_{n}(\omega )$, we solve Eq. (\ref{eq:1D-GPE}) numerically, using the
schemes outlined in Appendix~\ref{appendix:Numerical_Methods}, and then
identify amplitudes $\alpha _{n}(t)$ as per a truncated version of Eq. (\ref%
{eq:mode_expansion_NLS_general}).

\begin{figure*}[t]
\centering	
\includegraphics[width=\columnwidth]{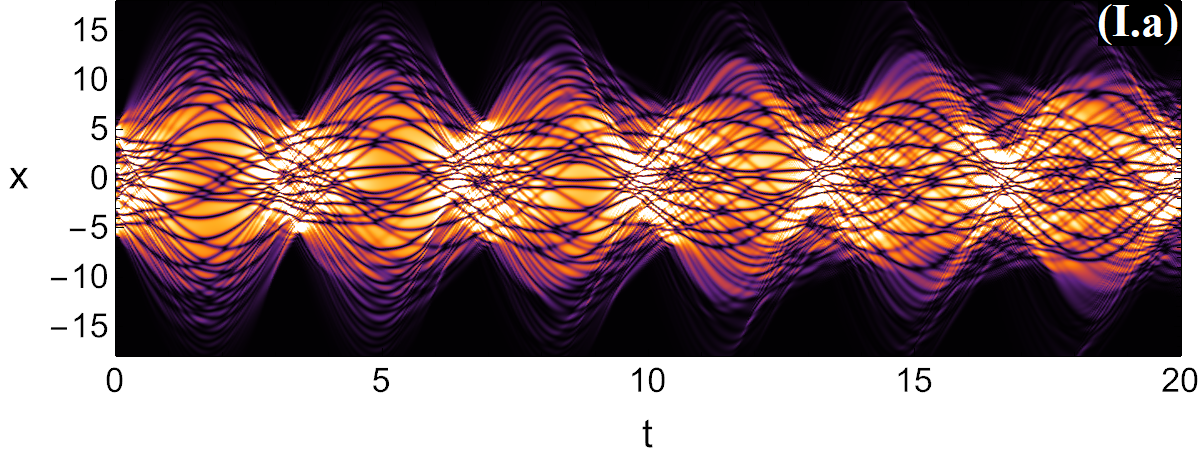}
\includegraphics[width=%
\columnwidth]{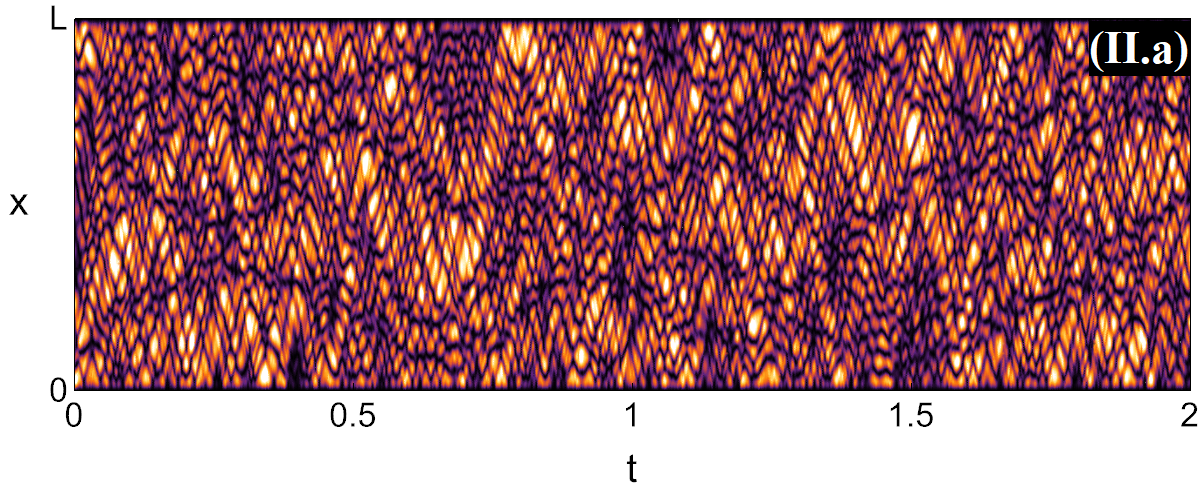} 

\includegraphics[width=\columnwidth]{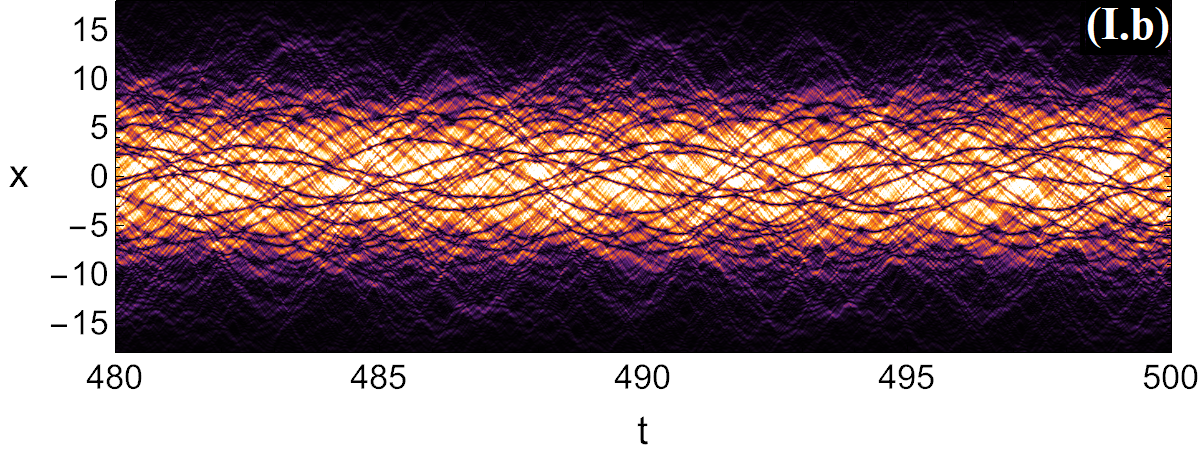}
\includegraphics[width=%
\columnwidth]{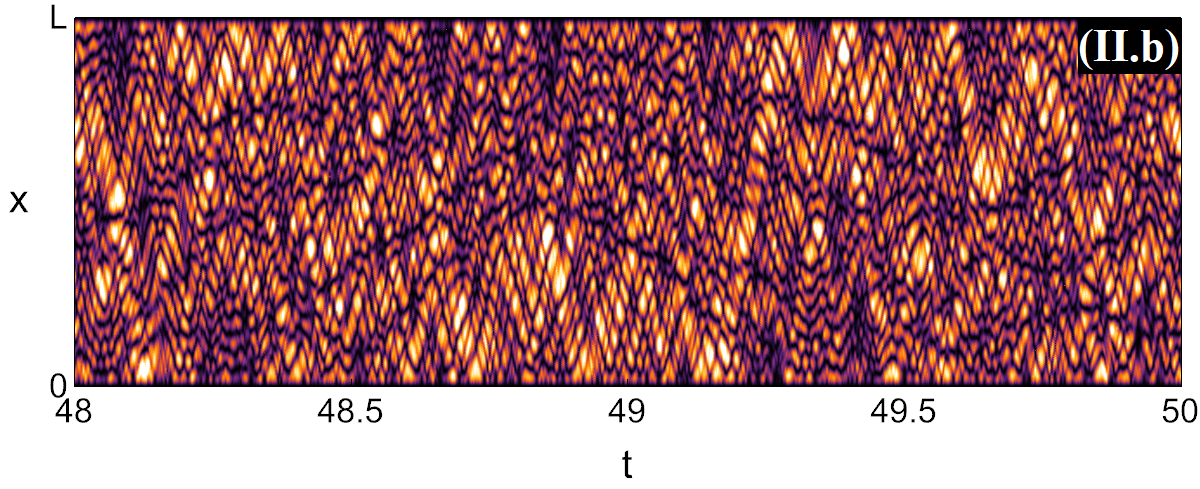} 

\includegraphics[width=\columnwidth]{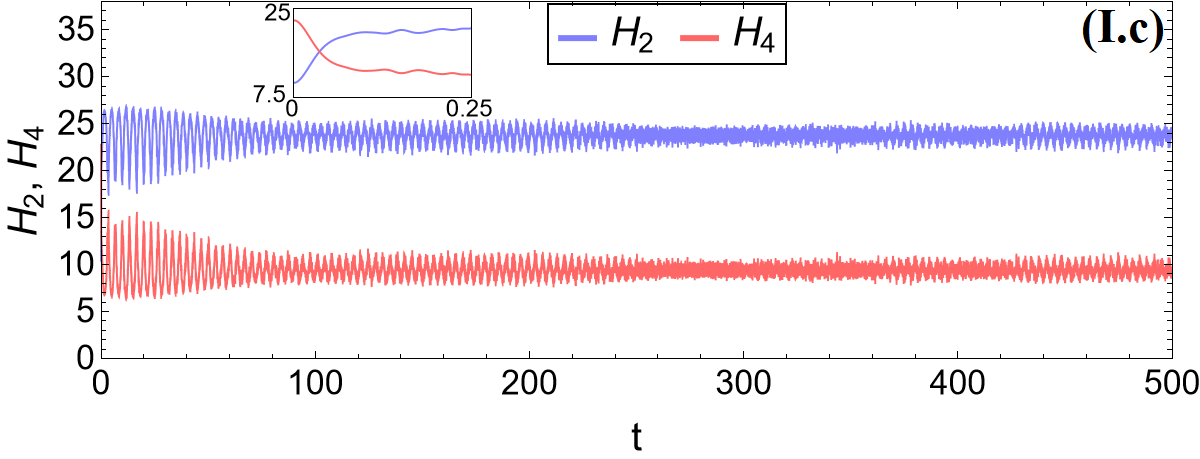}
\includegraphics[width=%
\columnwidth]{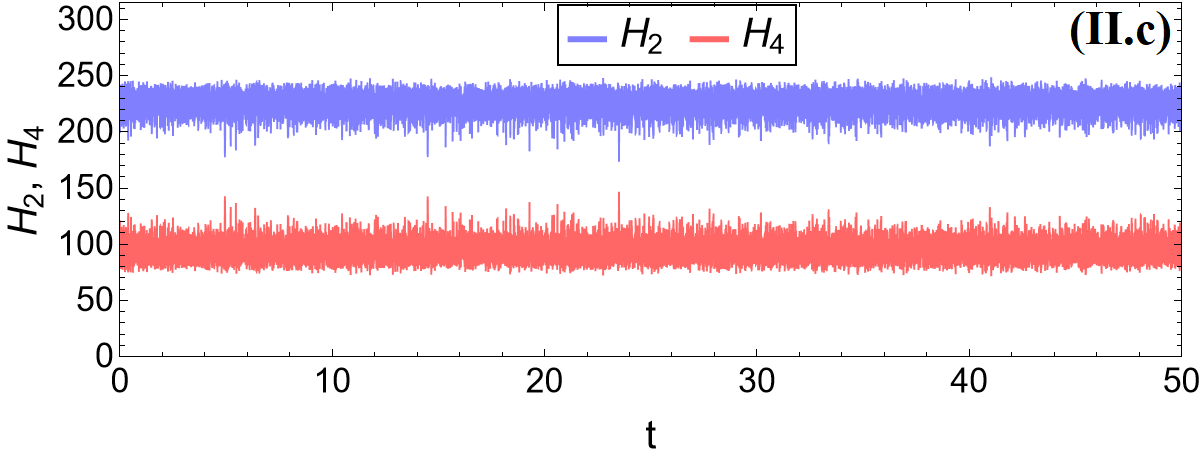} 

\includegraphics[width=\columnwidth]{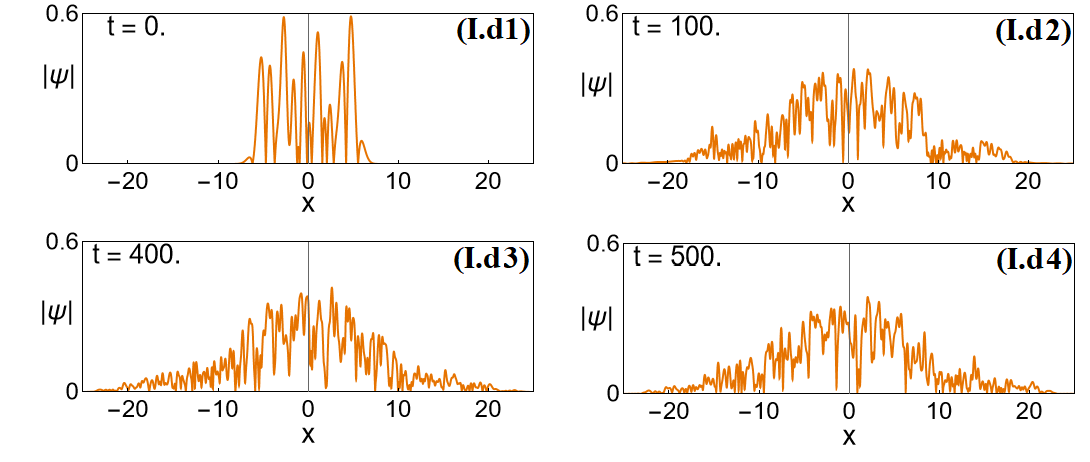}
\includegraphics[width=%
\columnwidth]{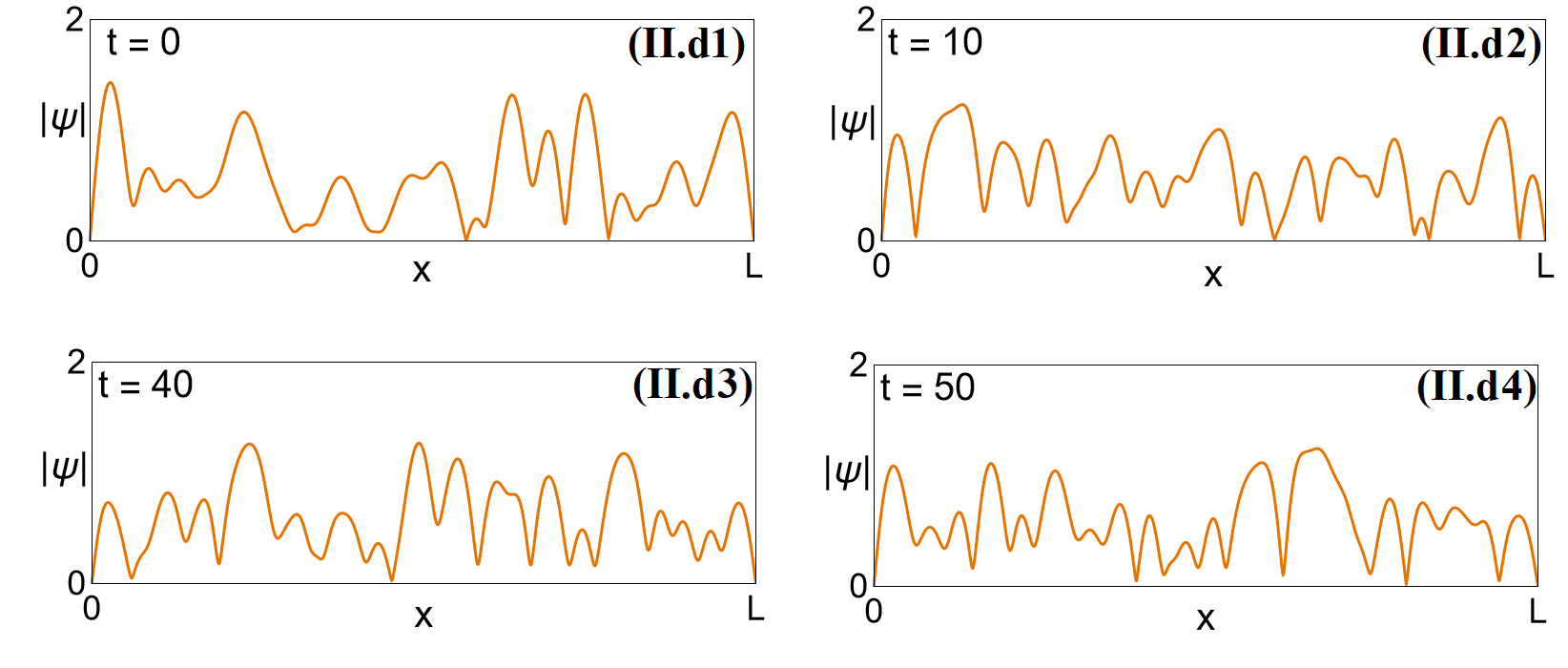} 

\includegraphics[width=\columnwidth]{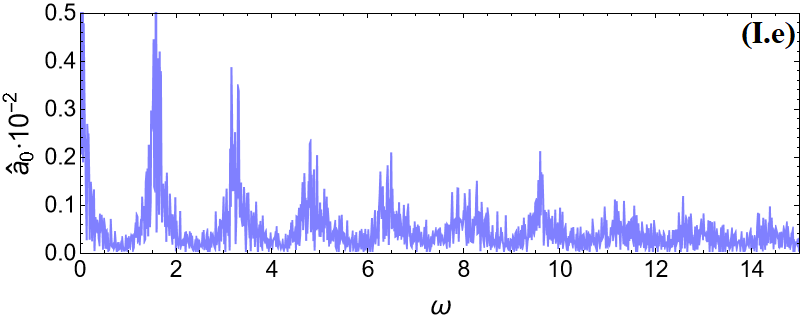}
\includegraphics[width=%
\columnwidth]{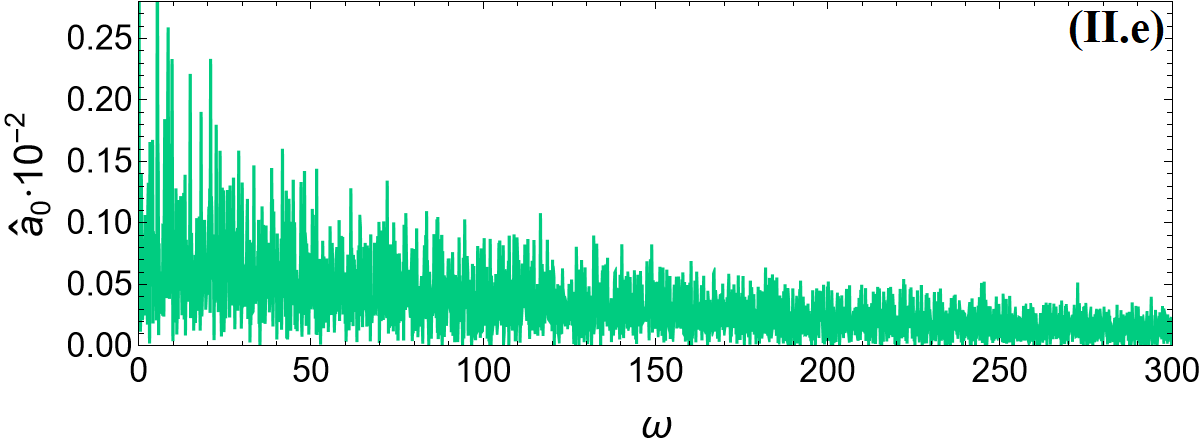}
\caption{The evolution of 1D defocusing random waves under the action of the
HO (left column, labeled I) and box-shaped potentials of size $L=\protect\pi %
/\protect\sqrt{2}$ (right column, labeled II), for a large nonlinearity
coefficient $g=250$ in Eq. (\protect\ref{eq:1D-GPE}). From top to bottom:
the initial stage of the spatiotemporal evolution (a); the evolution at an
advanced stage (b); the temporal evolution of the quadratic (\protect\ref%
{eq:E_L}) and quartic (\protect\ref{eq:E_NL}) parts of the energy (c); four
snapshots illustrating the shape of the profile in the course of the
evolution (d); and the power spectrum of the lowest-mode's amplitude, $%
\protect\alpha _{0}(t)$, (e), with higher modes displaying similar shapes.
The initial conditions are random waves prepared as per Eq. (\protect\ref%
{eq:Initial_Data_Random_Ncut_general}) with $\mathcal{N}=20$ and $\protect%
\beta =1$. Both cases, corresponding to the HO and box potentials, keep the
ratio $H_{4}/H_{2}\simeq 0.42$ at the established stage of the evolution.}
\label{fig:Evolution_1D-GPE_HP_Box}
\end{figure*}

As initial conditions we use waves prepared with random phases and
amplitudes, in the form of
\begin{equation}
\alpha _{n}(0)=%
\begin{cases}
\mathcal{A}_{n}e^{i\mathcal{P}_{n}} & \text{for }n\leq \mathcal{N} \\
\mathcal{A}_{n}e^{i\mathcal{P}_{n}}e^{-\beta (n-\mathcal{N})} & \text{for }n>%
\mathcal{N}%
\end{cases}
\label{eq:Initial_Data_Random_Ncut_general}
\end{equation}%
where $\mathcal{A}_{n}$ and $\mathcal{P}_{n}$ are random numbers uniformly
distributed in intervals $[0,1]$ and $[0,2\pi )$, respectively, $\mathcal{N}$
is the number of significantly excited modes, and $\beta >0$ determines the
suppression of higher modes. The value of $M$ is not fixed by $\mathcal{N}$ and $\beta$ in
(\ref{eq:Initial_Data_Random_Ncut_general}).
For this reason, the set of initial amplitudes $\alpha_{n}(0)$ is scaled so
as to satisfy the normalization, $M=1$. We use input (\ref%
{eq:Initial_Data_Random_Ncut_general}) because the exponential suppression
of the higher modes typically occurs in configurations arising in the course
of the dynamical evolution. Each realization of input (\ref%
{eq:Initial_Data_Random_Ncut_general}) features a different content of modes
and phases, yielding an adequate form of generic (``natural") initial
states. Therefore, they provide an appropriate arena for formulating generic
results. This approach brings in a broader perspective in comparison with
focusing on special solutions, such as single solitons. In this regard, our
simulations may actually be understood as the evolution of configurations
given by superpositions of a large number of dark solitons, corresponding to
notches in the pattern (the superposition also including other ingredients),
as Fig.~\ref{fig:Evolution_1D-GPE_HP_Box} suggests.

Random initial conditions similar to those defined by Eq. (\ref%
{eq:Initial_Data_Random_Ncut_general}) are used in studies of the wave
turbulence \cite{NazarenkoBook}, with the aim to produce a generic dynamical
picture, rather than focusing on specific solutions. In particular, the 1D
NLSE in a very broad box with periodic boundary conditions was used to study
the dynamics of random waves in integrable equations \cite%
{Randoux0,Randoux,Randoux2} (implementing the concept of the
\textquotedblleft integrable turbulence" introduced by Zakharov \cite%
{Zakharov2009}), the formation of rogue waves \cite%
{Zakharov,RogueWave2,RogueWave3}, etc.
Initial conditions of the same type have been also used in the
context of the 2D NLSE with a truncated HO potential in
connection with experiments on the light propagation in multimode optical fibers
\cite{Picozzi2011,Picozzi2019,Picozzi2020,Picozzi2023,Picozzi2023_V2}, and
in general, in the studies of optical wave turbulence \cite{ReviewOpticalWT}.
Thus, our use of random initial configurations in the presence of trapping
potentials follows the general framework adopted for the studies of
spatially confined random waves.

A detailed visualization of the evolution of random waves in the HO and box
potentials is produced, respectively, in the left and right columns of Fig.~%
\ref{fig:Evolution_1D-GPE_HP_Box}. In both cases, the evolution is affected
by the nonlinearity and broad wavelength spectrum of the initial excitation (%
$g=250$, $\mathcal{N}=20$, $\beta =1$).

Proceeding with the analysis, we first dwell on the case of the HO
potential. In this case, profile $|\psi (x)|$ is initially localized at the
center of the domain, exhibiting many notches. At the initial stage of the
subsequent evolution, the profile performs a sequence of alternating
expansion-compression cycles under the action of the HO potential (Fig.~\ref%
{fig:Evolution_1D-GPE_HP_Box}I.a), and then relaxes to a spread state (Fig.~%
\ref{fig:Evolution_1D-GPE_HP_Box}I.b) that keeps a nearly constant envelope
in time, together with a large number of notches shuttling from side to
side, resembling a gas of dark solitons \cite{Gennady}. The relaxation
process may be observed in Fig.~\ref{fig:Evolution_1D-GPE_HP_Box}I.c in the
evolution of the energy terms $H_{2}$ and $H_{4}$, defined as per Eqs. (\ref%
{eq:E_L}) and (\ref{eq:E_NL}). Their ratio, starting from $%
H_{4}/H_{2}\approx 2.2$ (inset in Fig.~\ref{fig:Evolution_1D-GPE_HP_Box}%
I.c.), initially oscillates with large amplitudes corresponding to expansion
and compression of the profile. After $t\approx 80$ the energy exchanges
significantly subside, with the energies oscillating around nearly constant
values in the course of the subsequent evolution, with the ratio $%
H_{4}/H_{2}\simeq 0.42$, which is essentially larger than in the weakly
nonlinear regime ($H_{4}/H_{2}\ll 1$). The power spectrum associated with
this evolution scenario features, in Fig.~\ref{fig:Evolution_1D-GPE_HP_Box}%
I.e, a comb-like shape similar to that exhibited above in Fig.~\ref%
{fig:harmonic_vs_box}(a). While one might assume that this shape originates
from the initial expansion-compression stage, the simulations are long
enough to guarantee the completion of the system's relaxation in the course
of $20\%$ of the total simulation time, while the established stage of the
evolution covers the remaining $80\%$ of the time. Moreover, omitting the
initial relaxation stage in the computation of the power spectrum, its shape
practically does not change.
The same happens if one performs extremely long simulations,
which also reveal the establishment of a comb-like structure, see
(Appendix~\ref{appendix:Long_Time}). As concerns the propagation of dark
solitons in the profile, Figs.~\ref{fig:Evolution_1D-GPE_HP_Box}I.a-I.b
exhibit their relatively smooth trajectories at both stages of the
evolution, the expansion-compression and established ones.

In the case of the box potential, Fig.~\ref{fig:Evolution_1D-GPE_HP_Box}II
shows that the random-phase-and-amplitude input (\ref%
{eq:Initial_Data_Random_Ncut_general}) fills the box from the beginning,
remaining in this state at all times. We have also explored the case where
the random-phase-and-amplitude input is localized at the center of the box.
In that case, following the initial expansion, the profile remains in the
spread state, without featuring expansion-compression cycles. In the course
of the evolution, the energies again keep the ratio $H_{4}/H_{2}\simeq 0.42$%
. Taking close-by values of this ratio in the cases of the HO and box
potentials is necessary, once the objective is to compare similar nonlinear
regimes. In spite of the proximity of the ratio $H_{4}/H_{2}\simeq 0.42$ in
both cases, the action of the box potential leads to the emergence of a
continuous (ergodic) power spectrum in Fig.~\ref{fig:Evolution_1D-GPE_HP_Box}%
, in contrast with its comb-shaped counterpart for the HO potential. It is
also worthy to note a significant difference in the range of excited
frequencies in the respective power spectra. We stress that the difference
from the case of the HO potential is not a mere consequence of the mismatch
in the box size, because we have set $L={\pi }/\sqrt{2}$ above precisely
with the purpose to match the linear energy spectra of both systems [$\left(
E_{n}\right) _{\mathrm{HO}}=n+{1}/{2}$ and $\left( E_{n}\right) _{\mathrm{box%
}}=(n+1)^{2}$], and, as we show in the next section, this value of $L$
provides matching of positions of the excited frequencies in the power
spectra of both systems. The ergodicity in case of the box-shaped potential
has been previously observed in Ref. \cite{StochasticNLS} for initial
conditions that expand from the center, in agreement with our inference that
there is no essential difference from the long-time evolution initialized by
the input filling the entire domain. Similar to the HO potential, it is
observed that dark solitons propagate throughout the box, but they do not
follow smooth trajectories even in its interior, because of the multiple
collisions between them, and shapes of individual solitons are identified
less clearly than in the case of the HO potential

We have also tested the presence of ergodicity in the case of non-HRPs{}
whose generically shaped spectra of energy eigenvalues do not admit
resonances. For instance, the 1D quartic potential, $V(x)=x^{4}/2$, is a
non-HRP{} one, as shown by lowest eigenvalues numerically computed with
accuracy $\Delta E_{n}\sim 10^{-4}$:
\begin{equation}
\begin{matrix}
E_{0}=0.5302,\  & E_{1}=1.8998,\  & E_{2}=3.7278, \\
E_{3}=5.8224,\  & E_{4}=8.1309,\  & E_{5}=10.6192, \\
E_{6}=13.2642,\  & E_{7}=16.0493,\  & E_{8}=18.9615.%
\end{matrix}%
\end{equation}%
In this case, the input provided by random waves gives rise to an initial
expansion-compression stage before relaxing to a spread state, apparently
similar to the dynamical scenario observed above under the action of the HO
potential, but the power spectrum is ergodic in the present case, see Fig.~%
\ref{fig:harmonic_vs_box}(c), like in the case of the box potential, cf.
Fig.~\ref{fig:Evolution_1D-GPE_HP_Box}II.e, in agreement with the general
picture outlined above.

\section{The analytical description in the weakly nonlinear regime}

\label{sec:Analytic_section}

In this section, we aim to provide an analytical form of the power spectrum
in the weakly nonlinear regime, with $|g|\ll 1$ in Eq. (\ref{eq:1D-GPE}), in
which the difference in the emergence of comb-like or ergodic spectra in
HRPs{} and non-HRPs{} can be understood explicitly. To do that, we again
address the 1D GPE with the HO and box potentials, which generate, as
mentioned above, the following commonly known equidistant and quadratic
spectra:
\begin{equation}
\text{\textbf{HO:}}\ E_{n}=n+\frac{1}{2};\quad \text{\textbf{box:}}\ E_{n}=%
\frac{\pi ^{2}}{2L^{2}}(n+1)^{2}.  \label{eq:Eigenvalues_HP_Box}
\end{equation}%
First, we are going to demonstrate that both potentials produce, in the case
of weak nonlinearity, a comb-like power spectrum composed of ``slender"
peaks. After that, we show how the eigenvalues determine interactions
between the eigenmodes, and how the equidistant eigenvalues in the case of
the HO potential arrange the interactions in a way that helps to preserve
the comb-like spectrum as the nonlinearity strengthens. On the other hand,
we demonstrate that the deviation from the equidistant structure of the
spectrum in the case of the box potential is responsible for erasing the
comb-like spectral shape, already for moderately weak nonlinearity. The
extension of the analysis to generic HRPs{} subject to condition (\ref%
{eq:equidistant_linear_spectrum}) is presented in Appendix~\ref%
{appendix:Decay_Snk}, where we show that our arguments developed for the HO
potential apply to generic HRPs as well{}, safeguarding the preservation of
the comb-like power spectra. The arguments are independent of the sign of $g$%
, thus being valid for both the defocusing and focusing signs of the
nonlinearity. For this reason, $g$ means $|g|$ in this section.


\subsection*{A slender comb-like spectrum}

For our analysis, it is useful to rewrite the 1D GPE (\ref{eq:1D-GPE}) as a
system of equations for mode amplitudes $\alpha _{n}$. To do that, one has
to insert $\psi (t,x)$, written in the form of expansion (\ref%
{eq:mode_expansion_NLS_general}), in Eq. (\ref{eq:1D-GPE}), and project the
result onto eigenmodes $f_{n}(x)$. This results in a system of ordinary
differential equations for the evolution of the amplitudes,
\begin{equation}
i\frac{d\alpha _{n}}{dt}=g\sum_{m=0}^{\infty }\sum_{i=0}^{\infty
}\sum_{j=0}^{\infty }C_{nmij}\bar{\alpha}_{m}\alpha _{i}\alpha
_{j}e^{i\Delta _{nmij}t},  \label{eq:EQUATION_Modes}
\end{equation}%
where the bar stands for the complex conjugate,
\begin{equation}
\Delta _{nmij}\equiv E_{n}+E_{m}-E_{i}-E_{j}
\label{eq:DeltaE_nmij_expression}
\end{equation}%
are the resulting frequencies of the four-wave interaction, and with the
respective couplings constants,
\begin{equation}
C_{nmij}=\int_{-\infty }^{+\infty }f_{n}(x)f_{m}(x)f_{i}(x)f_{j}(x)dx.
\label{eq:interaction_coefficients}
\end{equation}%
Expressions (\ref{eq:EQUATION_Modes})-(\ref{eq:interaction_coefficients})
are valid for any trapping potential, including the HO and box ones, the
distinction being in the values of $\Delta _{nmij}$ and $C_{nmij}$, when one
inserts specific eigenvalues $E_{n}$ and eigenmodes $f_{n}(x)$ into the
expressions.

\begin{figure}[t]
\centering	
\par
\includegraphics[width=\columnwidth]{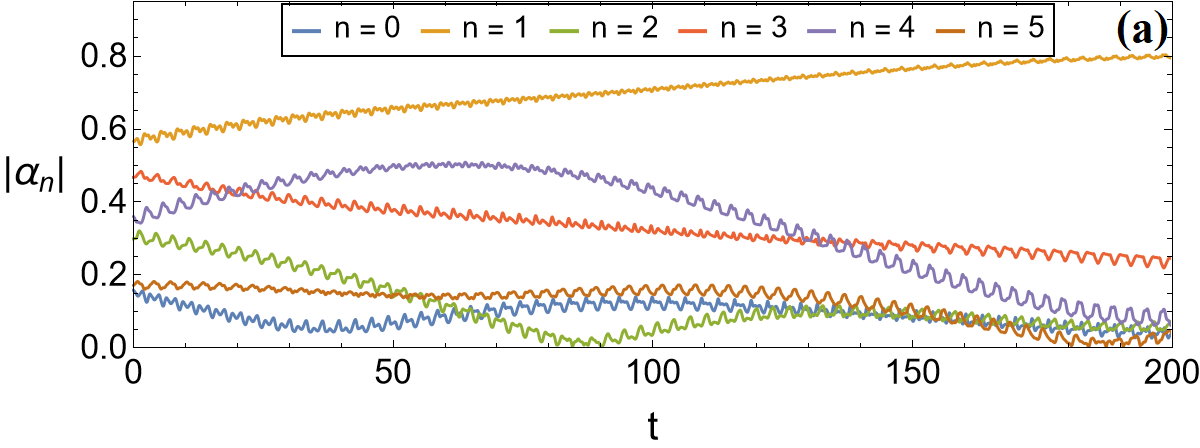}
\par
\includegraphics[width=\columnwidth]{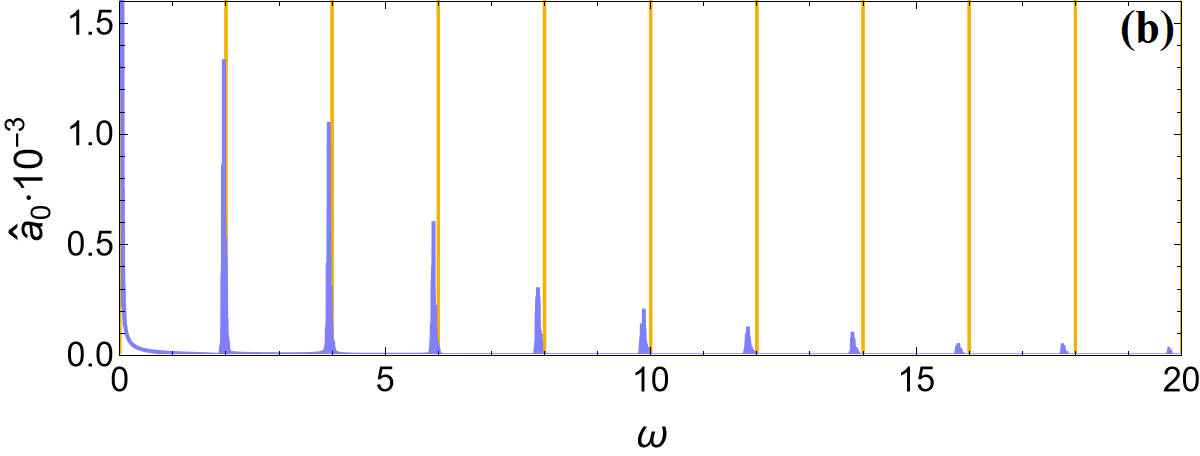}
\caption{The evolution of $\protect\alpha _{n}$ (a) and power spectrum of $|%
\protect\alpha _{0}|^{2}$ (b) governed by the 1D GPE with the HO potential
and defocusing sign of the nonlinearity. In (a) two constituents of the
evolution are observed: long-time modulations and small-amplitude
oscillations, which are associated with resonant and non-resonant
interactions, respectively. In (b) the effect of these terms on the power
spectrum of $|\protect\alpha _{0}|^{2}$ is observed. Vertical yellow lines
mark our analytic prediction, $\mathcal{W}_{2k}=2k$, for the location of the
excited frequencies in the case of the weak nonlinearity [see Eq. (\protect
\ref{2k})], which demonstrates very accurate agreement, up to a slight shift
originating from nonlinear corrections. These numerical results were
produced for $g=1$, to comprise the slow and fast constituents in the
evolution of $\protect\alpha _{n}$ in the framework of the same plot. The
picture demonstrates that the analytical prediction, originally obtained for
$g\ll 1$, works very well in this case too. }
\label{fig:Weakly_Nonlinear_Regime}
\end{figure}

Using Eqs. (\ref{eq:EQUATION_Modes}), we aim to demonstrate, first, that the
structure of the power spectrum is quite simple for the weak nonlinearity ($%
g\ll 1$). The equations give rise to two constituents of the evolution, as
seen in Fig.~\ref{fig:Weakly_Nonlinear_Regime}. On the one hand, there are
frequencies $\sim g$ and amplitudes $\sim 1$, which are associated with
resonances. On the other hand, there are contributions with small amplitude $%
\sim g$ corresponding to frequencies associated with non-resonant
interactions. The frequencies of the latter type are precursors of the
characteristic spikes in the comb-like spectrum which exist in the case of
strong nonlinearity, as shown in the next section. In view of their
relevance to the analysis, we introduce them by means of the following
definition.

\textbf{Definition:} \emph{$\mathcal{W}_{k}$ with $k\in \mathbb{Z}$
represent all different values of $\Delta _{nmij}$ with $n,m,i,j\in \mathbb{N%
}$, defined by Eq. (\ref{eq:DeltaE_nmij_expression}). and arranged in the
increasing order,}
\begin{equation}
\ldots <\mathcal{W}_{k-1}<\mathcal{W}_{k}<\mathcal{W}_{k+1}<\ldots \quad
\text{with}\quad k\in \mathbb{Z}.  \label{eq:W_k_definition}
\end{equation}%
\emph{When $\Delta _{nmij}$ take the same value for different sets of the
indices, there is single $\mathcal{W}_{k}$ associated with that value} (for
instance, $\Delta _{nnnn}=0$ for any $n$, hence there is single $k$ for
which $\mathcal{W}_{k}=0$.)


For generic systems, eigenvalues $E_{n}$ are irrational numbers, hence
frequencies $\mathcal{W}_{k}$ may form sets which are denser than the
underlying sets of eigenvalues $E_{n}$. As one proceeds to stronger
nonlinearity, further combinational harmonics arise, filling in frequency
axis still denser and leading to the emergence of generic continuous power
spectra. The situation is much more subtle for systems with integer
eigenvalues $E_{n}$, as is the case for the HO and box potentials, since $%
\mathcal{W}_{k}$ are then integers too. In this case, further analysis is
required to identify the shape of the power spectra, which arises from the
effect of the right-hand side (RHS) of Eq. (\ref{eq:EQUATION_Modes}) on the
evolution of $\alpha _{n}$. In the present context, two key ingredients are
prefactor $g$ and the complex exponential, which is an oscillatory term with
frequency $\Delta _{nmij}$ that vanishes in the resonant case, $\Delta
_{nmij}=0$. When $g$ is very small, the evolution splits in components
corresponding to the natural time scales, $t\sim \mathcal{O}(1)$, $\mathcal{O%
}(1/g)$, etc. \cite{Murdock, GHT}. For $t\sim \mathcal{O}(1)$, $\alpha _{n}$
remain constant up to nonlinear contributions of orders $\sim g$ and higher.
We focus on the contributions of order $g$ because they dominate in this
regime. On the one hand, resonant terms
with $\Delta _{nmij}=0$ generate contributions $\sim $ $gt$ (i.e., secular
terms in terms of the ``naive expansion" in powers of $g$), which induce
substantial modulations in $\alpha _{n}$ at times $t\sim \mathcal{O}(1/g)$,
(see the slow evolution of $|\alpha (t)|$ in Fig.~\ref%
{fig:Weakly_Nonlinear_Regime}). Therefore, such long-time modulations excite
frequencies $\sim g$ in the power spectrum. On the other hand, non-resonant
terms, with $\Delta _{nmij}\neq 0$, oscillate with frequencies $\Delta
_{nmij}$ (including corrections $\sim g$) and amplitudes $\sim g$ (see small
oscillations of $|\alpha (t)|$ in Fig.~\ref{fig:Weakly_Nonlinear_Regime}).
The latter terms excite frequencies $\Delta _{nmij}$ in the power spectrum
of $\alpha _{n}$ (with corrections $\sim g$), and have amplitudes $\sim g$.
From here, we conclude that the structure of the power spectrum in the
weakly nonlinear regime includes two kinds of excitation frequencies: those
determined by $\mathcal{W}_{k}$, and the frequencies forming a continuum in
a small region of width $\sim g$ around the origin. When the nonlinearity
strength grows, frequencies produced as combinations from these two sets
will emerge, being responsible for the broadening of the sharp peaks located
at various values of $\mathcal{W}_{k}$.

From the previous discussion, one can deduce the condition to display the
comb-like power spectrum in the regime of weak nonlinearity. This is just
the condition that $\mathcal{W}_{k}$ must be equidistant because the
spectrum is tightly localized around $\mathcal{W}_{k}$. The 1D GPE with the
HO and box potentials precisely satisfy this property because they give rise
to
\begin{equation}
\mathcal{W}_{2k}=2k  \label{2k}
\end{equation}%
and $\mathcal{W}_{2k}={\pi ^{2}}k/{L^{2}}$, respectively. This means that
both potentials give rise to a ``slender" version of the comb-like power
spectrum in the case of very weak nonlinearity. The expressions for $%
\mathcal{W}_{2k}$ follow from Eqs. (\ref{eq:Eigenvalues_HP_Box}) and (\ref%
{eq:DeltaE_nmij_expression}),
\begin{equation}
\Delta _{nmij}=(n+m-i-j)\quad \text{with}\quad n,m,i,j\in \mathbb{N},
\label{eq:DeltaE_nmij_HP}
\end{equation}%
\begin{equation}
\Delta _{nmij}=\frac{\pi ^{2}}{2L^{2}}\left[
(n+1)^{2}+(m+1)^{2}-(i+1)^{2}-(j+1)^{2}\right] ,  \label{eq:DeltaE_nmij_Box}
\end{equation}%
in the case of the HO and box potential, respectively. To make the structure
of expression (\ref{eq:DeltaE_nmij_Box}) more transparent, we set $m=i-1$, $%
j=n-1$, which yields $\Delta _{nmij}={\pi ^{2}(n-i)}/{L^{2}}$, so that any
integer is generated at times ${\pi ^{2}}/{L^{2}}$. We use index $2k$ in Eq.
(\ref{2k}), instead of $k$, to highlight the absence of interactions between
three modes with odd numbers and an even one, and vice versa, for parity
reasons [the respective couplings $C_{nmij}$ vanish according to Eq. (\ref%
{eq:interaction_coefficients}), hence $\mathcal{W}_{2k+1}$ are not present
in the power spectrum]. To ensure a meaningful comparison between the HO and
box potentials, we choose, as said above, $L={\pi }/\sqrt{2}$. Then, the
spike positions ($\mathcal{W}_{2k}$) in the power spectrum are the same for
the two cases in the weakly nonlinear regime. By means of such
identification of the frequency scales, a meaningful comparison is possible
between the HO and box-shaped potentials also for strong nonlinearity.


\subsection*{Departing from the weakly nonlinear regime}

It has been demonstrated above that the 1D GPE with the HO or box potentials
display a comb-like power spectrum for very weak nonlinearity. However, as
Figs.~\ref{fig:harmonic_vs_box} and \ref{fig:Evolution_1D-GPE_HP_Box} show,
this shape of the spectrum is not preserved in the case of the box-shaped
potential, turning into a generic ergodic spectrum with the increase of the
nonlinearity strength. Here, we aim to explain why, on the other hand, the
HO potential preserves the comb-like shape of the power spectrum even for
strong nonlinearity. We demonstrate that the key difference is due to the
linear and quadratic eigenvalue spectra (\ref{eq:Eigenvalues_HP_Box}) of
these systems. This is because the eigenvalues determine, through the
frequency combinations $\Delta _{nmij}$, which modes are involved in the
four-wave interactions, and then different structures of $\Delta _{nmij}$ in
Eqs. (\ref{eq:DeltaE_nmij_HP}) and (\ref{eq:DeltaE_nmij_Box}) produce
different predictions for the excitation of frequencies $\mathcal{W}_{k}$.
We show that, through this mechanism, equidistant eigenvalues produce a
strong suppression of large frequencies, while a large range of them are
excited in case of the quadratic eigenvalue spectrum in Eq. (\ref%
{eq:Eigenvalues_HP_Box}).

\begin{figure}[tbp]
\centering	
\includegraphics[width=\columnwidth]{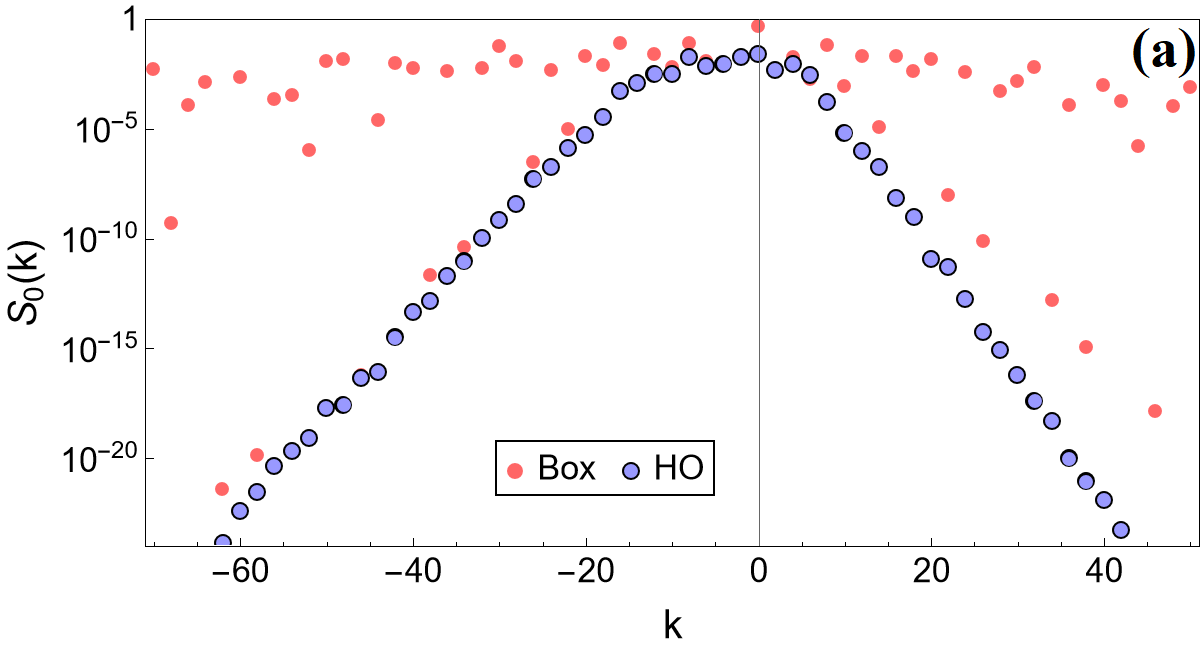}
\par
\includegraphics[width=\columnwidth]{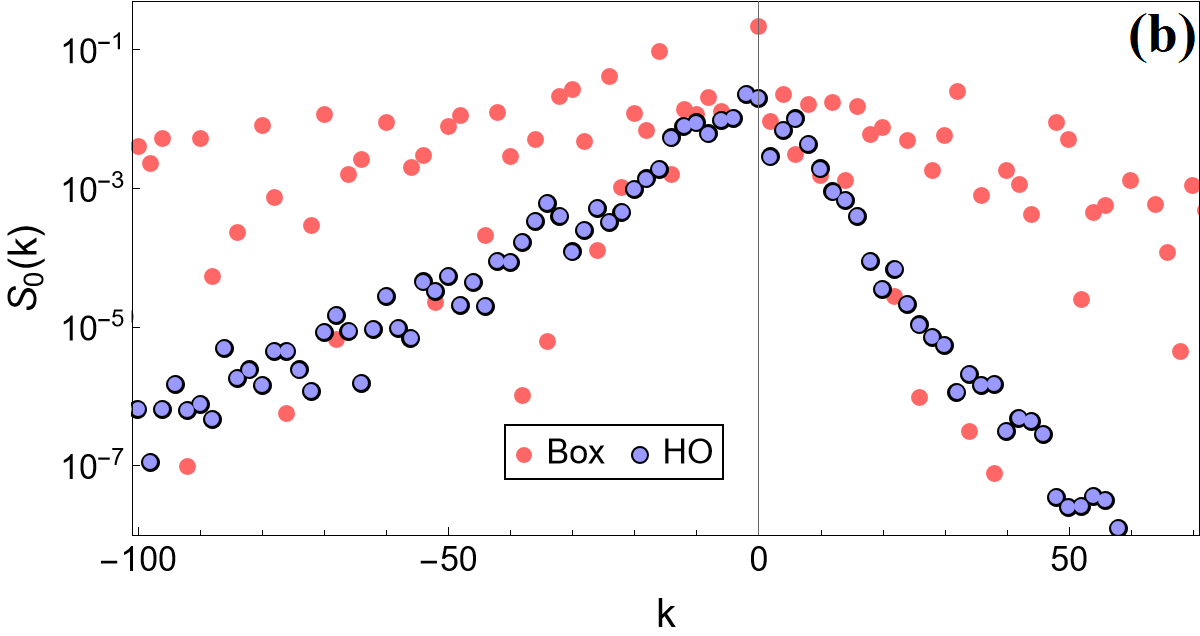}
\par
\includegraphics[width=\columnwidth]{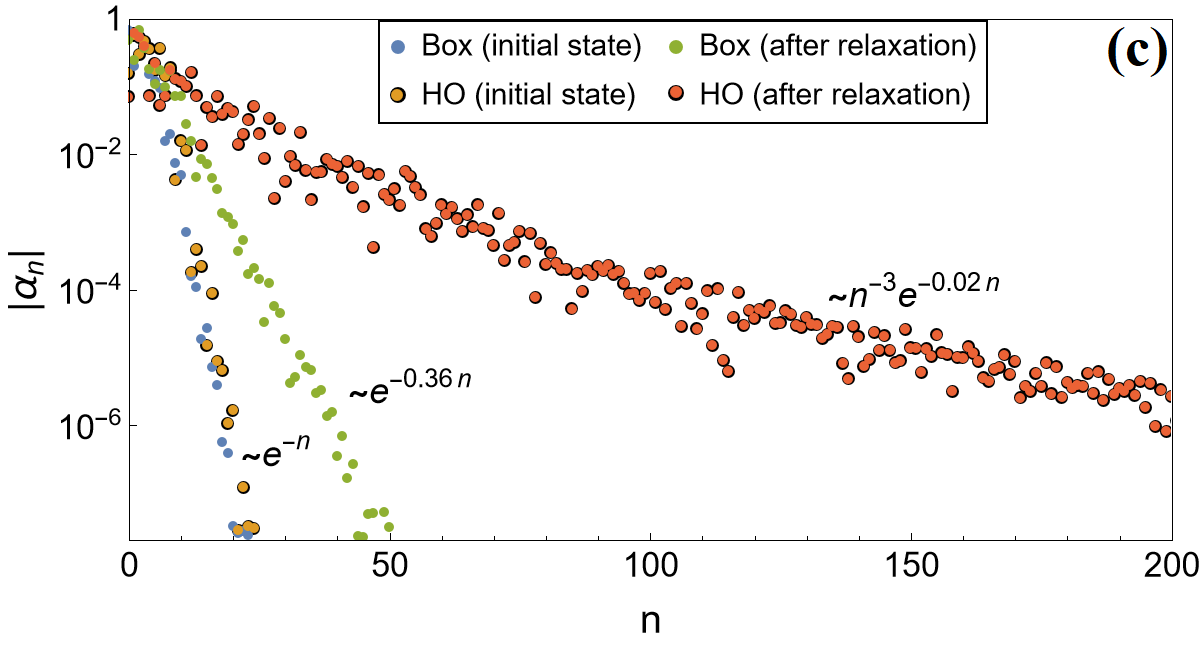}
\caption{The dependence of amplitude $\mathcal{S}_{0}(k)$ on $k$ in the
cases of the HO and box potentials, as produced by the numerical solutions
initialized by input (\protect\ref{eq:Initial_Data_Random_Ncut_general})
with random phases and amplitudes, for $\mathcal{N}=5$ and $\protect\beta =1$%
. (a) $\mathcal{S}_{0}(k)$, associated with the initial state, when $\protect%
\alpha _{n}$ feature the exponential decay $\sim e^{-n}$ in both systems.
(b) The same amplitudes after the relaxation of the systems, when $\protect%
\alpha _{n}$ demonstrate a stronger suppression with $n$ in the box ($\sim
e^{-0.36n}$) than in the case of the HO potential ($\sim n^{-3}e^{-0.02n}$),
while $\mathcal{S}_{0}(k)$ still decay faster in the latter case. (c) Values
of $|\protect\alpha _{n}|$ used in (a) and (b).}
\label{fig:Decay_Frequencies_Incoherent}
\end{figure}

To demonstrate this, one has to estimate the contribution of the $k$-th
frequency $\mathcal{W}_{k}$ to the $n$-th mode $\alpha _{n}$. For that
purpose, one gathers all terms oscillating with frequency $\mathcal{W}_{k}$
on the RHS of (\ref{eq:EQUATION_Modes}), writing the system of equations as
\begin{equation}
i\frac{d\alpha _{n}}{dt}=g\sum_{k=-\infty }^{\infty }\mathcal{S}_{n}(k)e^{i%
\mathcal{W}_{k}t},
\end{equation}%
\begin{equation}
\mathcal{S}_{n}(k)\equiv \underset{\Delta _{nmij}=\mathcal{W}_{k}}{%
\underbrace{\sum_{m=0}^{\infty }\sum_{i=0}^{\infty }\sum_{j=0}^{\infty }}}%
C_{nmij}\bar{\alpha}_{m}\alpha _{i}\alpha _{j}.  \label{eq:Snk_amplitude}
\end{equation}%
The ``sources" $\mathcal{S}_{n}(k)$ defined by Eq. (\ref{eq:Snk_amplitude})
determine the contribution of the $k$-th frequency $\mathcal{W}_{k}$ to the $%
n$-th mode $\alpha _{n}$. Numerical computations using values of $\alpha _{n}
$ extracted from our simulations reveal that $\mathcal{S}_{n}(k)$ decay with
$|k|$ considerably faster for the HO potential than the for its box-shaped
counterpart, as Fig.~\ref{fig:Decay_Frequencies_Incoherent} shows (except
for a few values of $k$ as, explained below). This picture is confirmed
analytically in Appendix~\ref{appendix:Decay_Snk} showing that the
amplitudes $\mathcal{S}_{n}(k)$ decay exponentially in the former case,
\begin{equation}
\left\vert \mathcal{S}_{n}(k)_{\mathrm{HO}}\right\vert <e^{-\beta
|k-n|}P_{n,k},  \label{eq:estimate_Snk_main_text}
\end{equation}%
while they exhibit a much slower decay for the box-shaped potential,
\begin{equation}
\left\vert \mathcal{S}_{n}(k)_{\mathrm{box}}\right\vert <e^{-\beta \sqrt{%
|k-(n+1)^{2}|}}D_{n,k},  \label{eq:estimate_Snk_BOX_main_text}
\end{equation}%
where $P_{n,k}$ and $D_{n,k}$ are polynomials in $n$ and $k$, and $\beta $
is a positive constant. To derive these results, we have used a
``phenomenological" analytical constraint for $\alpha _{n}$ that captures
the qualitative structure revealed by our simulations, see Fig.~\ref%
{fig:Decay_Frequencies_Incoherent}(c),
\begin{equation}
|\alpha _{n}|<p_{n}^{(s)}e^{-\beta n}\mathcal{A}_{n},
\label{eq:alpha_expression_polynomial_exponential_decay}
\end{equation}%
where $\beta >0$ is the same constant as in Eqs. (\ref%
{eq:estimate_Snk_main_text}) and (\ref{eq:estimate_Snk_BOX_main_text}), $%
p_{n}^{(s)}$ is a polynomial of degree $s\geq 0$, while $\mathcal{A}_{n}$ is
a random variable uniformly distributed in the interval of $[0,1]$.

Below, we explain that the difference between the HO and box potentials in
the decay of $\left\vert \mathcal{S}_{n}(k)\right\vert $ with $k$ has an
impact on the structure of the power spectrum in the cases of weak and
moderate nonlinearities, but, prior to that, we should clarify where this
difference comes from. One might conjecture that it is associated with the
couplings $C_{nmij}$, but the actual reason is the difference between the
equidistant (\ref{eq:eigensystem_HP}) and quadratic (\ref{eq:eigensystem_Box}%
) energy spectra, together with the rapid decay of $\alpha _{n}$ (\ref%
{eq:alpha_expression_polynomial_exponential_decay}). As we show in Appendix~%
\ref{appendix:Decay_Snk}, for HRPs{} satisfying condition (\ref%
{eq:equidistant_linear_spectrum}), such as the HO potential, and $\alpha
_{n} $ given by (\ref{eq:alpha_expression_polynomial_exponential_decay}), $%
\mathcal{S}_{n}(k)$ decays exponentially for large $|k|$, independent of
whether $C_{nmij}$ decay, remain constant, or grow with the increase of the
indices, while the quadratic spectrum, such as the one corresponding to the
box potential, features a much slower decay. The key point is in the
restriction on the indices necessary to get $\Delta _{nmij}=\mathcal{W}_{k}$
in Eq. (\ref{eq:Snk_amplitude}). Namely, fixing $k$, the modes involved in
the interactions that generate frequency $\mathcal{W}_{k}$ differ for the
spectra (\ref{eq:DeltaE_nmij_HP}) and (\ref{eq:DeltaE_nmij_Box}). In the
former case, large $k$ requires at least one high-order mode involved, while
in the latter case the quadratic eigenvalues make it possible to achieve
large $k$ easier, using low-order modes in most cases.

Thus, the exponential decay of high modes gives rise to the difference in
the magnitude of $\mathcal{S}_{n}(k)$.
The following examples illustrate this picture (Example I), and also explain the strong decay of some
amplitudes $\mathcal{S}_{n}(k)$ observed in the box in Fig.~\ref{fig:Decay_Frequencies_Incoherent} (Example II).

\textbf{Example I:} \emph{Frequency $\mathcal{W}_{35}$ contributes to $\alpha
_{0}$ via several combinations of modes $\{n,m,i,j\}$ in Eq.} (\ref{eq:Snk_amplitude})\emph{. For the sake of simplicity we use the following
expressions in this example:}
\begin{equation}
\alpha _{n}=e^{-n},\qquad \text{and}\qquad C_{nmij}=1,
\end{equation}\emph{while the conclusion is the same for other choices of $\alpha _{n}$
and $C_{nmij}$, as explained in Appendix~\ref{appendix:Decay_Snk}. We focus on
the largest contributions to $\mathcal{S}_{0}(35)$, which involves
the lowest possible modes, $\{0,35,0,0\}$ in the case of
the HO\ spectrum (\ref{eq:DeltaE_nmij_HP}), or $\{0,5,0,0\}$ in the case of the box spectrum (\ref{eq:DeltaE_nmij_Box}). Then, it follows from Eq. (\ref{eq:Snk_amplitude}) that the contribution of this interaction in the case of
the HO potential, $\bar{\alpha}_{35}\alpha_{0}\alpha_{0}=e^{-35}$, is many
orders of magnitude smaller than the one in the case of the box potential, $\bar{\alpha}_{5}\alpha _{0}\alpha _{0}=e^{-5}$, because they, respectively,
involve modes $m=35$ and $m=5$ to generate the same frequency $\mathcal{W}_{k}$.}

\textbf{Example II:} \emph{Reproducing the previous example, but with frequency $\mathcal{W}_{38}$ instead of $\mathcal{W}_{35}$, one finds that the largest contribution to $\mathcal{S}_{0}(38)$ in the box, $(\bar{\alpha}_{20}\alpha_{1}\alpha_{19}=e^{-40})$, is close to its counterpart in the case of the HO potential $(\bar{\alpha}_{38}\alpha_{0}\alpha_{0}=e^{-38})$. This difference from the common situation (see Fig.~\ref{fig:Decay_Frequencies_Incoherent}) happens because there are no relatively low-order modes that satisfy condition (\ref{eq:DeltaE_nmij_Box}) for special combinations of $(n,k)$. This is the explanation behind the strong suppression of a few
amplitudes $\mathcal{S}_{0}(k)$ in the box observed in Fig.~\ref{fig:Decay_Frequencies_Incoherent}. No essential contribution from these amplitudes is expected in the subsequent description of the population of the power spectrum because of their low presence and small values.}

For weak nonlinearity, the difference in the decay of amplitudes $\mathcal{S}%
_{n}(k)$ corresponding to the HO and box potentials has an impact on the
power spectrum because they determine the excitation of frequencies $%
\mathcal{W}_{k}$. The strong suppression of $\mathcal{S}_{n}(k)$ in the HO
case is translated into strong suppression of high frequencies $\mathcal{W}%
_{k}$ (rapid decay of peaks in the comb-like power spectrum), while the much
slower suppression of $\mathcal{S}_{n}(k)$ in the case of the box potential
facilitates excitation of higher frequencies (observed as spectral peaks at
higher frequencies). In the regime of moderate nonlinearity, amplitudes $%
\mathcal{S}_{n}(k)$ have an even stronger influence on the shape of the
power spectrum, as we aim to explain now. In this regime, the evolution of $%
\alpha _{n}$ no longer features solely two motions contributed to by
resonances ($\Delta _{nmij}=0$) and oscillations with frequencies $\mathcal{W%
}_{k}$, like in Fig.~\ref{fig:Weakly_Nonlinear_Regime}(a), as subdominant
oscillatory terms start to appear as relevant ones. Thereby, the equidistant
structure of $\mathcal{W}_{k}$ is no longer sufficient to maintain the
comb-like spectrum. Subdominant components emerge from the combination of
the resonant and non-resonant terms, as mentioned above. Namely, these
combined terms act as sources driving the generation of subdominant
components (similar to the usual principle that, in any perturbative
expansion, higher-order terms are sourced by lower-order ones). Frequencies
of the terms that emerge in this way result from combinations of $\mathcal{W}%
_{k}$ and those $\sim g$ around the origin. They produce contributions in
the power spectrum that slightly deviate from $\mathcal{W}_{k}$, broadening
in this way the ``slender" spikes in the power spectrum in the
weak-nonlinearity regime. This set of subdominant contributions is naturally
extended to higher orders in $g$, producing more and more frequencies in the
power spectrum which are originally sourced by $\mathcal{S}_{n}(k)$.
Therefore, the behavior of these amplitudes determines how the power
spectrum is populated when the system departs from the weakly nonlinear
regime. In the case of the box potential, we have demonstrated above that $%
\mathcal{S}_{n}(k)$ slowly decay with $|k|$ [see Eq. (\ref%
{eq:estimate_Snk_BOX_main_text})], thus giving rise to a broad range of high
frequencies $\mathcal{W}_{k}$, and triggering the rise of a large number of
high-frequency subdominant peaks, which dress the basic power spectrum with
a complicated structure. In this way, the comb-like spectral shape, which
exists in the weakly nonlinear regime, quickly gets destroyed, a spectral
tail of high frequencies arises, and individual peaks broaden considerably,
absorbing multiple combinational contributions arising from already excited
peaks. In the case of the HO potential, higher-order contributions are, of
course, produced as well, but the exponential suppression of
high-frequencies $\mathcal{W}_{k}$, as seen in Eq. (\ref%
{eq:estimate_Snk_main_text}), ensures that a majority of subdominant terms
are suppressed as well. This mechanism drastically reduces the number of
significant subdominant contributions the power spectrum receives,
preventing its ``wild" population and protecting its comb-like structure.
Note that, while our analysis is performed in the framework of the weakly
nonlinear regime, the picture produced by it correctly captures the shapes
of the power spectra for the strong nonlinearity, as observed in Figs.~\ref%
{fig:harmonic_vs_box} and \ref{fig:Evolution_1D-GPE_HP_Box}: a disordered
distribution of many spikes in the case of the box-shaped potential, and the
nearly equidistant array of spikes in the HO case, confined to the
low-frequency range.

Before concluding the analysis of the weakly nonlinear
regime, we aim to highlight differences between the present analysis and works
on quasi-periodic solutions and FPU recurrences in the 2D GPE with the HO potential.
In both cases, the modal decomposition (\ref{eq:EQUATION_Modes})
has been used, but Refs. \cite{BBCE1,BBCE2,BEM} focused
on resonant interactions, namely, long-time dynamics at $|g|\ll1$,
exploiting the specific structure of $C_{nmij}$ and neglecting non-resonant
interactions. On the other hand, our analysis considers both
resonant and nonresonant interactions, while the specific form of $C_{nmij}$
for each system was not used. It was done with the purpose of getting a
description of the power spectrum of HRPs in the weakly nonlinear regime,
that helps to guide the intuition for moderate values of $|g|$. It is worth
mentioning that analytic solutions obtained in Refs. \cite{BBCE1,BBCE2} are not
generic among the class of HRPs, as they rely on a special structure of
$C_{nmij}$ \cite{BBE1,Evnin}, although they have a significant presence
in the class of NLSEs \cite{BBCE1,BBCE2,BBE2,Evnin,BEF}. On the other hand, a part of
the study of FPU recurrences performed in \cite{BEM} relied on less restrictive property of $C_{nmij}$, leaving open the possibility that this kind of
dynamics is generic for NLSEs with HRPs.

\section{Numerical results in the fully nonlinear regime}

\label{secnum}


\begin{figure*}[t]
\centering	
\par
\includegraphics[width=4.2cm]{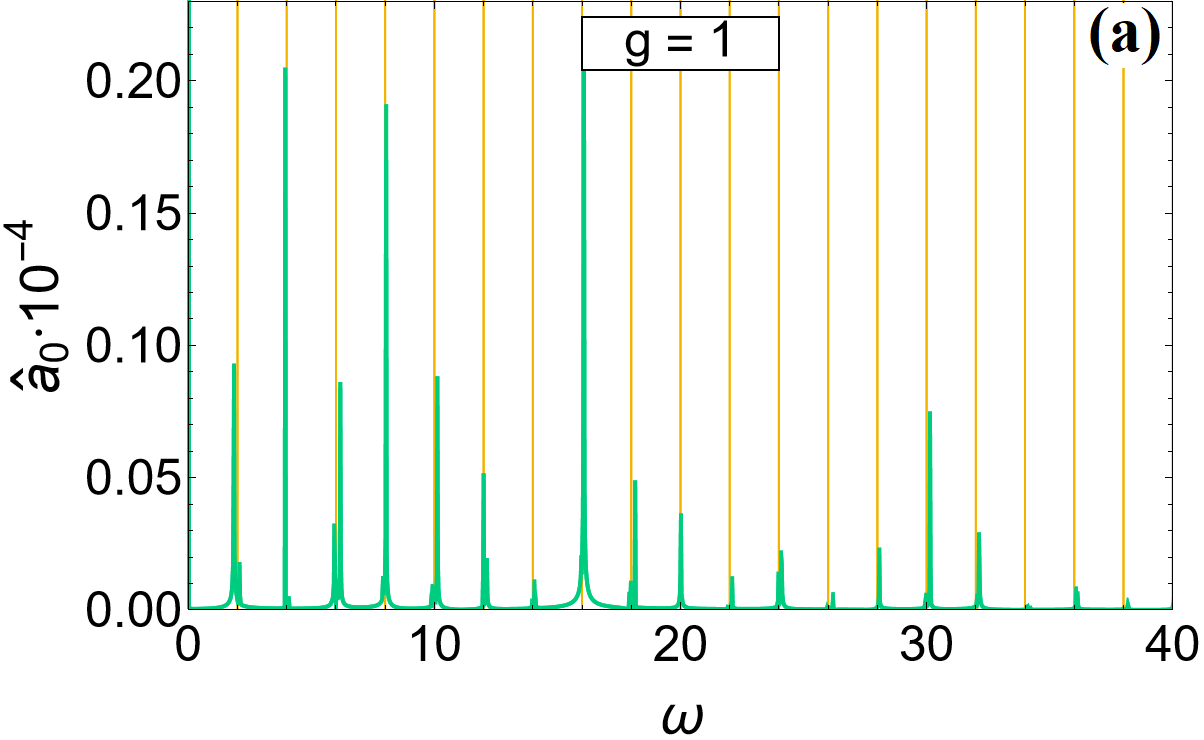} %
\includegraphics[width=4.2cm]{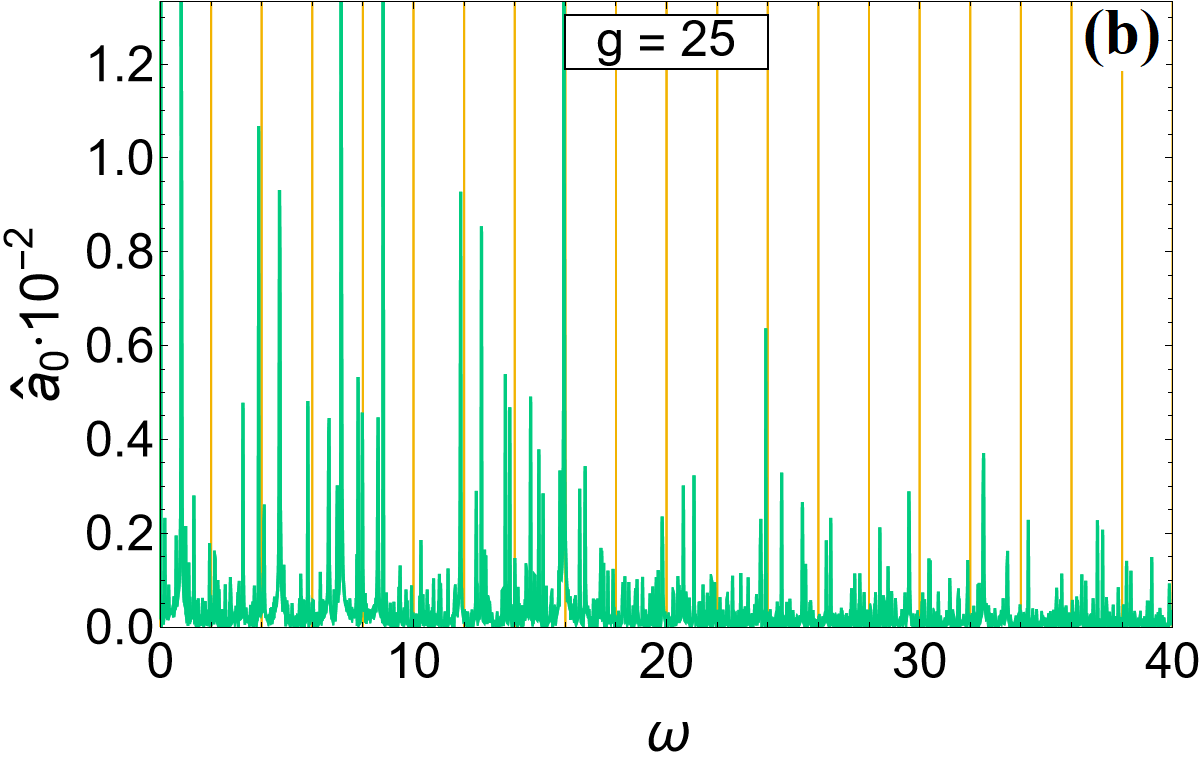} %
\includegraphics[width=4.2cm]{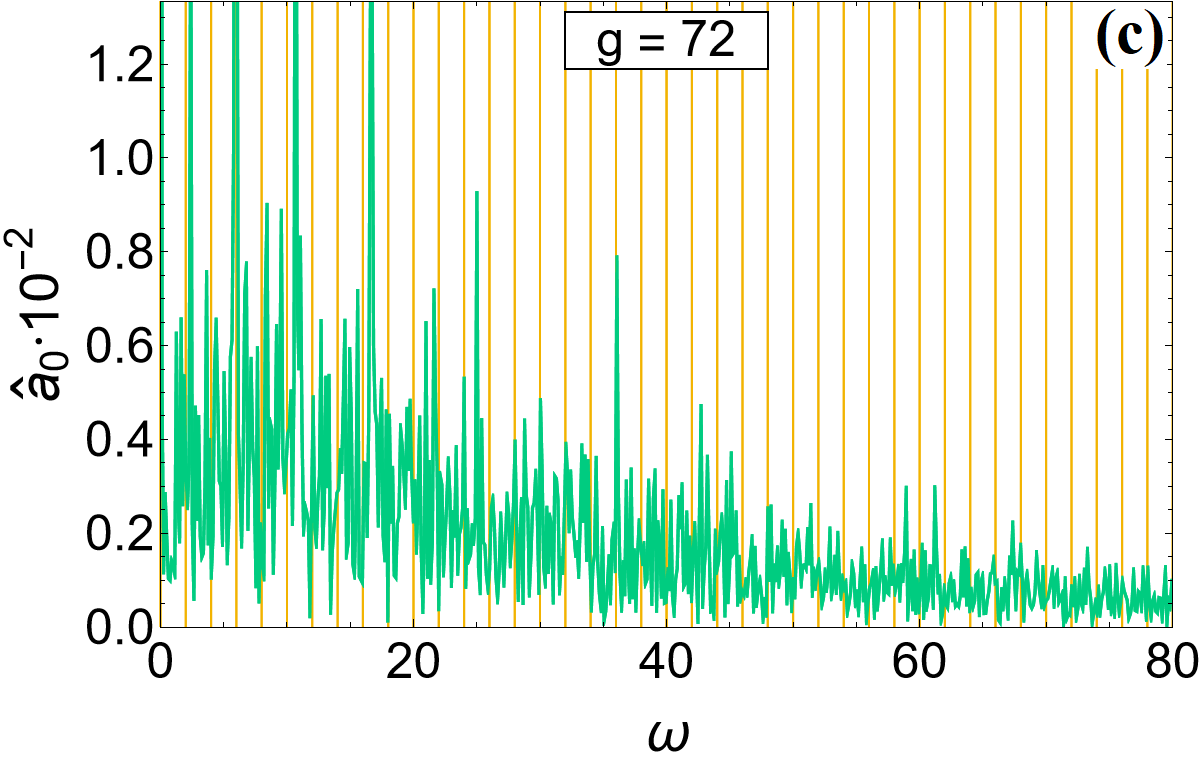} %
\includegraphics[width=4.2cm]{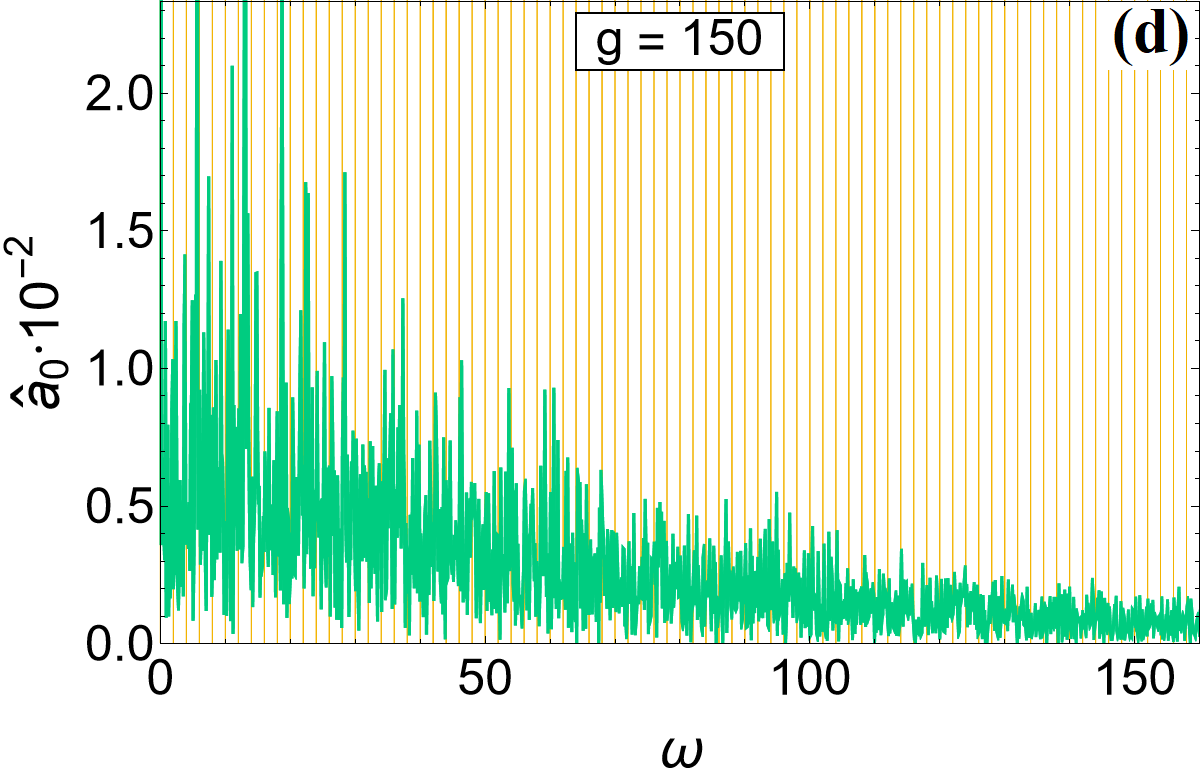}
\par
\includegraphics[width=4.2cm]{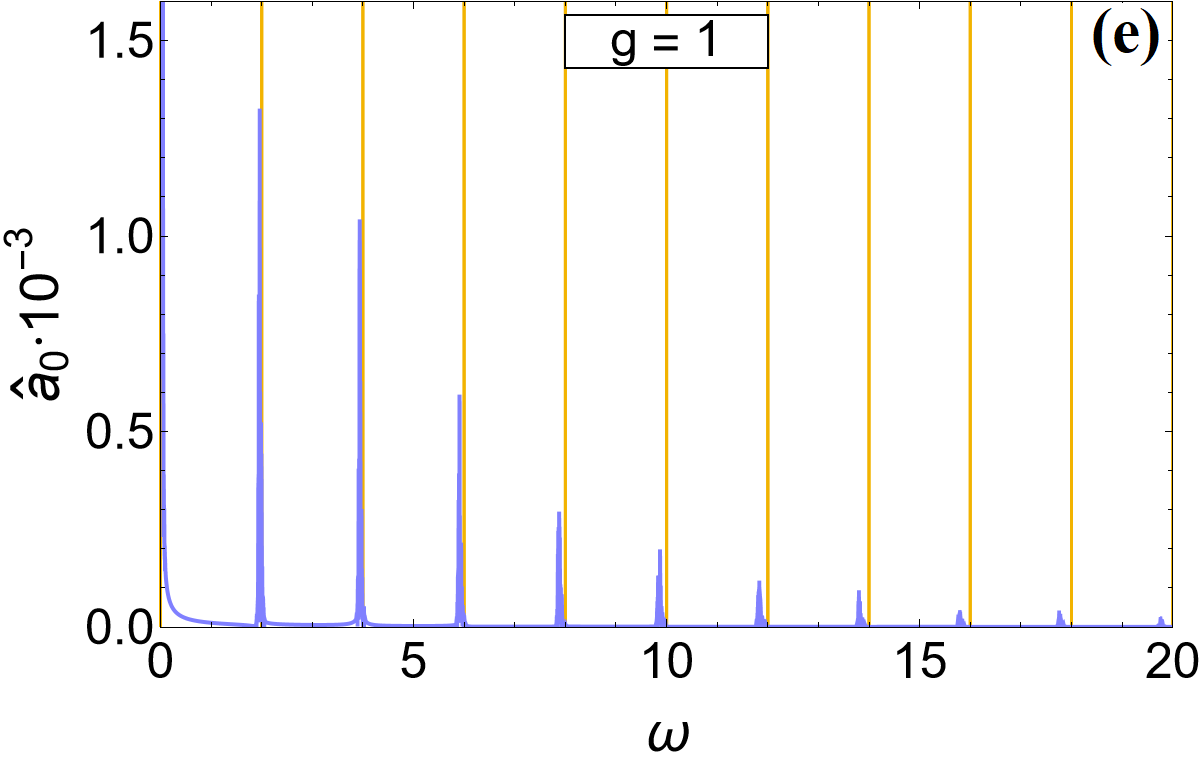} %
\includegraphics[width=4.2cm]{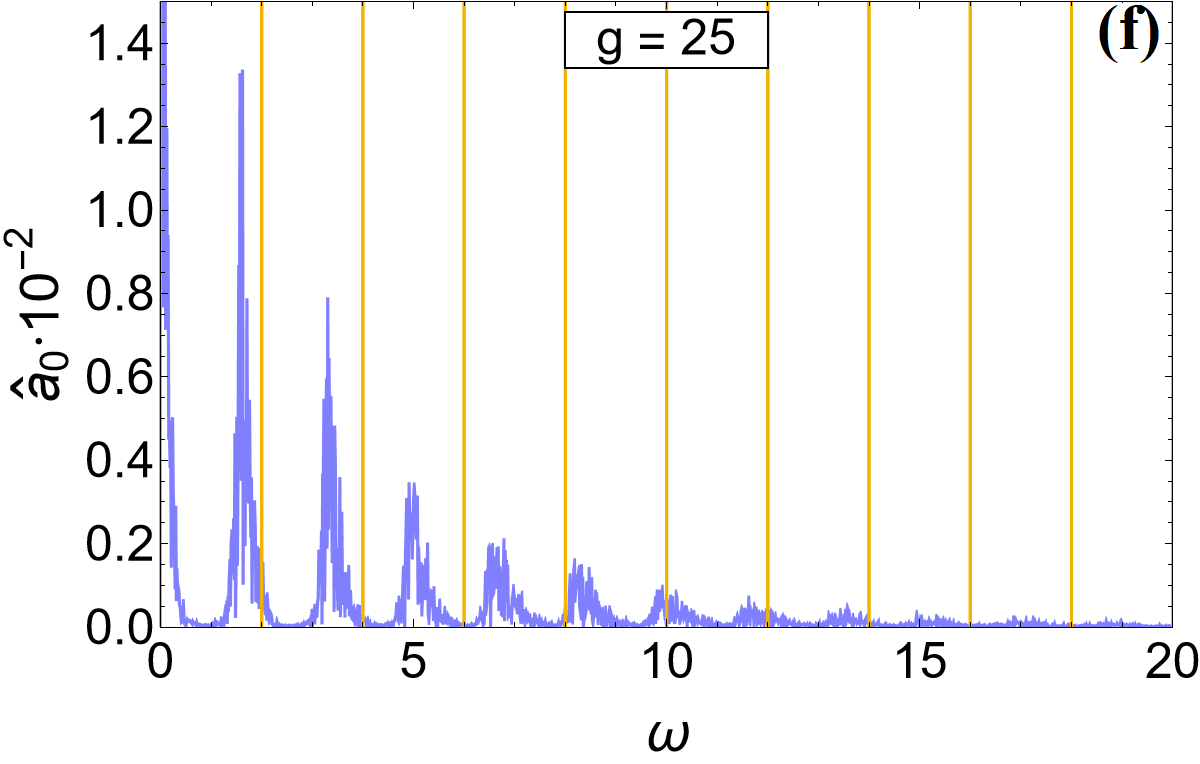} %
\includegraphics[width=4.2cm]{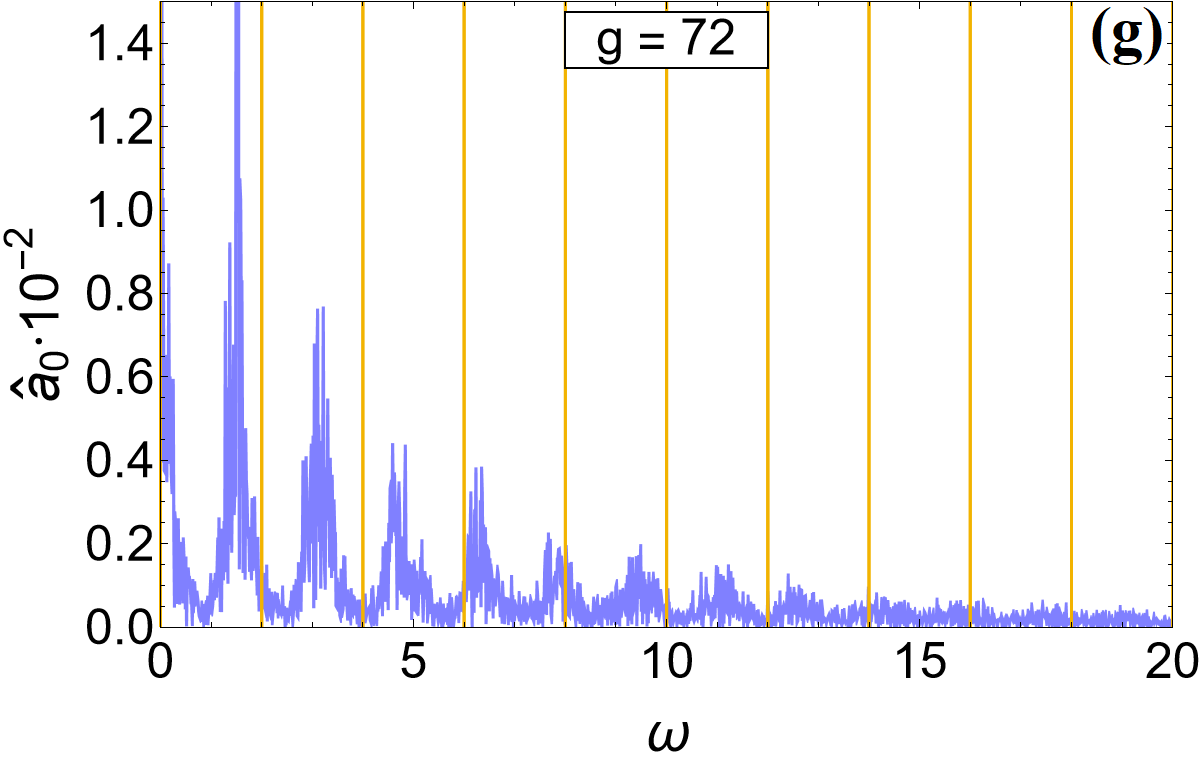} %
\includegraphics[width=4.2cm]{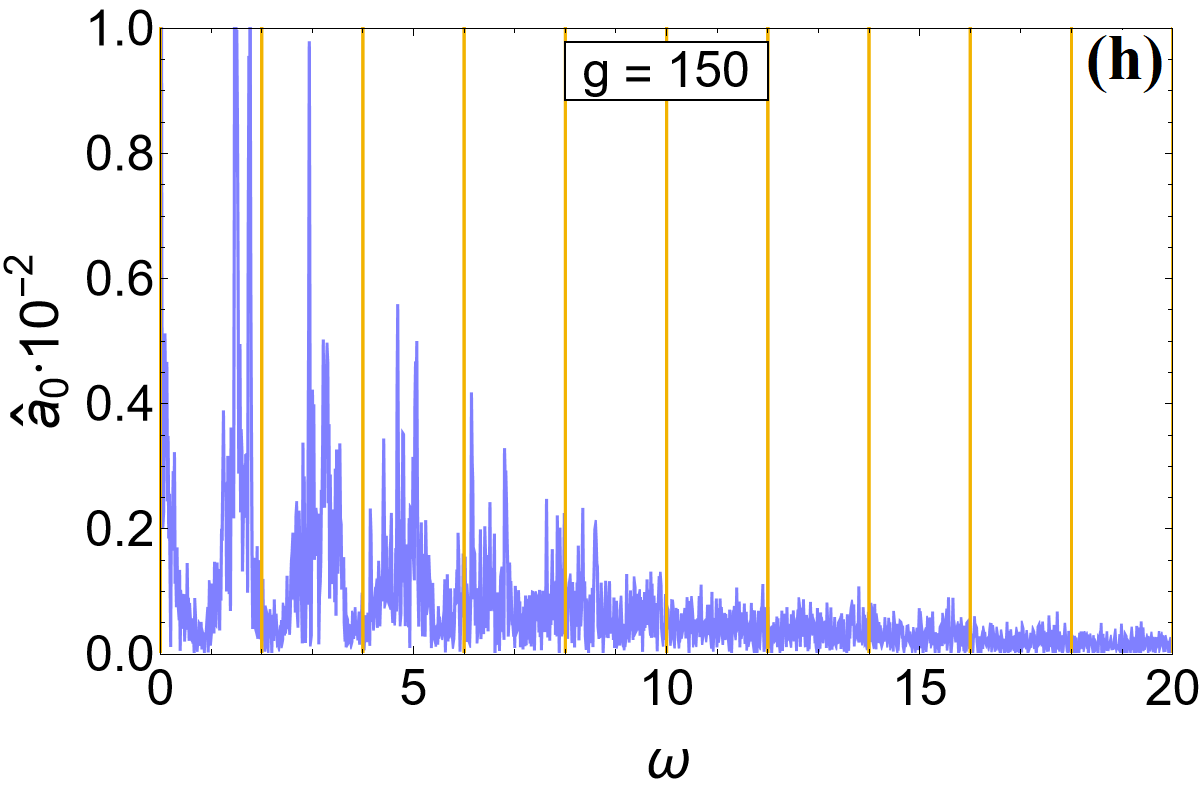}
\par
\includegraphics[width=4.2cm]{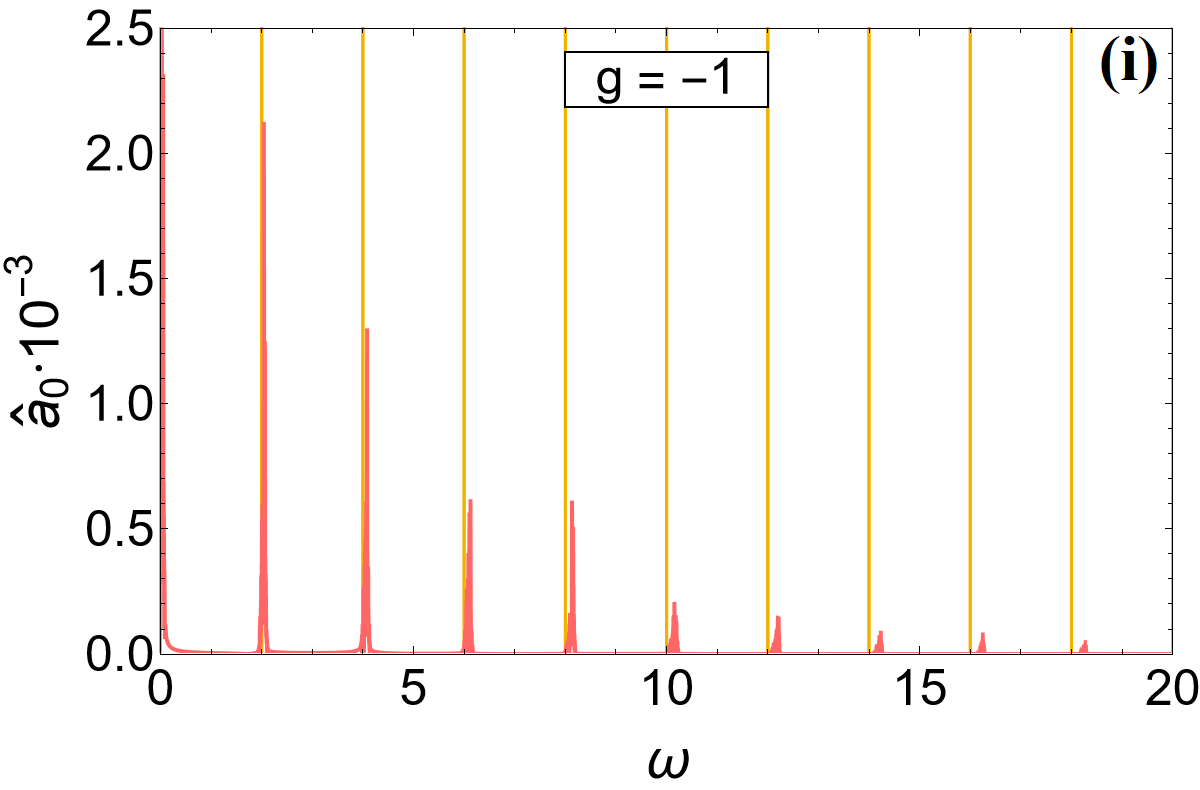} %
\includegraphics[width=4.2cm]{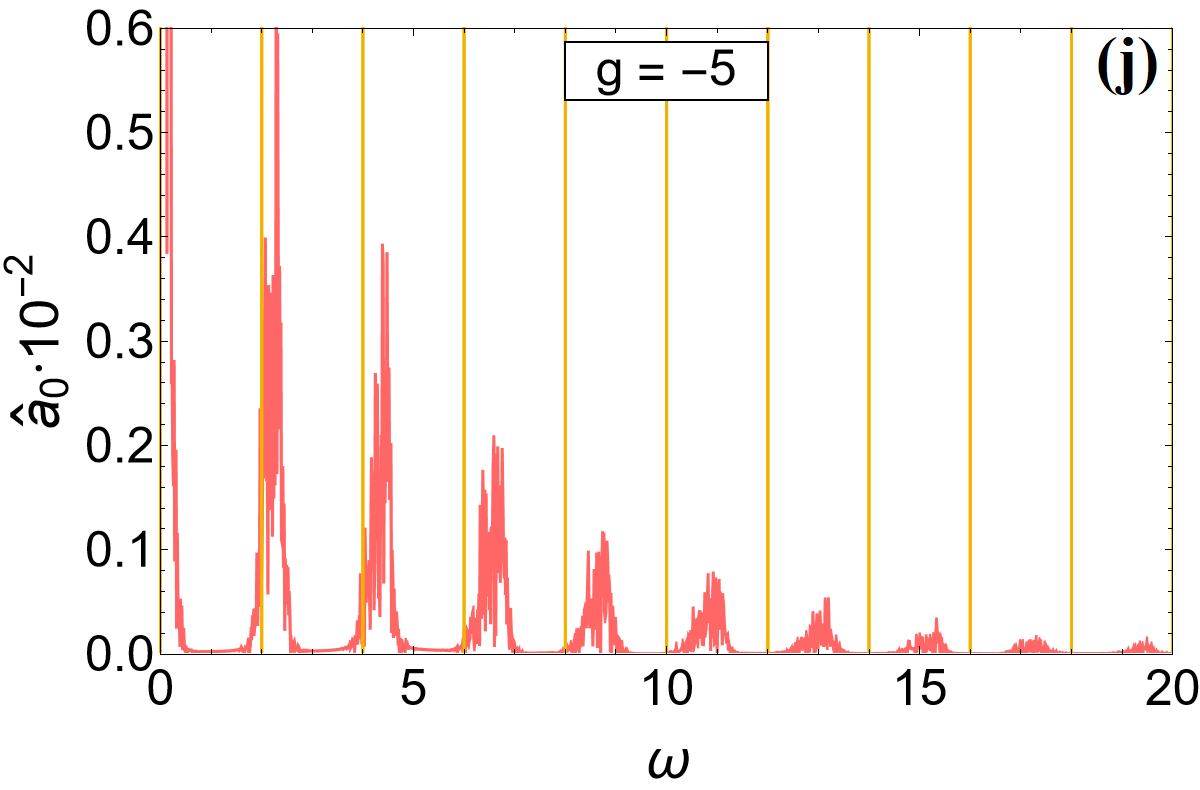} %
\includegraphics[width=4.2cm]{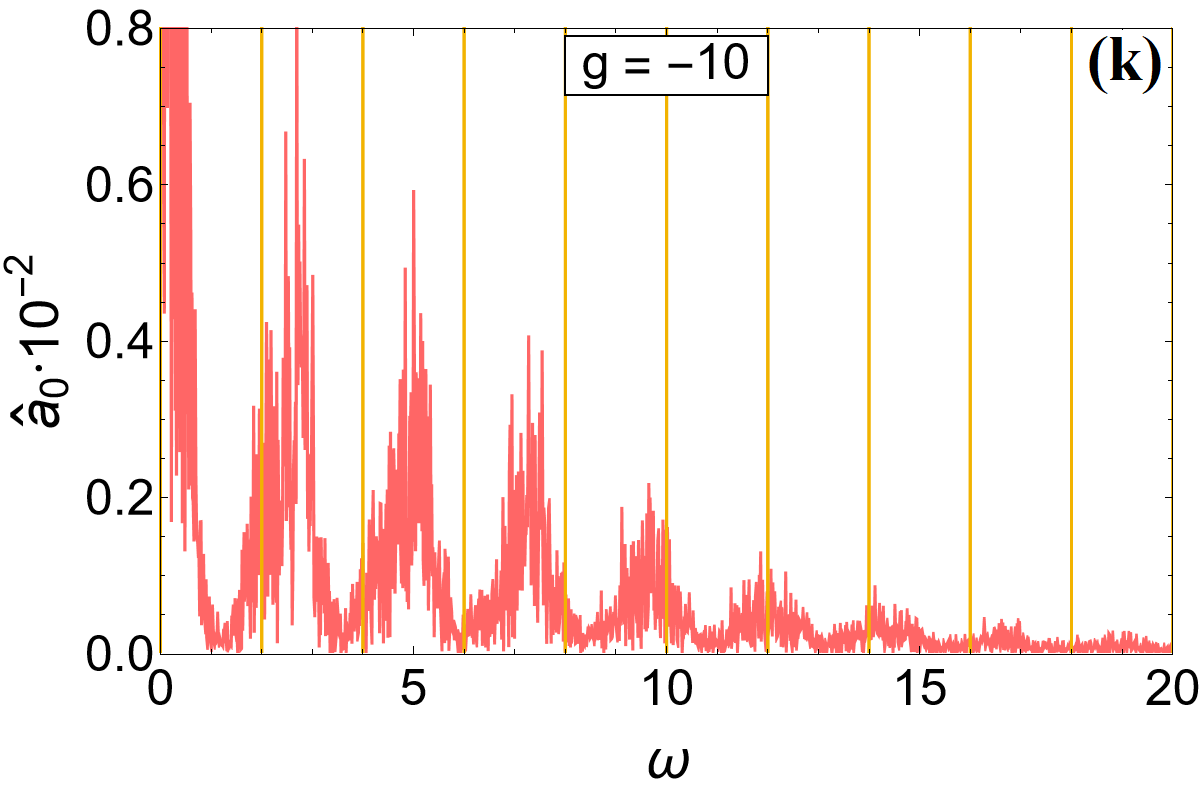} %
\includegraphics[width=4.2cm]{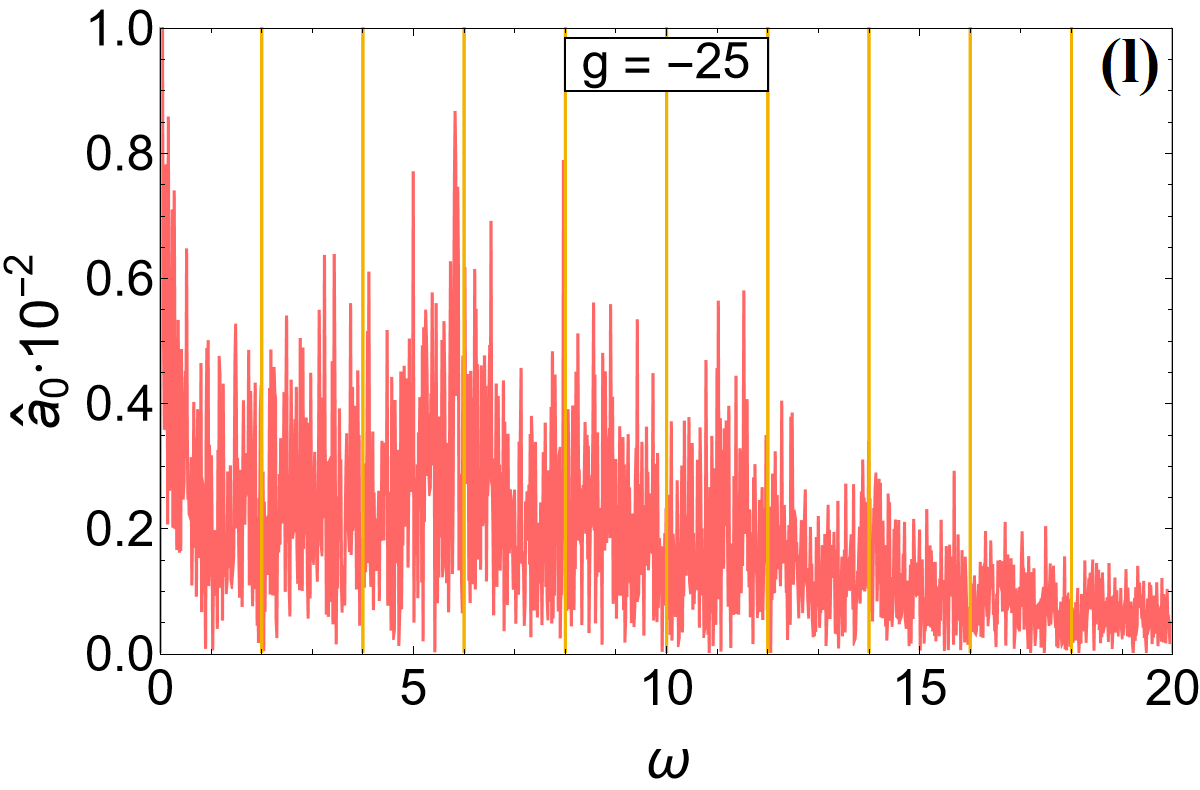}
\caption{The power spectrum produced by the defocusing 1D GPE with the
box-shaped potential (a)-(d), and by the defocusing and focusing [(e)-(h)
and (i)-(l), respectively] 1D GPE with the HO potential. The computations of
the spectra for increasing values of the nonlinearity strength, $|g|$, are
performed with random-wave initial conditions. The initial data are the same
for all plots in the case of the HO potential. Vertical yellow lines mark
the location of $\mathcal{W}_{2k}$, see Eq. (\protect\ref{2k}).}
\label{fig:1DNLS_Harmonic_defocusing_vs_focusing}
\end{figure*}

Having explained the emergence of the comb-like power spectra for weak
nonlinearity, it is natural to explore how the picture changes towards
strong nonlinearity. Specifically, it is relevant to find out how the
structure gradually deviates from the above prediction for the weakly
nonlinear regime with the increase of $|g|$, and how it depends on the sign
of the nonlinearity, self-defocusing ($g>0$) or focusing ($g<0$).

Figure~\ref{fig:1DNLS_Harmonic_defocusing_vs_focusing} provides answers to
these questions. One observes how the comb-like spectrum evolves away from
the \textquotedblleft slender" version as the nonlinearity strength grows,
for the 1D GPEs with the box and HO potentials. The comparison between these
potentials demonstrates that the main predictions of the weakly nonlinear
analysis developed above still hold qualitatively in the fully nonlinear
regime. The power spectrum for the box potential transits from the comb-like
shape to an ergodic one, which includes a conspicuous high-frequency
component. On the other hand, the spectrum corresponding to the HO potential
still keeps a comb-like spectral shape for large values of $|g|$. This
contrast between the different potentials reflects the fact that the
equidistant linear energy spectrum (\ref{eq:equidistant_linear_spectrum})
plays a central role in the strongly nonlinear regime too. Nevertheless,
these results are not explained by proximity to the linear regime. Indeed,
while the \textquotedblleft slender" power spectra exist at small $g$ in
both cases of the HO and box potentials, extending them to a comparable
level of the nonlinearity, characterized by the ratio $H_{4}/H_{2}$ of the
energy terms, see Eqs. (\ref{eq:E_L}) and (\ref{eq:E_NL}), the GPE with the
HO potential still maintains a comb-like spectrum, while its counterpart
with the box potential displays an ergodic spectral distribution, totally
different from the weakly-nonlinear regime.

For the case of the HO potential in the GPE with focusing and defocusing
nonlinearities, Fig.~\ref{fig:1DNLS_Harmonic_defocusing_vs_focusing} shows
two characteristic effects involving spikes of the comb-like spectra. The
first is a gradual deviation from locations $\mathcal{W}_{k}$, that were
predicted in the weak-nonlinearity regime, towards smaller (larger)
frequencies for the defocusing (focusing) sign of the nonlinearity (in
agreement with the usual definitions of the self-defocusing and focusing),
while keeping their nearly-equidistant structure. The second effect,
observed with the growth of $|g|$, is that the spectral spikes get wider,
and at some point they start to overlap with each other, compromising the
comb-like shape. The magnitude of $|g|$ at which this happens depends on the
sign of the nonlinearity. In the case of the self-attraction, the transition
happens at much lower values of $|g|$.
This trend can be easily explained
too, noting that the focusing nonlinearity enhances the interaction and
mixing between different modes, while the defocusing suppresses the
interaction.

Note that the above analysis is presented for the nonlinearity magnitude, $g$%
, treated as the control parameter. An alternative way to quantify the
strength of the nonlinearity is, as mentioned above, to use the ratio
between the quadratic and quartic energies, $|H_{4}|/H_{2}$. We observe
that, with the increase of $|g|$, the self-focusing GPE rapidly accumulates
energy in the nonlinear terms, which is translated into larger values of $%
|H_{4}|/H_{2}$, in comparison to the defocusing case, which requires much
higher values of $g$ to reach the same ratio.

The transition from the comb-like power spectrum to ergodicity in the case
of very strong nonlinearity is not surprising. What is nontrivial in these
results, is the great impact the equidistant structure of the linear
spectrum on the nonlinear regime and the persistence of the non-ergodic
spectrum even for strong nonlinearity.


\section{Other nonlinear Schr\"{o}dinger equations with HRPs
(highly-resonant potentials)}

\label{sec:Other_highly_resonant_equations}

\begin{figure}[t]
\centering
\par
\includegraphics[width=4.2cm]{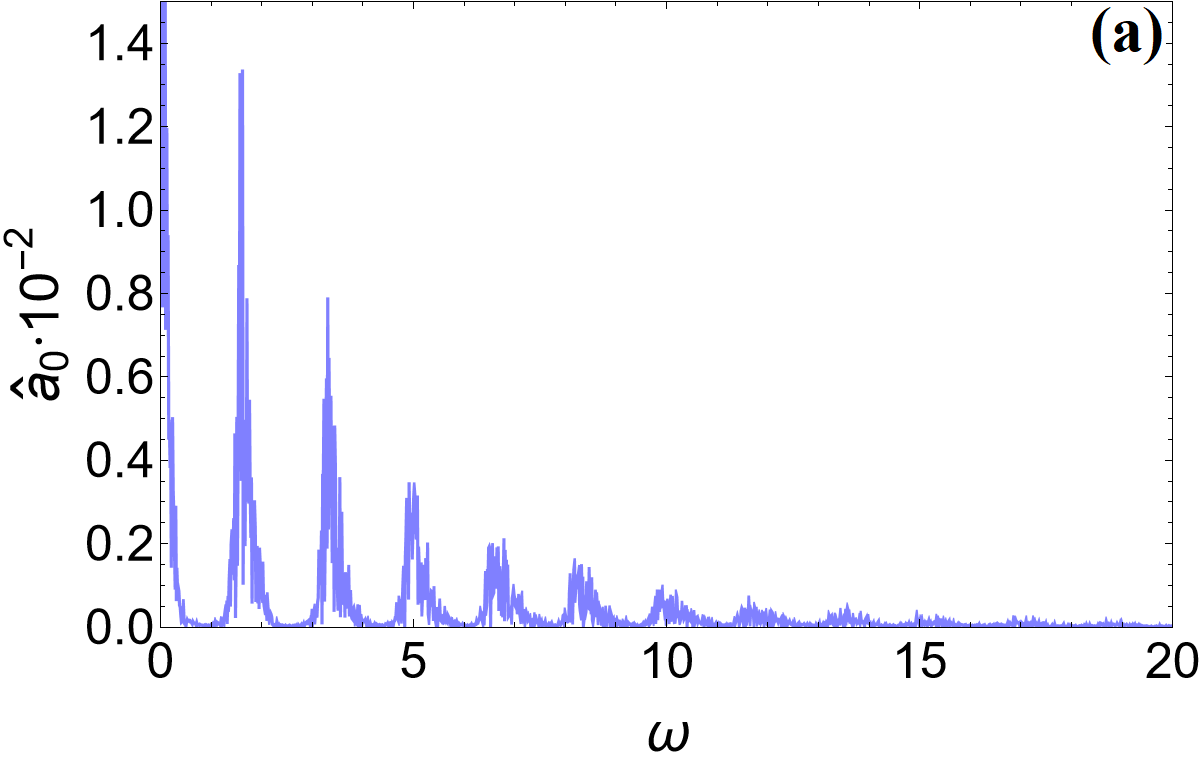} \includegraphics[width=4.2cm]{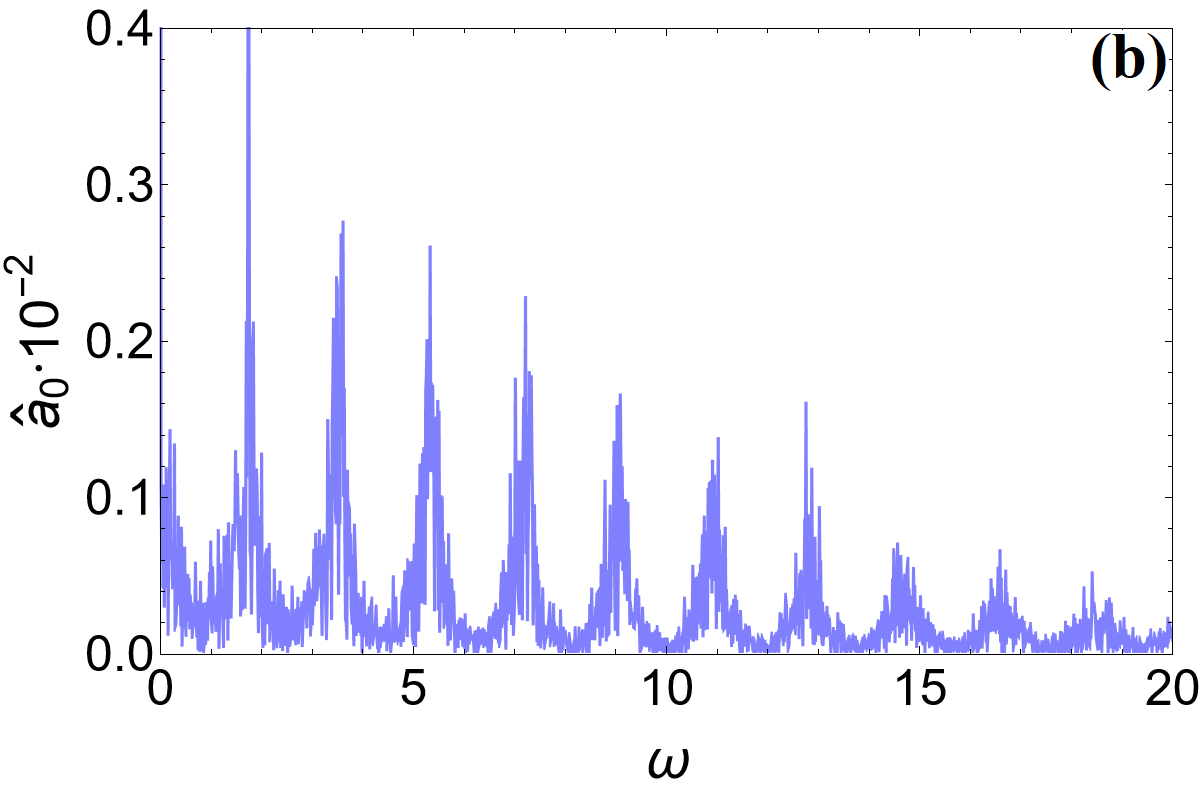} %
\includegraphics[width=4.2cm]{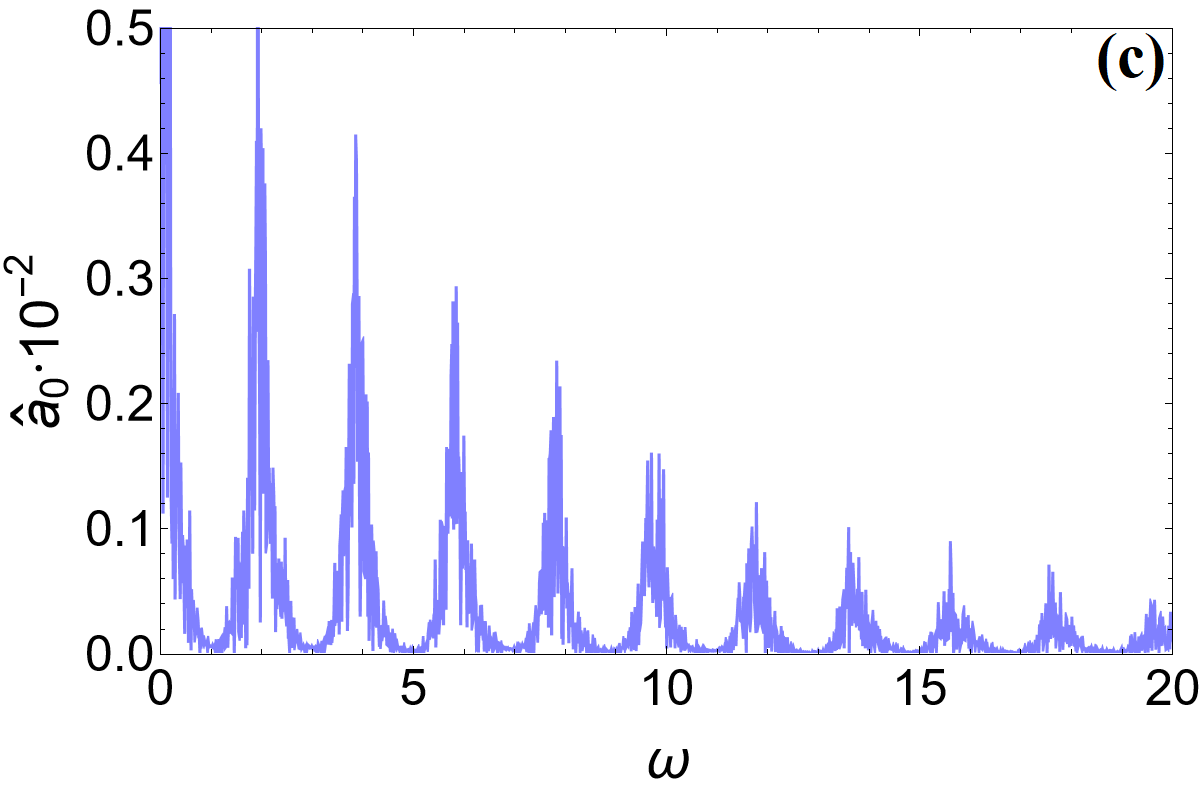} \includegraphics[width=4.2cm]{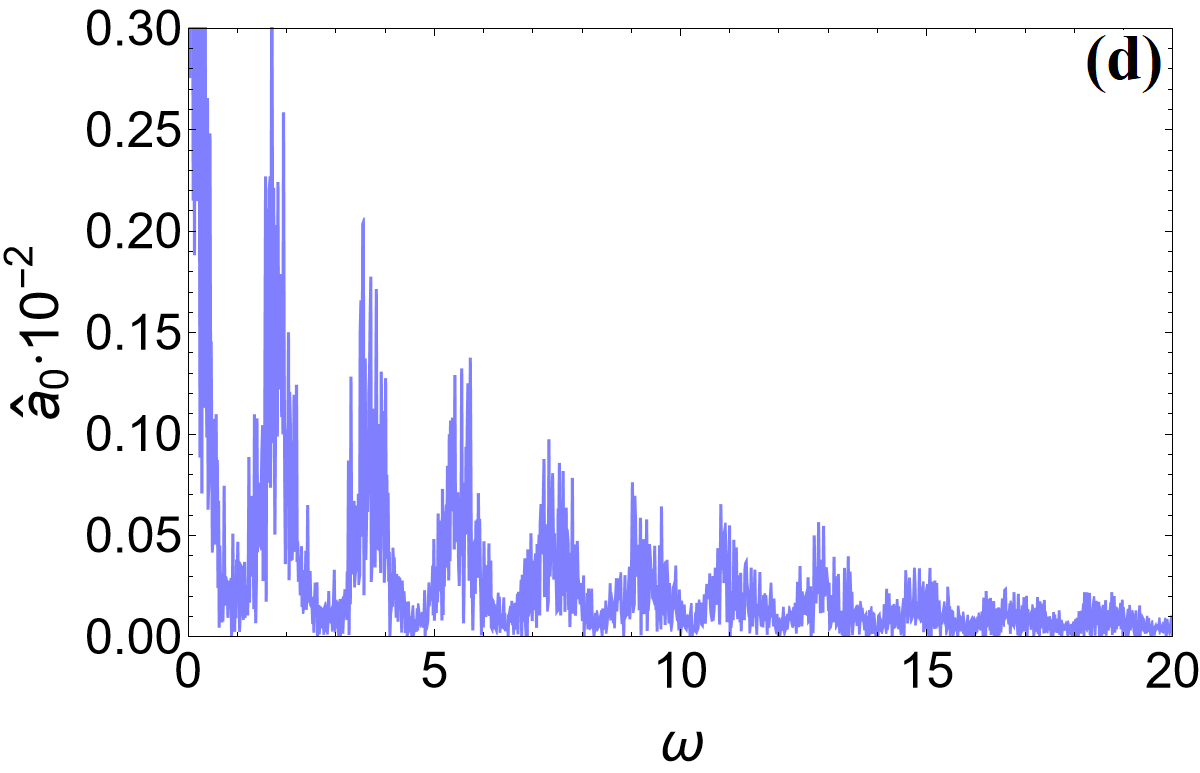}
\par
\includegraphics[width=4.2cm]{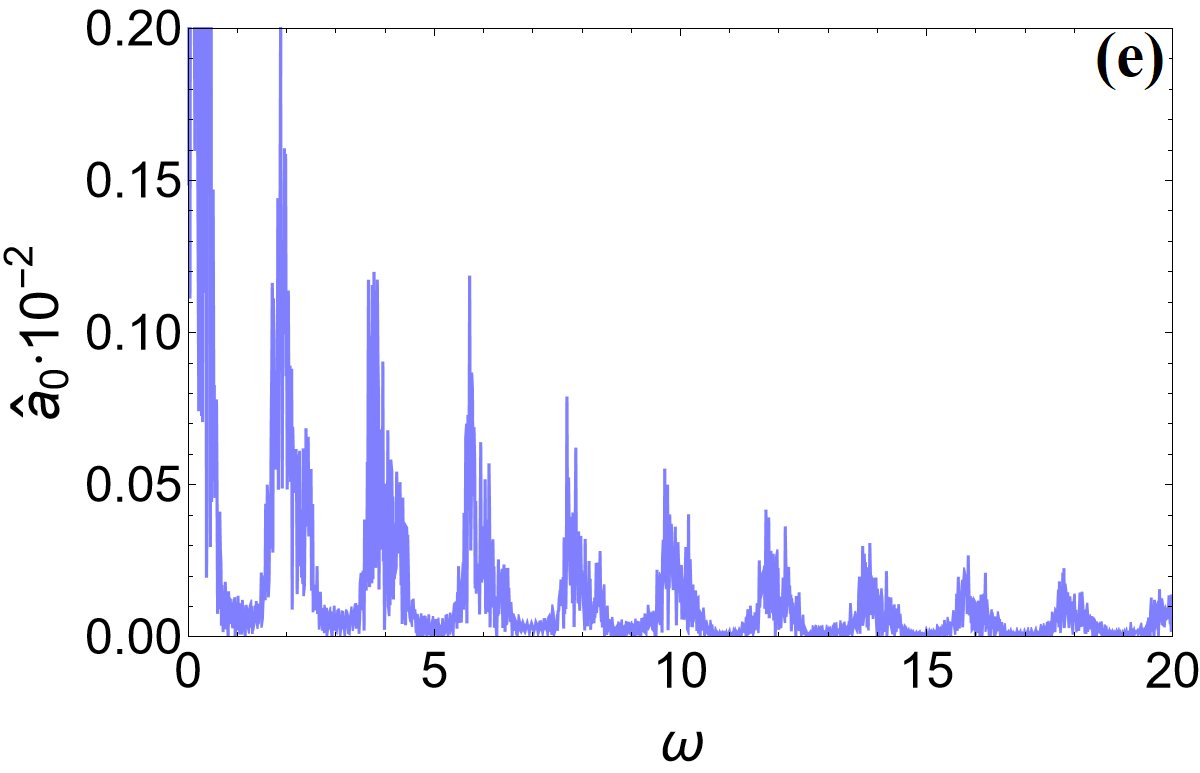} \includegraphics[width=4.2cm]{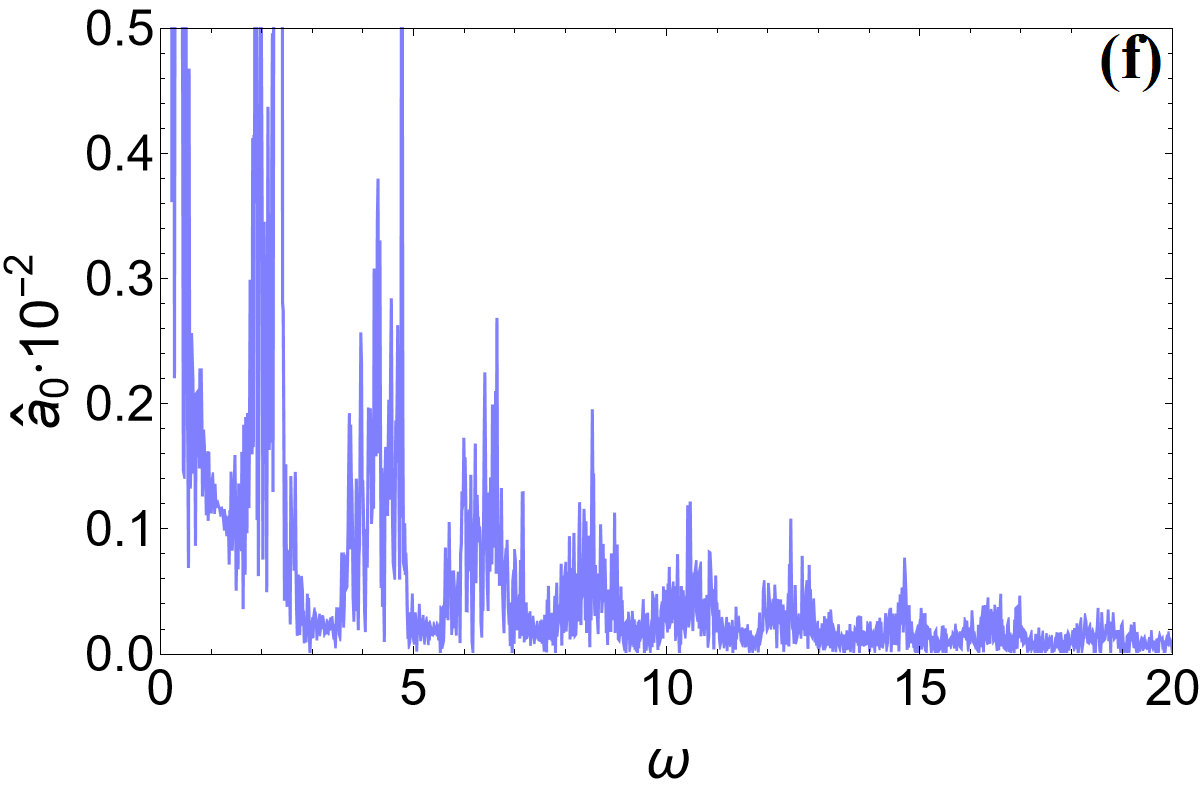} %
\includegraphics[width=4.2cm]{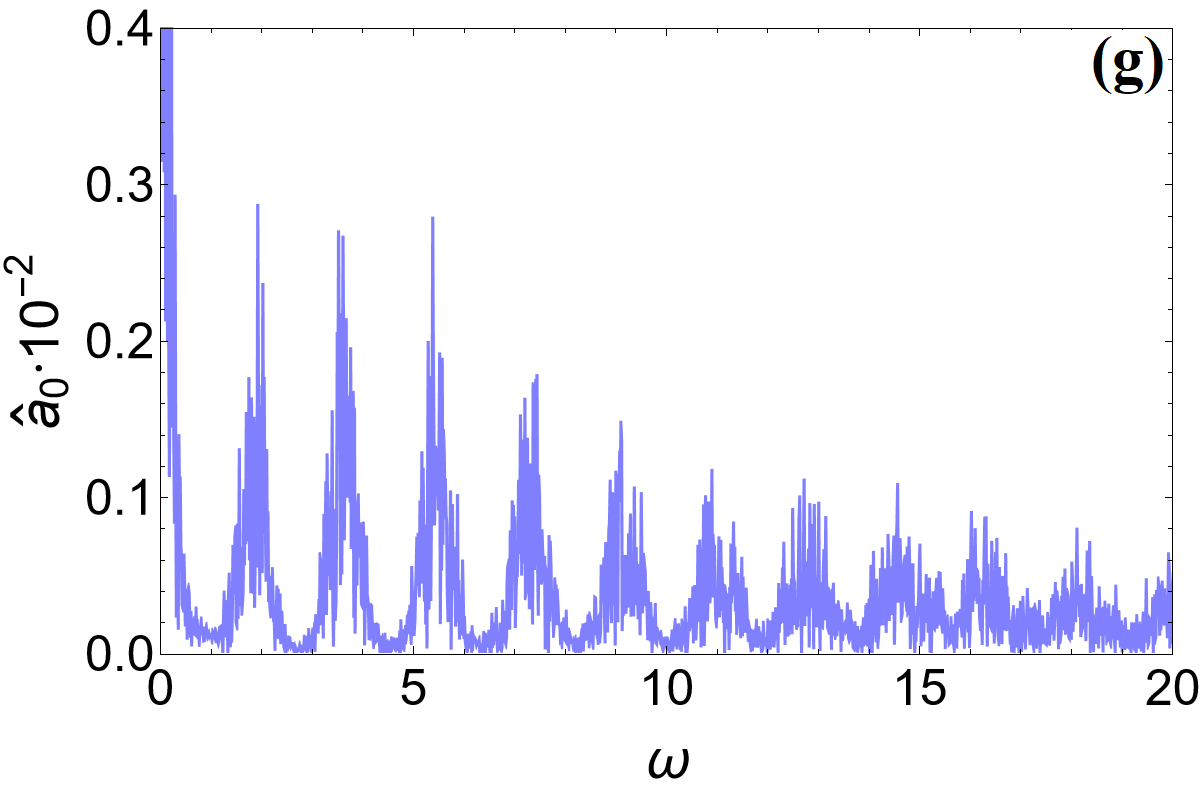} \includegraphics[width=4.2cm]{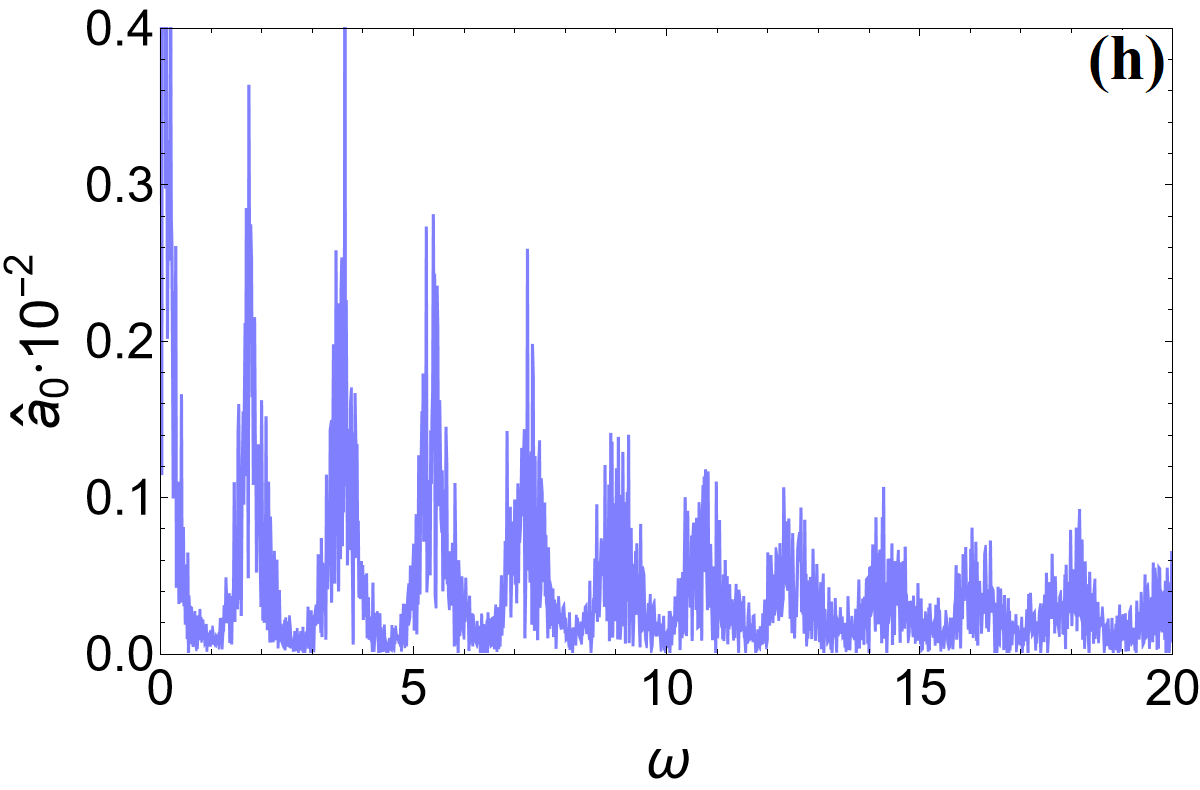}
\par
\includegraphics[width=4.2cm]{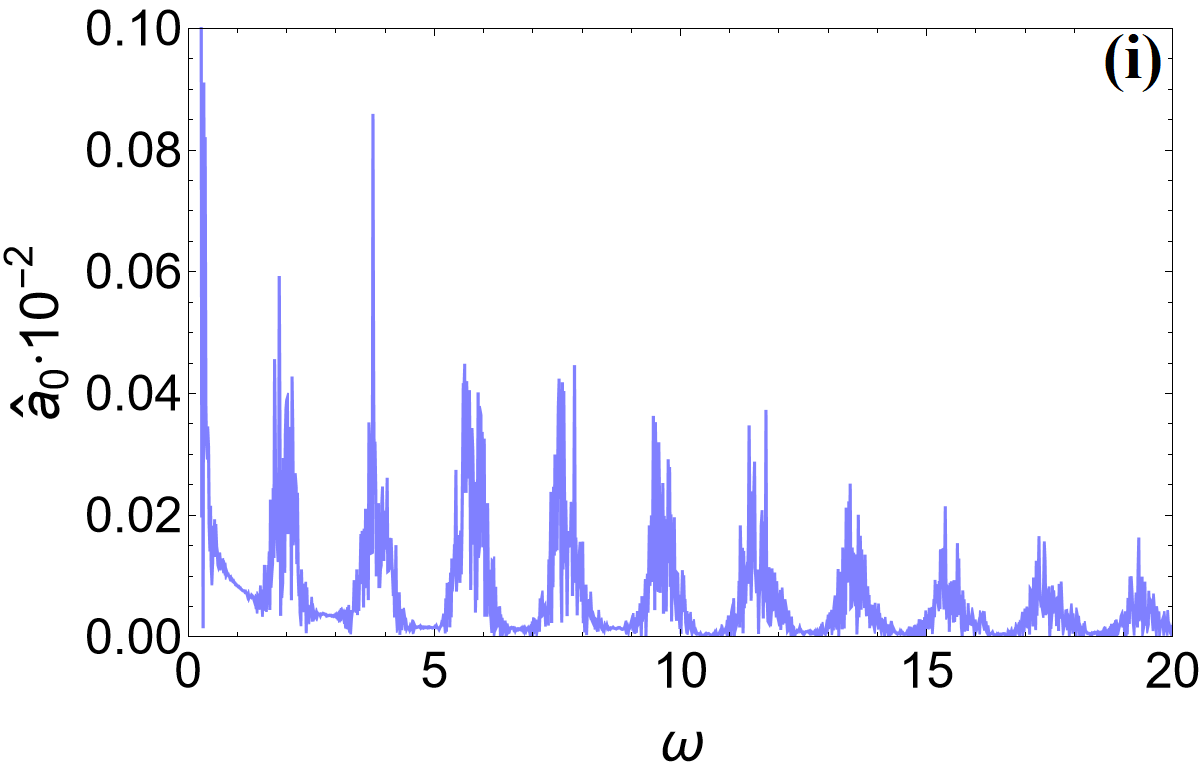} \includegraphics[width=4.2cm]{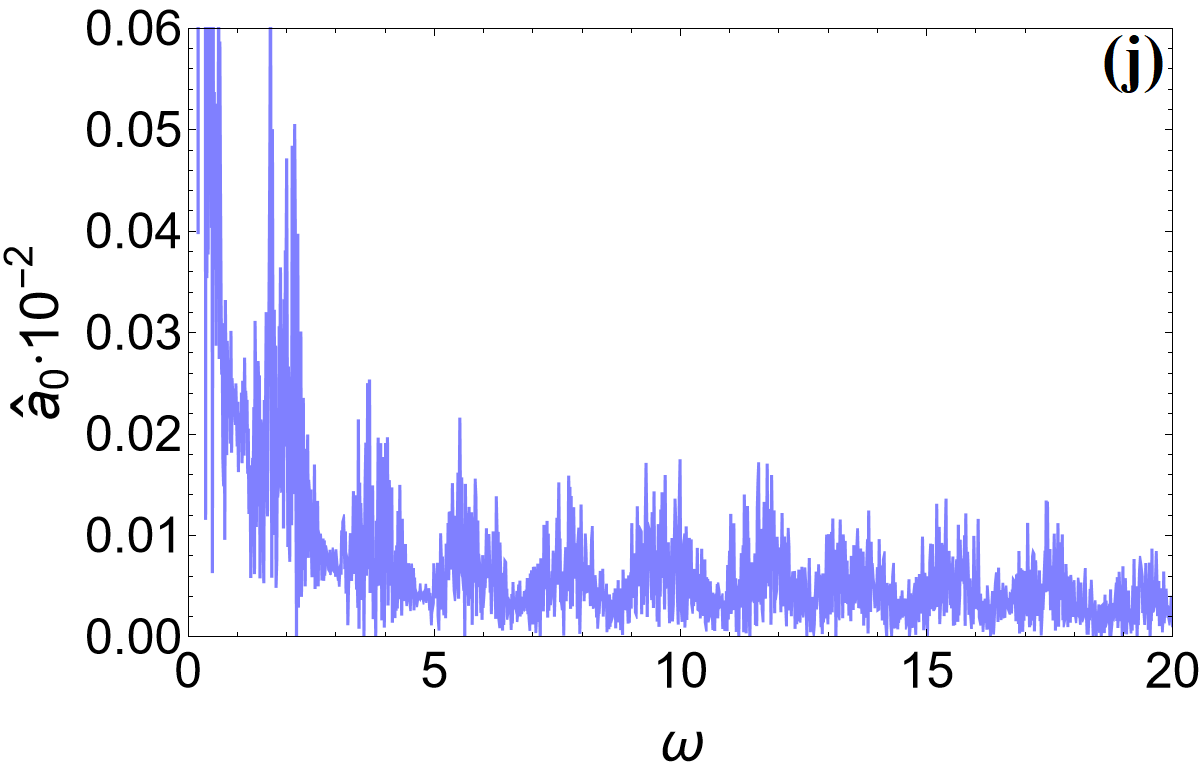} %
\includegraphics[width=4.2cm]{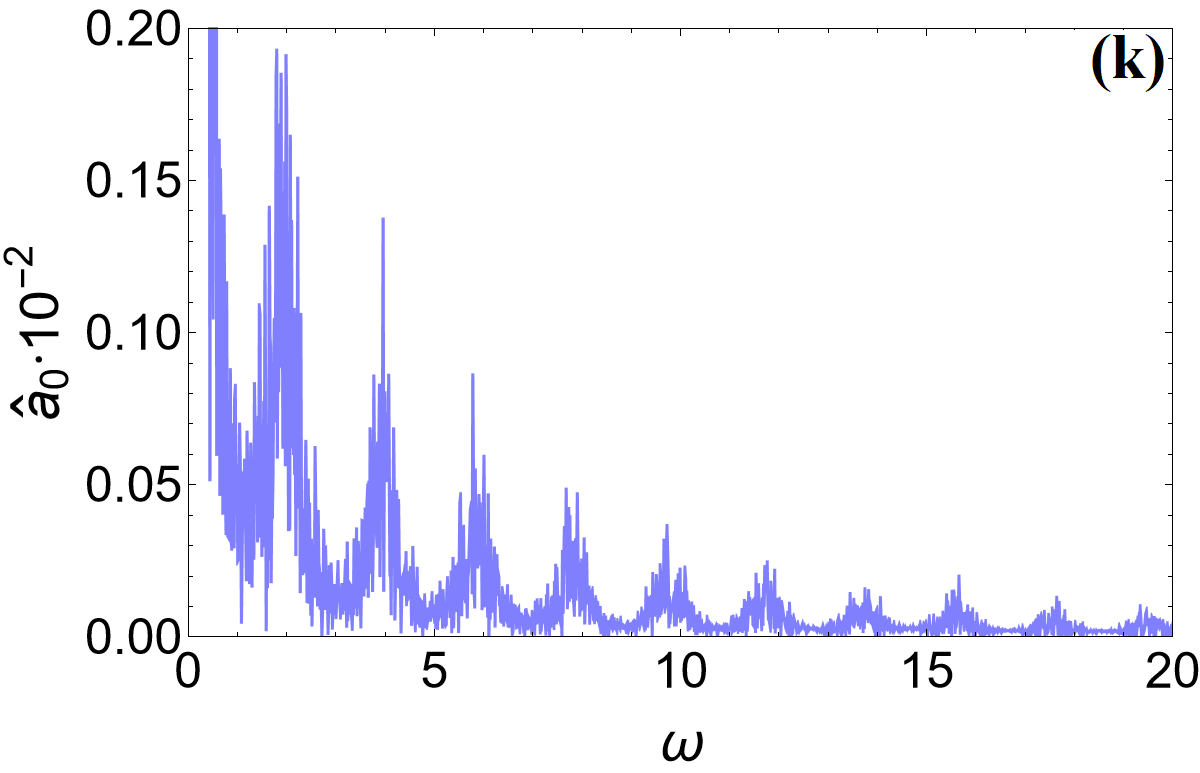} \includegraphics[width=4.2cm]{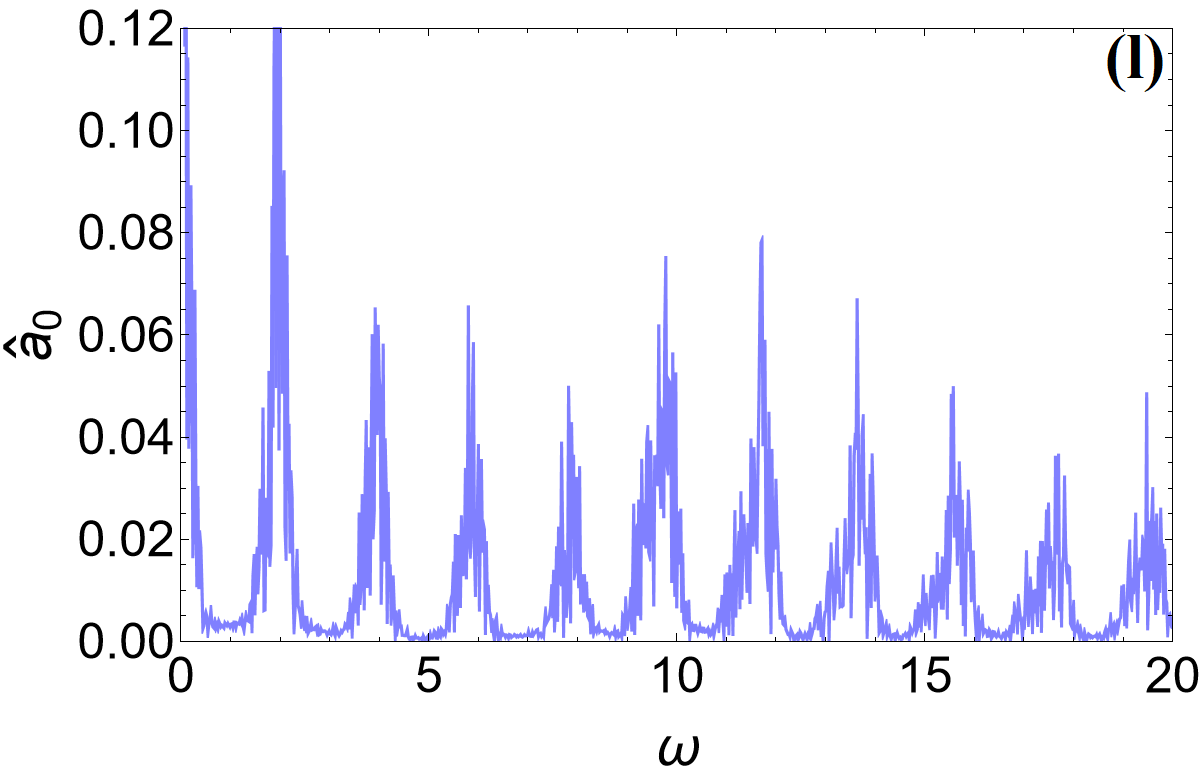}
\caption{Comb-like power spectra produced by the 1D, 2D, and 3D GPEs (a-c),
by the 1D, 2D, and 3D quintic NLSEs with the HO potential (d-f), by the 1D
NLSE with potentials $V^{(1)}$, $V^{(2)}$, $V^{(3)}$, $V^{(4)}$ defined by
Eqs. (\protect\ref{eq:V1}) -- (\protect\ref{eq:V1}) [panels (g-j),
respectively], by the two-component NLSE (\protect\ref{eq:Two-Component_NLS}%
) (k), and by the relativistic real wave equation (\protect\ref{eq:AdS})
(l). }
\label{fig:All_NLSEs}
\end{figure}

In the above analysis, we addressed the 1D GPE with the HO potential as the
guiding example to present the characteristic features of HRPs{} and observe
how their resonance structure hinders the onset of the ergodicity. Here, we
proceed to demonstrate that this effect is generic for other resonant
potentials, which cover a wide range of interesting models. To this end, we
have explored the dynamics of NLSEs with different nonlinearities, including
HRPs{} in different spatial dimensions, a two-component NLSE, and even a
related relativistic wave equation. Below, we present detailed results for
these equations. In Fig.~\ref{fig:All_NLSEs}, one can see that all of them
display comb-like power spectra, confirming the genericity of the principles
formulated above. Actually, these findings imply that the form of the
nonlinearities plays a secondary role, as it determines the values of $g$ at
which comb-like spectra transit to ergodic ones, but not the overall
presence of the effect.

\noindent \textbf{1) The quintic 1D NLSE with the HO potential:} A natural
modification of the original 1D GPE with the HO potential is to replace the
cubic nonlinear term by the quintic one. The equation has the form
\begin{equation}
i\partial _{t}\psi =-\frac{1}{2}\partial _{xx}\psi +\frac{1}{2}x^{2}\psi
+g|\psi |^{4}\psi ,  \label{eq:Quintic_1D-NLSE_HP}
\end{equation}%
keeping the equidistant spectrum{}, $E_{n}=n+{1}/{2}$. This modification
provides a new setting because the cubic 1D NLSE in the free space is
integrable, while the quintic one is not, and gives rise to 1D Townes
solitons and critical collapse \cite{1D Townes}. This, in particular, rules
out the integrability of the underlying equation in the free space as a
reason for the emergence of comb-like power spectra.
With $t$ replaced by
the propagation distance, $z$, Eq. (\ref{eq:Quintic_1D-NLSE_HP}) is a
natural model for a planar waveguide in optics, where the purely quintic
nonlinearity may be realized in colloidal suspensions of metallic
nanoparticles \cite{Cid}.

\noindent \textbf{2) D-dimensional cubic and quintic NLSEs with the HO
potential:} It is also natural to explore the existence of comb-like power
spectra in higher dimensions (here we restrict the consideration to the case
of spherical symmetry). We did that for the cubic and quintic NLSEs with the
HO potential:
\begin{equation}
i\partial _{t}\psi =\frac{1}{2}\left( -\partial _{rr}-\frac{D-1}{r}\partial
_{r}+r^{2}\right) \psi +g|\psi |^{p-1}\psi ,  \label{eq:D-NLSE_cubic_quintic}
\end{equation}%
where $r\in \lbrack 0,\infty )$ is the radial coordinate, $D=2,3,...$ is the
spatial dimension, and $p=3$ or $5$ is the power of the nonlinear term. For
any combination of these parameters and $g>0$ (self-repulsion, otherwise the
multidimensional NLSE gives rise to the collapse \cite{BMP}), Eq. (\ref%
{eq:D-NLSE_cubic_quintic}) has the commonly known equidistant linear energy
spectrum of the multidimensional HO, $E_{n}=2n+{d}/{2}$.

\noindent \textbf{3) Anharmonic potentials:} Another way to test the
robustness of our findings is by modification of the trapping potential,
keeping its equidistant spectral structure. Some special 1D potentials which
maintain this property can be found in Ref. \cite{EquidistantPotentials}:
\begin{align}
& V^{(1)}(x)=\frac{x^{2}}{2}+\frac{s^{2}-1}{8x^{2}},  \label{eq:V1} \\
& V^{(2)}(x)=\frac{x^{2}}{2}+\frac{3}{x^{2}}\frac{4x^{4}+3}{(2x^{2}+3)^{2}}+%
\frac{4}{3},  \label{eq:V2} \\
& V^{(3)}(x)=\frac{x^{2}}{2}+\frac{8x^{2}-4}{(2x^{2}+1)^{2}}+\frac{2}{3},
\label{eq:V3} \\
& V^{(4)}(x)=\frac{x^{2}}{2}+8\frac{(8x^{6}+12x^{4}+18x^{2}-9)}{%
(4x^{4}+12x^{2}+3)^{2}}+2,  \label{eq:V4}
\end{align}%
where $s>1$ is a constant, $x\in (0,\infty )$ for the first two potentials,
and $x\in (-\infty ,\infty )$ for the last two.
In particular, potential $V^{(1)}(x)$ represents the so-called ``superselection",
\textit{viz}., the interaction of a particle, confined by the HO potential
and carrying a permanent dipole electric moment, with an electric charge
placed at $x=0$ \cite{supersel}, while $V^{(2)}(x)$ is a modified version of
the same potential, in the case when the quasi-1D (cigar-shaped) trap is
embedded in partly screening host medium. Moreover, the 1D GPE with potential $V^{(1)}(x)$ is identical to the radial reduction of the D-dimensional NLSE with the HO potential and nonlinear term $r^{D-1}|\psi |^{2}\psi $ (see appendix~\ref%
{appendix:eigensystems} for the derivation). In any case, this equation is
truly different from Eqs. (\ref{eq:1D-GPE}) and (\ref%
{eq:D-NLSE_cubic_quintic}), constituting a new element for our study of HRPs. Potentials $V^{(3)}(x)$ and $%
V^{(4)}(x)$ do not have a straightforward physical interpretation, but they
provide additional relevant realizations of equidistant spectra.

The energy eigenvalues are fully equidistant for $V^{(1)}$ and $V^{(2)}$,%
\begin{equation}
E_{n}^{(1)}=2n+1+\frac{s}{2},\quad E_{n}^{(2)}=2n+\frac{23}{6},
\end{equation}%
with $n\geq 0$. On the other hand, for $V^{(3)}$ and $V^{(4)}$ there is a
gap between the ground-state eigenvalue and ones corresponding to the
excited states, which form equidistant arrays (``towers"):
\begin{align}
& E_{0}^{(3)}=-\frac{5}{6},\quad E_{n\geq 1}^{(3)}=n+\frac{7}{6}, \\
& E_{0}^{(4)}=-\frac{3}{2},\quad E_{n\geq 1}^{(4)}=n+\frac{5}{2}.
\end{align}

\noindent \textbf{4) A two-component NLSE system:} Another possibility \cite%
{OutEquiVVVV-1,TwoComponentNLSE0,TwoComponentNLSE1,TwoComponentNLSE2,TwoComponentNLSE3}
to realize the comb-like (non-ergodic) power spectra is offered by a
two-component 1D NLSE,
\begin{equation}
\begin{cases}
& i\partial _{t}u=-\frac{1}{2}\partial _{xx}u+\frac{x^{2}}{2}%
u+cv+g_{u}|u|^{2}u, \\
& i\partial _{t}v=-\frac{1}{2}\partial _{xx}v+\frac{x^{2}}{2}%
v+cu+g_{v}|v|^{2}v,%
\end{cases}
\label{eq:Two-Component_NLS}
\end{equation}%
where $g_{u},\ g_{v}$ and $c$ are constants. The linear version of the
system decouples into two single-component equations for $\psi _{+}=u+v$ and
$\psi _{-}=u-v$, which gives rise to two ``towers" of equidistant energy
eigenvalues, $E_{n}^{(\pm )}=n+\frac{1}{2}\pm c$.

\noindent \textbf{5) A wave equation in anti-de Sitter spacetime:} Our
considerations of the comb-like spectra are based on the equidistant energy
spectrum (\ref{eq:equidistant_linear_spectrum}), and depend little on
peculiarities of the NLSEs. We have further tested the validity of the
principles formulated here for the case of a relativistic wave equation
whose normal-mode frequencies also fit condition (\ref%
{eq:equidistant_linear_spectrum}). This choice is motivated by the
connection between the GPE and the following equation for a real scalar
field $\phi $ in the anti-de Sitter spacetime \cite{BMP,BEF,Evnin_review}:
\begin{equation}
\partial _{tt}\phi =\cot ^{2}x\ \partial _{x}(\tan ^{2}{x}\,\partial
_{x}\phi )+g\phi ^{3},  \label{eq:AdS}
\end{equation}%
which is subject to boundary condition $\phi (t,\pi /2)=0$, where $x\in
\lbrack 0,\pi /2)$ is the radial coordinate. This equation gives rise to an
equidistant spectrum, $E_{n}=2n+3$.


\section{Discussion}

Our analysis has revealed that NLSEs with HRPs (highly resonant potentials)
pose a barrier to the emergence of ergodic power spectra in weakly- and
strongly-nonlinear regimes alike. While usually the consideration of
non-ergodic dynamics is restricted to small deformations of integrable
equations \cite{GHD1,GHD2,GHD3,GHD4,GHD5,PreTherm0,
PreTherm1,PreTherm2,PreTherm3,ErgodicBreakingTrap,AtomicLosses,TDK}, our
focus has been on mechanisms that do not directly rely on proximity to
integrability. The potentials in question, namely, the ones with equidistant
linear spectra of energy eigenvalues [in particular, the HO
(harmonic-oscillator) potential], produce a strong impact on the power
spectra of the full nonlinear system, which remain concentrated in comb-like
arrays of spikes. This pattern is captured by our analytical consideration
for weak nonlinearity, performed in Section~\ref{sec:Analytic_section}, and
numerical simulations of the strongly nonlinear regime in Section~\ref%
{secnum}. These spectra are in clear contrast with the continuous ones
produced by generic potentials, and resemble quasi-discrete spectra
associated with integrable dynamics.

While the difference between the HRPs and generic potentials without any
resonances is obvious, the difference is more subtle when comparing HRPs to
potentials that feature some resonances in their spectra, but the energy
levels do not fit the rigid pattern defined by Eq. (\ref%
{eq:equidistant_linear_spectrum}). In the case of generic potentials,
normal-mode frequencies are incommensurate, and combinational frequencies
created by nonlinearities quickly populate the real line, creating a generic
continuum power spectrum. For that reason, much of our study has been
focused on the peculiar but physically motivated case of the infinitely deep
box potential. In that case, the linear normal-mode frequencies and all of
their combinations are integers, which, however, does not preclude the
emergence of the continuum power spectrum at a finite nonlinearity strength,
in contrast to what is seen in the case of the HO and other HRPs with linear
energy spectra in the form of Eq. (\ref{eq:equidistant_linear_spectrum}).
The analytical consideration carried out in Section~\ref{sec:Analytic_section}, together with numerical experiments reported in Sections~\ref{secnum}~and~\ref{sec:Other_highly_resonant_equations}, make it clear that a central role is played by the spectrum of linear
energy eigenvalues, even in the case of strong nonlinearity. We have seen in the weakly nonlinear regime
how they determine the interactions between the modes in the system, which
translates into the structure of the power spectrum. Equidistant energy
eigenvalues, like those in the case of the HO potential, distribute the
interactions in such a special way that a reduced set of frequencies
dominate in the power spectrum, providing strong suppression of high
frequencies and ensuring the protection of the non-ergodic comb-like power
spectra in the regime of stronger nonlinearity. On the other hand, the
quadratic energy eigenvalues produced by the box-shaped potential do not
provide for the suppression of higher frequencies, and give rise to truly
continuous ergodic spectra. This analysis is extended to a broad class of
HRPs in Appendix \ref{appendix:Decay_Snk}, leading to the same conclusion.
Further, we have made use of simulations to study the dependence of the
comb-like power spectrum on the nonlinearity strength, and tested the
genericity of our conclusions, checking them for NLSEs with various HRPs.
Random waves were used as initial conditions to capture the evolution of a
wide range of inputs. Our numerical results corroborates that the analysis
developed in the weak-nonlinearity limit correctly forecasts the qualitative
shape of the power spectra in the strongly nonlinear regime as well. We also
inspected the distinction between the cases of focusing and defocusing
nonlinearities, concluding that the comb-like power spectra degrade faster
with the growth of the nonlinearity strength in the former case.

In general, linear features tend to get rapidly overwhelmed by nonlinear
effects when the system departs from the weakly nonlinear regime, although
some models for 1D random waves demonstrate regimes where dynamical features
of weak and strong nonlinearities coexist \cite{ZakharovCoexisit} (i.e.,
random waves and coherent modes, such as solitons, exist in both regimes).
HRPs admit similar coexistence between the features of weak and strong
nonlinearity: while solitons (and other essentially nonlinear modes) are
involved in the dynamics, it is still heavily influenced by the
weak-nonlinearity features, such as the structure of the spectrum of energy
eigenvalues. As a result, the comb-like power spectra, which are directly
associated with weakly nonlinear dynamics, persist for stronger nonlinearity.

After producing the basic results with the help of the guiding example of
the 1D GPE with the HO potential, we have demonstrated that the same
mechanism of the obstruction to ergodicity is maintained by generic HRPs. To
do that, in addition to the analytical description developed in Section~\ref%
{sec:Analytic_section} and Appendix~\ref{appendix:Decay_Snk}, we have
explored several NLSEs with this class of potentials. We observed comb-like
power spectra in the presence of different nonlinear terms, different
potentials (belonging to the HRP class), different spatial dimensions, and
in the two-component GPE as well. The presence of the multidimensional
models in the class of highly resonant NLSEs, such as the 2D and 3D GPEs,
and the quintic NLSE with the HO potential, are noteworthy findings. This is
in contrast to studies of non-ergodic dynamics that rely on proximity to
integrability, as a vast majority of integrable equations are
one-dimensional. In this work, we have studied the obstruction to ergodicity
in the multidimensional equations under the assumption of the spherical
symmetry. It would be interesting to lift this condition, addressing fully
multidimensional spectra for states carrying angular momentum.

The presence of the 2D-GPE with the HO in the class of HRPs
suggests a potential connection between our results and experiments on wave
thermalization in multimode optical fibers, a topic of numerous ongoing experiments
\cite{Picozzi2011,Picozzi2019,Picozzi2020,Picozzi2023,Picozzi2023_V2,Mangini,Pourbeyram}. Light propagation in graded-index multimode fibers, studied in these experiments, is modeled by the finite-mode version of the 2D-GPE with the HO potential \cite{Picozzi2011,Picozzi2019,Picozzi2020,Picozzi2023,Picozzi2023_V2}. In this setup, Refs.~\cite{Picozzi2019,Picozzi2023_V2} have studied the role played by structural disorder (addition of a random term to the HO) on the dynamics of weakly interacting random waves. Rapid thermalization has been observed in the presence of disorder, while the thermalize was hindered in the absence of this element (i.e., in the case of
the pure HO potential). The thermalization is commonly explained in terms of the wave-turbulence theory, and no hindrances were expected when the experiments started. It is plausible that the anomaly pointed out in Refs. \cite{Picozzi2019,Picozzi2023_V2} may be caused by the influence of the equidistant energy eigenvalues inherent to the model that governs the observed dynamics, agreeing in this way with our inference that the specific structure of eigenvalues (in this case, produced by the 2D NLSE with the HO potential) may account for deviations from the ordinary principles of non-equilibrium dynamics. It would be interesting to investigate whether the phenomenology of the effective non-ergodicity observed in Refs. Ref.~\cite{Picozzi2019,Picozzi2023_V2} is generic for other models with HRPs.

Finally, our study suggests an extension of the concept of
quasi-integrability, identified in the form of quasi-discrete power spectra
in the 1D GPE with the HO potential and self-defocusing cubic nonlinearity
in Ref. \cite{BorisQuasiIntegrability}. In that context, the case of
self-focusing remained unexplored till now. We have tackled it here too,
demonstrating the presence of the comb-like power spectra in this case as
well, although they degrade faster with the growth of the nonlinearity
strength. We have provided an analytical description of this effect in the
regime of weak nonlinearity, while previously reported results were purely
numerical. Finally, we have broadened the understanding of the
quasi-integrability by showing that its characteristic quasi-discrete power
spectrum, produced by the evolution of random-wave initial conditions, is
shared by a large class of the nonlinear models including HRPs. Thus, our
results imply that the 1D GPE with the HO potential is not exceptional in
this regard, although it is worthwhile to mention the large range of values
of the strength of the defocusing nonlinearity for which this physically
relevant model produces well-defined comb-like power spectra. While we have
mostly focused on NLSEs, our weakly nonlinear analytics suggests that the
obstruction-to-ergodicity mechanism should be present in equations of other
types, such as nonlinear wave equations. We have briefly demonstrated the
latter possibility by presenting the comb-like power spectrum generated by
the highly-resonant real wave equation (\ref{eq:AdS}).


\section*{Acknowledgments}

AB thanks A.~Picozzi, Z.~Hani, G.~Staffilani, and J.~Amette for useful
discussions. In the course of this work. AB has been supported by the Polish
National Science Centre grant No. 2017/26/A/ST2/00530 and by the LabEx
ENS-ICFP: ANR-10-LABX-0010/ANR-10-IDEX-0001-02 PSL*. OE has been supported
by Thailand NSRF via PMU-B (grant numbers B01F650006 and B05F650021) and by
FWO-Vlaanderen through project G006918N. BAM has been supported, in part, by
the Israel Science Foundation (grant No. 1695/22).



\appendix

\section{Numerical methods}

\label{appendix:Numerical_Methods}

Numerical simulations of NLSEs have been performed using two schemes. One is
based on a pseudo-spectral decomposition of the spatial coordinate similar
to that used in Refs. \cite{Zakharov, Randoux, Julian} and the fourth-order
Runge-Kutta (4RK) method to advance in time. When the spatial coordinate is
unbounded, $x\in (-\infty ,\infty )$, such as in the case of the 1D GPE with
the HO potential, we truncate the domain to a finite one $x\in \lbrack
-R_{\max },R_{\max }]$ with $R_{\max }$ large enough to guarantee that $%
|\psi (t,\pm R_{\max })|$ is exponentially suppressed. This interval is
discretized into $N$ points of the form $x_{n}=R_{\max }(\frac{2n}{N}-1)$
with $n=0,1,...,N-1$. The goal of this procedure is to compute the second
derivative on the RHS of the equation by using the Fast Fourier transform
(FFT), see \cite{BookSpectralMethods} for a deep description. For this
purpose, we decompose function $\psi (t_{j},x_{n})$ at time $t_{j}$ over the
truncated set of the lowest $N/2$ Fourier modes propagating to the left and
to the right,
\begin{equation}
\psi (t_{j},x_{n})\approx \sum_{k=0}^{N/2-1}\beta _{k}^{(-)}e^{-i\frac{\pi }{%
R_{\max }}k(x_{n}+R_{\max })}
\end{equation}%
\begin{equation}
+\sum_{k=1}^{N/2}\beta _{k}^{(+)}e^{i\frac{\pi }{R_{\max }}k(x_{n}+R_{\max
})}
\end{equation}%
where $\beta _{k}^{(\pm )}$ are the Fourier amplitudes at time $t_{j}$. We
apply the FFT to $\psi (t_{j},x_{n})$ to compute the amplitudes, and use the
inverse FFT to compute the second derivative,
\begin{equation}
\partial _{xx}\psi (t_{j},x_{n})\approx \sum_{k=1}^{N/2-1}-\left( \frac{\pi
}{R_{\max }}k\right) ^{2}\beta _{k}^{(-)}e^{-i\frac{\pi }{R_{\max }}%
k(x_{n}+R_{\max })}
\end{equation}%
\begin{equation}
+\sum_{k=1}^{N/2}-\left( \frac{\pi }{R_{\max }}k\right) ^{2}\beta
_{k}^{(+)}e^{i\frac{\pi }{R_{\max }}k(x_{n}+R_{\max })}.
\end{equation}%
Note that the boundary conditions $|\psi (t_{j},\pm R_{\max })|\ll 1$
require that $\beta _{0}^{(-)}\approx 0$ and $\beta _{k}^{(+)}\approx -\beta
_{k}^{(-)}$. We use these conditions as a quality check in our simulations.
Terms on the RHS of the equation that do not involve differentiation are
computed using $\psi (t_{j},x_{n})$.

Our second scheme to simulate NLSEs is similar to that employed in Ref. \cite%
{BMP,BorisQuasiIntegrability}. It truncates the spatial domain to $x\in
\lbrack -R_{\max },R_{\max }]$ as well, and discretizes it to $x_{n}=R_{\max
}\left( \frac{2n}{N}-1\right) $ with $n=0,1,...,N$. We use, in this case,
the finite-difference method to compute spatial derivatives like in Ref.
\cite{BMP,BorisQuasiIntegrability}, while the 6RK algorithm is used to
advance in time. The two schemes have shown an excellent agreement,
conserving the norm $M$ (\ref{eq:M}) and energy $H$ (\ref{eq:energy}), with
maximum deviations at the level of the numerical precision $\sim 10^{-13}$
for the first scheme, and $\sim 10^{-13}$ for $M$, $\sim 10^{-8}$ for $H$ in
the second scheme in the HO, while $\sim 10^{-9}$ for $M$, $\sim 10^{-5}$
for $H$ in the box. The codes have been implemented in C++, running parallel
computations on a GPU to speed up the simulation. The number of points that
we used varies depending on the initial data and the setup -- typically, $N$
ranges from $2^{13}$ to $2^{17}$ in the case of the HO potential, and from $%
2^{11}$ to $2^{13}$ in the case of the box potential.
It is relevant to
mention that the same results can be produced by dint of the split-step
integration method implemented in the usual numerical shell, cf. Ref. \cite{Lush}.


\section{The decay of $\mathcal{S}_{n}(k)$}

\label{appendix:Decay_Snk}

\begin{figure}[t]
\centering
\includegraphics[width=\columnwidth]{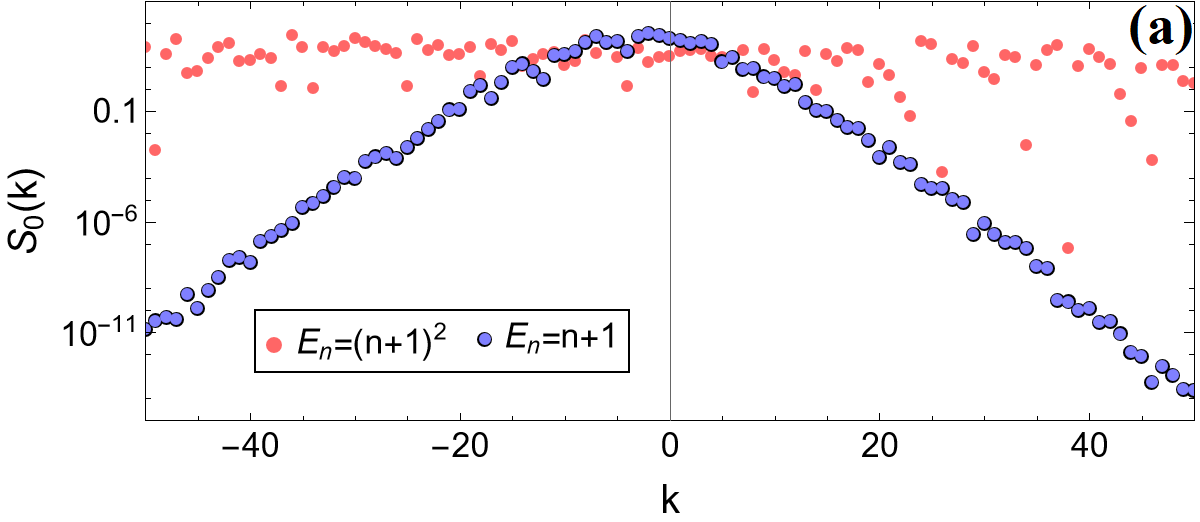} \includegraphics[width=%
\columnwidth]{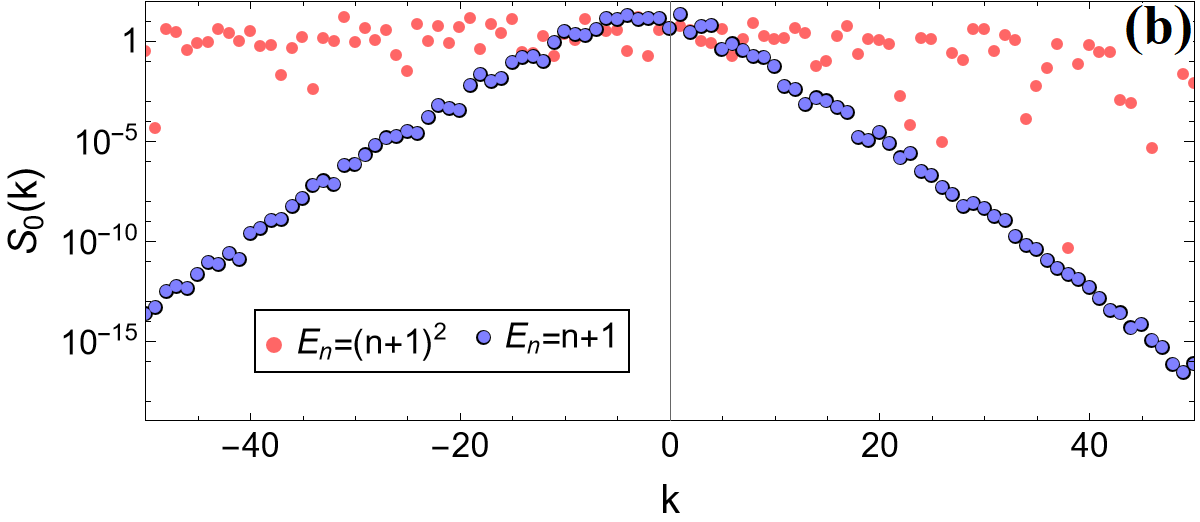} \includegraphics[width=\columnwidth]{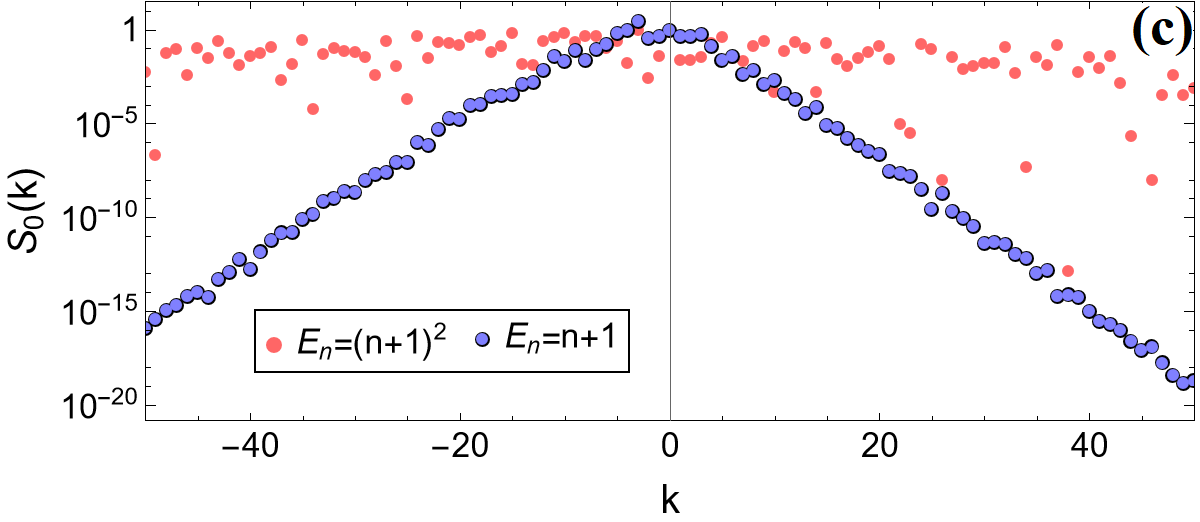} %
\includegraphics[width=\columnwidth]{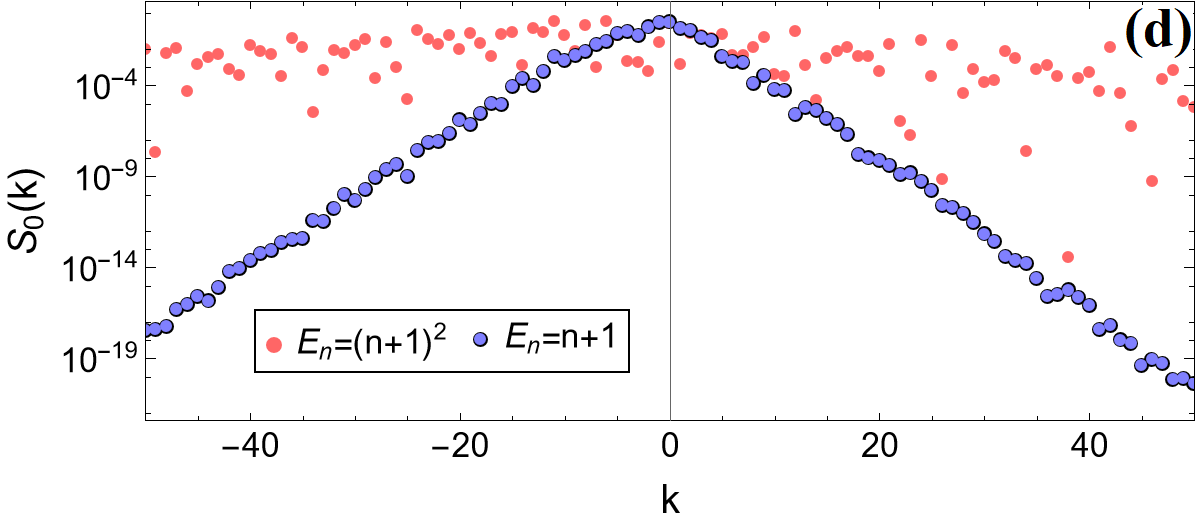}
\caption{Comparison between amplitudes $\mathcal{S}_{0}(k)$ for the linear $%
E_{n}=n+1$ and quadratic $E_{n}=(n+1)^{2}$ spectra of the energy
eigenvalues. In each plot, we have used the same values of $\protect\alpha %
_{n}$ and $C_{nmij}$ for both sets of eigenvalues, ad per Eqs. (\protect\ref%
{eq:Initial_data_PLOTS_generic_Snk_APPENDIX})-(\protect\ref%
{eq:Cnmij_PLOTS_generic_Snk_APPENDIX}). The difference between the plots is
the power of $C_{nmij}$, $r=3,\ 1.5,\ 0,\ -1$, from top to bottom. }
\label{fig:Snk_GENERIC_NLSEs}
\end{figure}

We show here that amplitudes
\begin{equation}
\mathcal{S}_{n}(k)\equiv \underset{\Delta _{nmij}=\mathcal{W}_{k}}{%
\underbrace{\sum_{m=0}^{\infty }\sum_{i=0}^{\infty }\sum_{j=0}^{\infty }}}%
C_{nmij}\bar{\alpha}_{m}\alpha _{i}\alpha _{j},
\label{eq:Snk_amplitude_APPENDIX}
\end{equation}%
are strongly suppressed at large $|n-k|$ in highly resonant systems, with
spectra
\begin{equation}
E_{n}=an+b\qquad \text{with}\quad n\in \mathbb{N},\ a,b\in \mathbb{R},
\label{eq:equidistant_linear_spectrum_APPENDIX}
\end{equation}%
for configurations of $\alpha _{n}$ that actually occur in the course of the
evolution (with an exponential suppression at large $n$). First, Fig.~\ref%
{fig:Snk_GENERIC_NLSEs} visually illustrates the fact that the suppression
of $\mathcal{S}_{n}(k)$ at large $|n-k|$ depends very little on the
couplings $C_{nmij}$, irrespective of the decay or growth with the variation
of the indices, but, in contrast, it strongly depends on the equidistant
relation of eigenvalues (\ref{eq:equidistant_linear_spectrum_APPENDIX}). In
each plots of Fig.~\ref{fig:Snk_GENERIC_NLSEs} we have numerically
calculated the values of $\mathcal{S}_{n}(k)$, using the same set of $%
C_{nmij}$ and $\alpha _{n}$ but two different choices of eigenvalues,
equidistant $E_{n}=n+1$ and quadratic $E_{n}=(n+1)^{2}$ ones. This may seem
as a minor difference because $E_{n}$ only affects the computation of
expression (\ref{eq:Snk_amplitude_APPENDIX}) through
\begin{equation}
\Delta _{nmij}=E_{n}+E_{m}-E_{i}-E_{j},  \label{eq:Delta_nmij_Appendix}
\end{equation}%
to restrict the interactions to frequency $\mathcal{W}_{k}$. However, it
leads to a dramatic difference between the behavior of $\mathcal{S}_{n}(k)$
at moderate and large values of $|n-k|$ for the equidistant and quadratic
energy eigenvalues. As we observe in all plots of Fig.~\ref%
{fig:Snk_GENERIC_NLSEs}, in the equidistant case these amplitudes rapidly
decay with\ the increase of $|n-k|$ (resembling an exponential decay), while
this does not happen in the quadratic case. For those specific plots we have
used
\begin{align}
& \alpha _{n}=(n+1)^{2}e^{-n}\mathcal{A}_{n}e^{i\mathcal{P}_{n}},
\label{eq:Initial_data_PLOTS_generic_Snk_APPENDIX} \\
& C_{nmij}=(n+m+i+j+1)^{r},  \label{eq:Cnmij_PLOTS_generic_Snk_APPENDIX}
\end{align}%
where $\mathcal{A}_{n}$, and $\mathcal{P}_{n}$ are random variables
uniformly distributed on $[0,1]$ and $[0,2\pi )$, respectively, while power $%
r$ gave the opportunity to find out if the coefficients decay $(r<0)$,
remain constant $(r=0)$, or grow $(r>0)$ with the variation of the indices.
Note that expression (\ref{eq:Initial_data_PLOTS_generic_Snk_APPENDIX})
captures the qualitative behavior of $\alpha _{n}$ in our numerical
simulation, as explained in Section~\ref{sec:Analytic_section}. Other
choices from this class of conditions lead to the same conclusion. For $%
C_{nmij}$ we have used the power law in Eq. (\ref%
{eq:Cnmij_PLOTS_generic_Snk_APPENDIX}) because they exhibit, at most, a
polynomial growth with the increase of the indices in all physically
relevant systems we are aware of. For instance, the $D$-dimensional GPE with
the HO potential has asymptotic values $C_{nnnn}\sim n^{\frac{D}{2}-2}$ for
large $n$ \cite{BMP}, and similar asymptotics have been found for
relativistic wave equations \cite%
{EvninAsymptotics1,EvninAsymptotics2,EvninAsymptotics3}. Other choices of $%
C_{nmij}$ in this class lead to the same conclusion as well.


Next, we aim to show that the exponential decay of $\mathcal{S}_{n}(k)$ for
the highly resonant systems (\ref{eq:equidistant_linear_spectrum_APPENDIX})
may be derived analytically, arriving at the result
\begin{equation}
|\mathcal{S}_{n}(k)|<e^{-\beta |n-k|}P_{n,k}
\label{eq:estimate_Snk_main_text_Appendix}
\end{equation}%
with $\beta >0$, where $P_{n,k}$ is a polynomial in $n$ and $k$ that depends
on the detailed form of $\alpha _{n}$ and $C_{nmij}$. To derive this bound,
we use an estimate for the exponential decay of $\alpha _{n}$ at large $n$,
which is what actually happens in our simulations, and an arbitrary
polynomial $p_{n}^{(s)}$ in $n$ of degree $s>0$ to bound different values of
$\alpha _{n}$. Thus, this estimate takes the form of $|\alpha
_{n}|<p_{n}^{(s)}e^{-\beta n}$ where $\beta >0$ is not specified and appears
in Eq. (\ref{eq:estimate_Snk_main_text_Appendix}), as we show below. For the
couplings, we are going to use an estimate based on an arbitrary polynomial $%
q_{n}^{(r)}$ in $n$ of degree $r>0$ for each index, $%
|C_{nmij}|<Q_{nmij}^{(r)}\equiv q_{n}^{(r)}q_{m}^{(r)}q_{i}^{(r)}q_{j}^{(r)}$%
. This estimate comes from the observation that the couplings have a
polynomial growth at most for large values of the indices, as mentioned
above. An estimate admitting each index to have a different power may be
used, but it can be covered by the present choice, simply setting $r$ equal
to the largest power. Plugging the estimates for $\alpha _{n}$ and $C_{nmij}$
into the expressions for $\mathcal{S}_{n}(k)$ (\ref%
{eq:Snk_amplitude_APPENDIX}), we obtain
\begin{align}
& |\mathcal{S}_{n}(k)|=\bigg{|}\underset{n+m-i-j=k}{\underbrace{%
\sum_{m=0}^{\infty }\sum_{i=0}^{\infty }\sum_{j=0}^{\infty }}}C_{nmij}\bar{%
\alpha}_{m}\alpha _{i}\alpha _{j}\bigg{|} \\
& <\underset{n+m-i-j=k}{\underbrace{\sum_{m=0}^{\infty }\sum_{i=0}^{\infty
}\sum_{j=0}^{\infty }}}|C_{nmij}||\bar{\alpha}_{m}||\alpha _{i}||\alpha _{j}|
\notag \\
& <\underset{n+m-i-j=k}{\underbrace{\sum_{m=0}^{\infty }\sum_{i=0}^{\infty
}\sum_{j=0}^{\infty }}}Q_{nmij}^{(r)}p_{m}^{(s)}p_{i}^{(s)}p_{j}^{(s)}e^{-%
\beta (m+i+j)}.  \notag
\end{align}

The constraint on the indices, $n+m-i-j=k$, may be used to remove the
summation in $j$. Two cases must be distinguished, \textit{viz}., $k<n$ and $%
k\geq n$, to guarantee that $j\geq 0$.

\textbf{Case} $\mathbf{k<n}$: Substituting $j=n+m-k-i$ one gets
\begin{align}
&|\mathcal{S}_n(k)| < e^{\beta (k-n)}\sum_{m=0}^{\infty}e^{-2\beta
m}p_m^{(s)} \\
&\times \sum_{i=0}^{n-k+m} Q_{nmi(n-k+m-i)}^{(r)}
p_i^{(s)}p_{(n-k+m-i)}^{(s)}.  \notag
\end{align}
We know that the sum $\sum_{i=1}^{M}i^a$ with $a\geq0$ is a polynomial of
degree $a+1$ in $M$ \cite{BookIntegralsSeries}. Then, the summation in index
$i$ is a polynomial $n$, $m$, and $k$, denoted by $F_{n,m,k}$. We use now
that $\sum_{m=0}^{\infty} e^{-2\beta m} (m+1)^b$ with $b\in\mathbb{R}$ and $%
\beta>0$ is a finite number to get
\begin{equation}
|\mathcal{S}_n(k)| < e^{\beta(k-n)}\sum_{m=0}^{\infty}e^{-2\beta m}p_m^{(s)}
F_{n,m,k} < P_{n,k}e^{\beta(k-n)}  \label{eq:expresion_k<n}
\end{equation}
where $P_{n,k}$ is a polynomial in $n$ and $k$ respectively.


\textbf{Case} $\mathbf{k\geq n}$: In this case one has to be careful with
the ranges of $m$ and $i$ to guarantee that $j\geq 0$. Taking this into
account, the expression for $\mathcal{S}_{n}(k)$ is
\begin{align}
& \qquad \qquad |\mathcal{S}_{n}(k)|<e^{\beta (k-n)}\sum_{m=k-n}^{\infty
}e^{-2\beta m}p_{m}^{(s)} \\
& \qquad \qquad \times
\sum_{i=0}^{n-k+m}Q_{nmi(n-k+m-i)}^{(r)}p_{i}^{(s)}p_{(n-k+m-i)}^{(s)}
\notag \\
& =e^{\beta N}\sum_{M=0}^{\infty }e^{-2\beta
M}p_{M+N}^{(s)}%
\sum_{i=0}^{M}Q_{n(M+N)i(M-i)}^{(r)}p_{i}^{(s)}p_{(M-i)}^{(s)}.  \notag
\end{align}%
where we have made the following changes, $N=n-k$ and $M=m+N$, in order to
remove the dependence of the lowest value of $m$ on $k-n$. Note that these
changes flipped the sign in the first exponential. The resulting expression
is similar to the one for the case of $k<n$, and we proceed using the same
properties to conclude that
\begin{equation}
|\mathcal{S}_{n}(k)|<e^{\beta (n-k)}P_{n,k},  \label{eq:expresion_k>n}
\end{equation}%
where in this case $P_{n,k}$ is a polynomial in $n$ and $k$. The combination
of Eqs. (\ref{eq:expresion_k<n}) for $k<n$ and (\ref{eq:expresion_k>n}) for $%
k\geq n$ results in Eq. (\ref{eq:estimate_Snk_main_text_Appendix}).


One may also derive an estimate similar to Eq. (\ref{eq:Snk_amplitude_APPENDIX}) for case of the box potential, but the
quadratic eigenvalues (\ref{eq:eigensystem_Box}) make this process a bit more involved than in the case of the equidistant eigenvalues. The difficulties
appear in the constraint imposed on the indices (\ref{eq:Delta_nmij_Appendix}),
\begin{equation}
k=(n+1)^{2}+(m+1)^{2}-(i+1)^{2}-(j+1)^{2}.
\label{eq:Constraint_indices_Box_Appendix}
\end{equation}Using the
above-mentioned estimates $|\alpha _{n}|<p_{n}^{(s)}e^{-\beta n}$, and $|C_{nmij}|<q_{n}^{(r)}q_{k}^{(r)}q_{i}^{(r)}q_{j}^{(r)}$ one may see that
\begin{equation}
|\mathcal{S}_{n}(k)|< D_{n,k}e^{-\beta \sqrt{|k-(n+1)^2|}},
\label{eq:Snk_estimate_box_APPENDIX}
\end{equation}where $D_{n,k}$ is a polynomial in $n$ and $k$, and $\beta $ is again the
exponent of $|\alpha _{n}|$. Note that in this case the suppression of
frequencies is much weaker than for the equidistant energy
spectrum. This estimate comes from assessing the dominant contribution of $N \equiv k-(n+1)^2$
in $|\mathcal{S}_{n}(k)|$ according to Eq. (\ref{eq:Snk_amplitude_APPENDIX}),
\begin{equation}
|C_{nmij}||\bar{\alpha}_{m}||\alpha _{i}||\alpha
_{j}|<Q_{nmij}^{(r)}p_{m}^{(s)}p_{i}^{(s)}p_{j}^{(s)}e^{-\beta (m+i+j)}.
\label{eq:QQQQQQ}
\end{equation}The key part of this expression is the $\exp \left[ -\beta (m+i+j)\right] $, which must be studied in two parts.

	{\bf Case  $\mathbf{N\geq 0}$:} We use the
	relation (\ref{eq:Constraint_indices_Box_Appendix}) between the indices, to
	obtain
	\begin{equation}
		m=\sqrt{k-(n+1)^{2}+(i+1)^{2}+(j+1)^{2}}-1.  \label{eq:RRRRRRRRRRRR}
	\end{equation}	Plugging this expression in the exponent one arrives to the following expression
	\begin{align}
	& -\beta(m+i+j) = \\
	&-\beta \left(\sqrt{N+(i+1)^{2}+(j+1)^{2}}-1\right) - \beta (i+j). \nonumber
	\end{align}
	As $i,j\geq 0$, the first term is directly bounded by $-\beta \sqrt{N}$, getting
	\begin{align}
		&
		\sum_{i=0}\sum_{j=0}Q_{nmij}^{(r)}p_{m}^{(s)}p_{i}^{(s)}p_{j}^{(s)}e^{-\beta
			(m+i+j)}  \label{eq:ZZZZZ} \\
		& \leq e^{-\beta \sqrt{N}}\sum_{i=0}		\sum_{j=0}Q_{nmij}^{(r)}p_{m}^{(s)}p_{i}^{(s)}p_{j}^{(s)}ce^{-\beta (i+j)}
		\notag \\
		& \leq D_{n,k}e^{-\beta \sqrt{N}},  \notag
	\end{align}	where $D_{n,k}$ is a polynomial in $n,k$, and we have used the following
	properties to reach the last expression. First, we used constant $c$ large
	enough to bound the independent term in the exponent. We also used the fact that $\sum_{i=0}^{\infty
	}e^{-\beta i}(i+1)^{b}$ with $b\in \mathbb{R}$ and $\beta >0$ takes a finite
	value to bound the sums in the intermediate expression independently of the
	upper limit. Finally, we used a polynomial $D_{n,k}$ of high enough degree
	to bound the terms involving $n$ and $k$.

	{\bf Case  $\mathbf{N< 0}$:} We use $j$ instead of $m$ in this case
	\begin{equation}
		j=\sqrt{-N + (m+1)^2 -(i+1)^{2}}-1,
	\end{equation}	We define $I=i+1$ and $M = (m+1)^2-N$ to write the exponent in the following form
	\begin{equation}
	-\beta(m+i+j) = -\beta \left(\sqrt{M-I^2} + I\right) +2\beta - \beta m.
	\label{eq:SSSSSSSS}
	\end{equation}
	To guarantee that $j\geq0$ index $I$ take the integer values in $[1,\sqrt{M-1}]$. Because of the symmetry $i\leftrightarrow j$ we just need to consider the values of $I$ in $[1,\sqrt{M/2}]$, namely, $i$ from $0$ to $j$. In this interval, we are going to obtain an upper bound for the first term on the RHS of (\ref{eq:SSSSSSSS}). Adding and subtracting $\beta\sqrt{M}$ yields
	\begin{align}
		&-\beta \left(\sqrt{M-I^2} + I\right)= \\
		&-\beta \sqrt{M} +\beta \left(\sqrt{M}-I - \sqrt{M-I^2}\right) \nonumber\\
		& \underset{I\in[0,\sqrt{M/2}]}{\leq} -\beta \sqrt{M} -(2-\sqrt{2})I \nonumber
	\end{align}
	The later inequality is obtained by showing that the first derivative of the LHS is negative in $[0,\sqrt{M/2})$, the second derivative is positive in the same interval, and the RHS and LHS coincide at the edges of the interval. A final step is $-\beta\sqrt{M} < -\beta\sqrt{|N|}$ because $m\geq0$. From this point one repeats the same argument as before to bound the sums in $m$ and $i$ obtaining the estimate (\ref{eq:Snk_estimate_box_APPENDIX}) for $|\mathcal{S}_{n}(k)|$
	in the case of the box potential.


\section{Power spectrum from longer-time evolution}

\label{appendix:Long_Time}

\begin{figure}[t]
\centering
\includegraphics[width=\columnwidth]{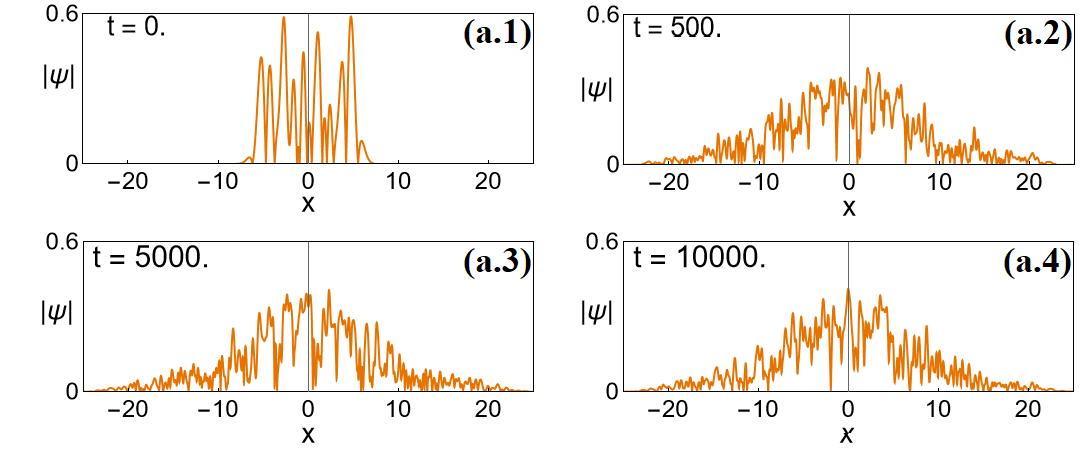} \includegraphics[width=0.95%
\columnwidth]{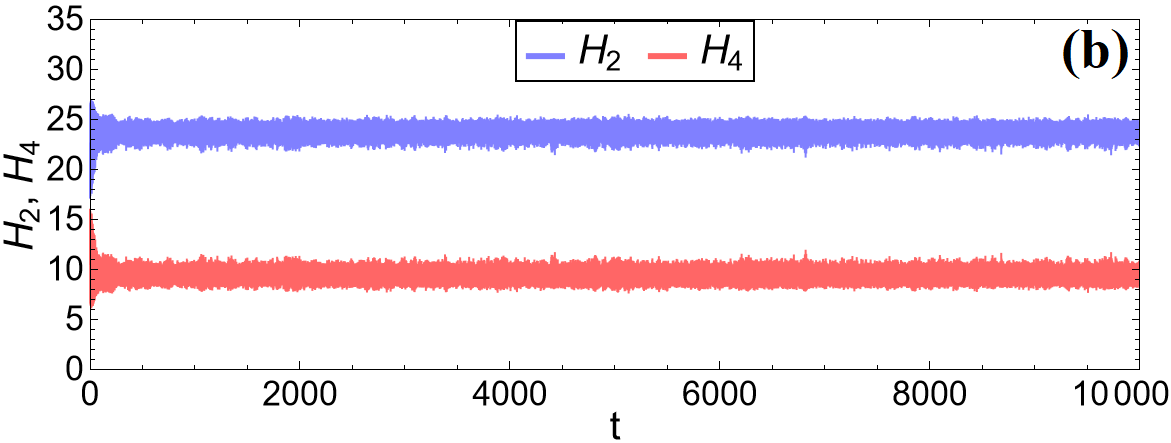} \includegraphics[width=0.95\columnwidth]{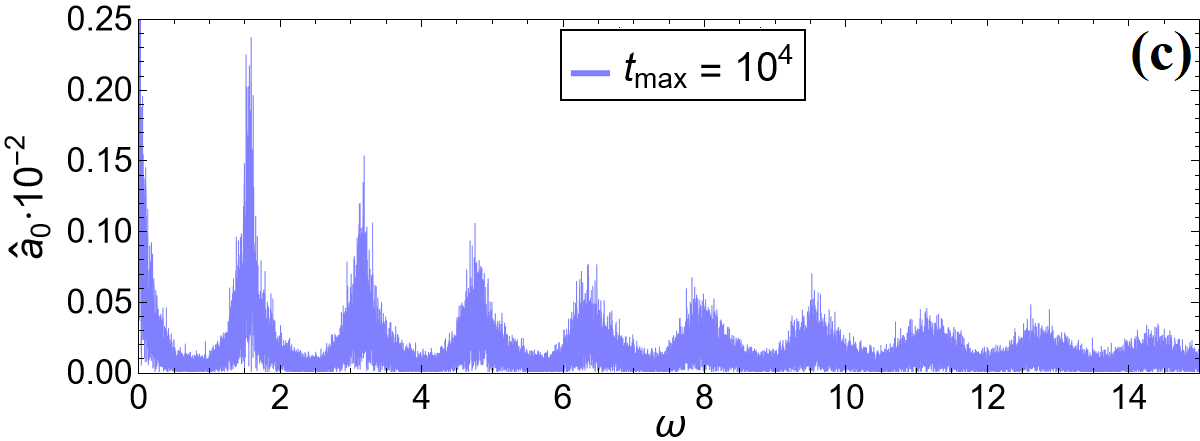} %
\includegraphics[width=0.95\columnwidth]{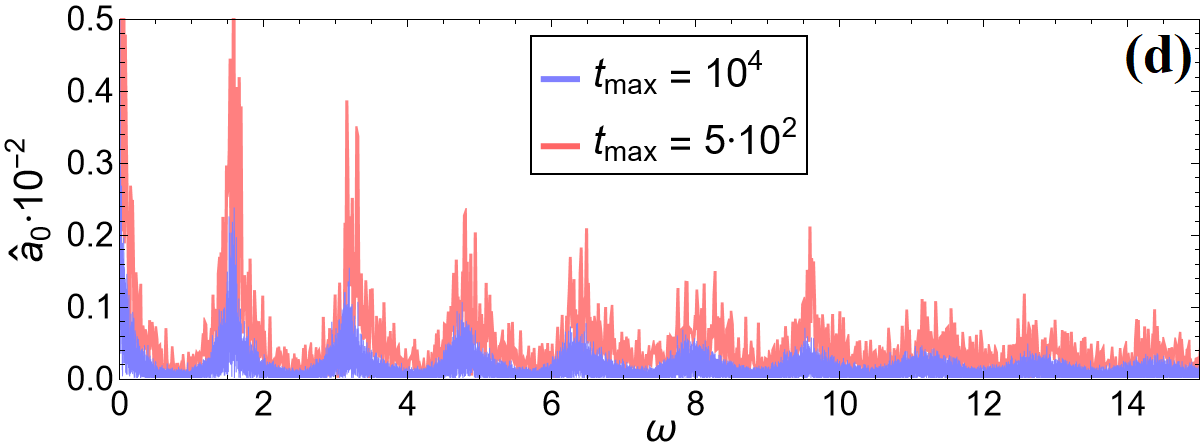} %
\includegraphics[width=0.95\columnwidth]{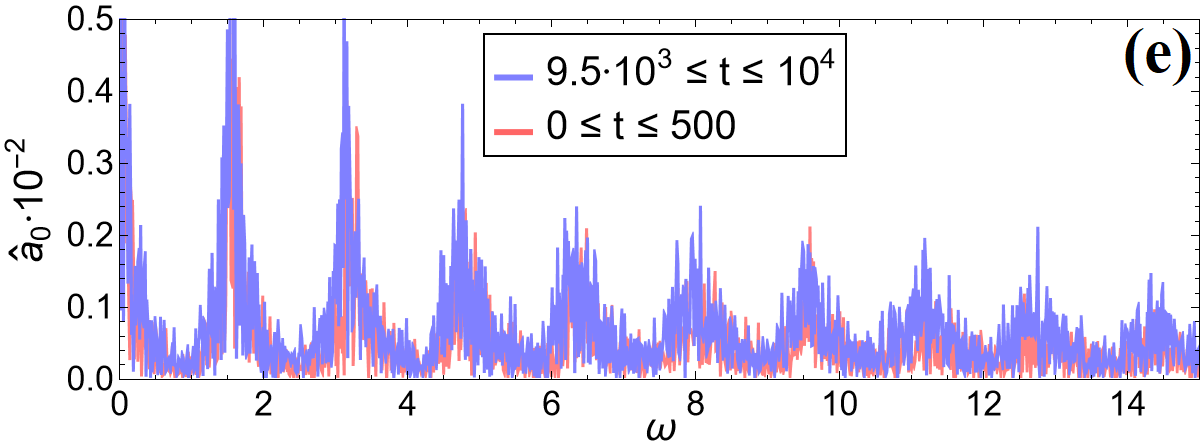}
\caption{Long-time evolution ($t_{\max}=10^{4}$) of the initial data shown
in Fig.~\protect\ref{fig:Evolution_1D-GPE_HP_Box}I. From top to bottom: four
snap-shots illustrating the shape of the profile in the course of the
evolution (a); temporal evolution of the quadratic (\protect\ref{eq:E_L})
and quartic (\protect\ref{eq:E_NL}) energies (b); power spectrum of the
lowest-mode's amplitude calculated in the window of time $\left[0,10^4\right]
$ (c); comparison of the power spectra calculated in the windows of time $%
\left[0,500\right]$ and $\left[0,10^4\right]$ (d); comparison of the power
spectra calculated in the windows of time $\left[0,500\right]$ and $\left[%
9500,10^4\right]$ (e).}
\label{fig:Long-Time_evolution_Power_Spectrum}
\end{figure}

\begin{figure}[t]
	\centering
	\includegraphics[width=\columnwidth]{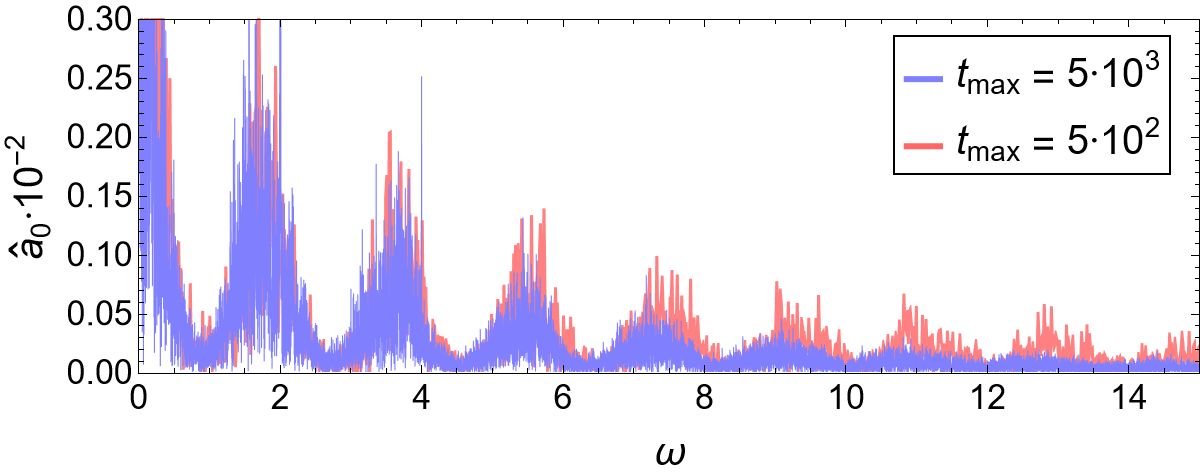}
	\caption{Comparison between the power spectra of the lowest-mode's amplitude calculated in the window of time  $\left[0,500\right]$ (the one shown in Fig.~\ref{fig:All_NLSEs}(d)) and a larger window $\left[0,5000\right]$  displayed by the quintic 1D NLSE with the HO potential (\ref{eq:Quintic_1D-NLSE_HP}).}
	\label{fig:Long-Time_evolution_Power_Spectrum_1D-NLS-Quintic-HO}
\end{figure}

We demonstrate that the comb-like structure of the power spectrum is present at times longer than those used in the main text. To do so, we have simulated the initial data presented in Fig.~\ref{fig:Evolution_1D-GPE_HP_Box}I for a total time of $t_{\max} = 10^4$, much larger than the time used before $t_{\max} = 500$. The results are presented in Fig.~\ref{fig:Long-Time_evolution_Power_Spectrum}. No significant difference is appreciated with respect to Fig.~\ref{fig:Evolution_1D-GPE_HP_Box} in the global shape of the profile and the evolution of the energies in this new scale. We also observe that the comb-like structure of the power spectrum is present as illustrated by Fig.~\ref{fig:Long-Time_evolution_Power_Spectrum}(c). For $t_{\max} = 10^{4}$ the amplitude of the spectrum is smaller than for $t_{\max}=500$ (both spikes and valleys). This effect does not have impact on the comb-like structure, it is just a matter of improvement in the resolution of frequencies ($\Delta\omega = 2\pi/t_{max}$) which allows to discern now ($t_{\max}=10^4$) slightly different frequencies interpreted as a single one before ($t_{\max}=500$). Note that both spikes and valleys in the power spectrum are affected by the same effect Fig.~\ref{fig:Long-Time_evolution_Power_Spectrum}(d). Furthermore, one may observe in Fig.~\ref{fig:Long-Time_evolution_Power_Spectrum}(e) the agreement between the power spectrum calculated from the first and the last $500$ units of time (without adjustable parameters), indicating no change in the governing dynamics. Fig.~\ref{fig:Long-Time_evolution_Power_Spectrum_1D-NLS-Quintic-HO} also confirms the presence of the comb-like power spectrum displayed by the quintic 1D-NLSE with the HO potential (\ref{eq:Quintic_1D-NLSE_HP}) at different time scales.

These results suggest the presence of the comb-like power spectrum at times larger than $t_{\max}\sim 10^{4}$. Dynamics acting at very large time-scales may be present, and in such a case, most of the effects on the power spectrum should be concentrated at small frequencies ($\omega$ proportional to the inverse of the scale) or slowly get present as $t_{\max}$ grows. For instance, it is expected that for generic wave systems the exponential suppression of high-modes $\alpha_n$ is gradually replaced by a power-law at very long times \cite{Pomeau1,Pomeau2,UV_catastrophe_2}. In that case, the specific values of the couplings between modes $C_{nmij}$ in (\ref{eq:EQUATION_Modes}) should play an important role in the preservation or erosion of any structure in the power spectrum.

\section{Eigenstates}

\label{appendix:eigensystems}

We collect here results for the Schr\"{o}dinger equations introduced in the
main text. The results for the 1D Schr\"{o}dinger equation with the HO and
box potentials are provided above in Eqs. (\ref{eq:eigensystem_HP})-(\ref%
{eq:eigensystem_Box}).

\noindent \textbf{Schr\"{o}dinger equation (\ref{eq:D-NLSE_cubic_quintic})
with the HO potential in $D$ dimensions:}
\begin{equation}
E_{n}=2n+\frac{D}{2},
\end{equation}%
\begin{equation}
f_{n}(r)=\sqrt{\frac{n!\Gamma (D/2)}{\pi ^{d/2}\Gamma (n+D/2)}}L_{n}^{\left(
\frac{D-2}{2}\right) }(r^{2})e^{-r^{2}/2},
\end{equation}%
where $L_{n}^{(\alpha )}$ are the generalized Laguerre polynomials.

\noindent \textbf{1D Schr\"{o}dinger equation with the ``superselection"
potential $V^{(1)}(x)$ (\ref{eq:V1}):}
\begin{equation}
E_{n}=2n+\frac{\delta }{2},
\end{equation}%
\begin{equation}
f_{n}(x)=\sqrt{\frac{n!\Gamma (\delta /2)}{\pi ^{\delta /2}\Gamma (n+\delta
/2)}}L_{n}^{\left( \frac{\delta -2}{2}\right) }(r^{2})e^{-x^{2}/2},
\end{equation}%
where $\delta =2+\sqrt{1+4s}$ and $L_{n}^{(\alpha )}(x)$ are the generalized
Laguerre polynomials.

We also show here how to derive the 1D-GPE with potential $V^{(1)}$, (\ref%
{eq:V1}), from the dimensional reduction of the following D-dimensional NLSE
with the HO potential:
\begin{equation}
i\partial _{t}\psi =\frac{1}{2}\left( -\partial _{rr}-\frac{D-1}{r}\partial
_{r}+r^{2}\right) \psi +gr^{D-1}|\psi |^{2}\psi ,
\end{equation}%
with $r\in \lbrack 0,\infty )$, and the nonlinear term has the factor $%
r^{D-1}$. First, one has to plug the change $\psi (t,x)=r^{\frac{1-D}{2}}%
\tilde{\psi}(t,r)$ into the equation to get rid of the first derivative in
the radial Laplacian. Then, one cancels factor $r^{\frac{1-D}{2}}$ on the
RHS and LHS to get
\begin{equation}
i\partial _{t}\tilde{\psi}=-\frac{1}{2}\partial _{rr}\tilde{\psi}+\left(
\frac{r^{2}}{2}+\frac{D^{2}-4D+3}{8r^{2}}\right) \tilde{\psi}+g|\tilde{\psi}%
|^{2}\tilde{\psi},
\end{equation}%
which is the 1D-GPE with the anharmonic potential $V^{(1)}$ (\ref{eq:V1}) on
the half-line.

\noindent \textbf{1D-Schr\"{o}dinger equation with potential $V^{(2)}(x)$ (%
\ref{eq:V2}):}
\begin{equation}
E_{n}=2n+\frac{23}{6},
\end{equation}%
\begin{equation}
f_{n}(x)=\frac{2^{n+1/2}n!}{\pi ^{1/4}\sqrt{(2n+5)(2n+1)(2n)!}}e^{-x^{2}/2}
\end{equation}%
\begin{equation}
\times \left( \frac{3(1+2x^{2})}{(3+2x^{2})}L_{n}^{\left( \frac{1}{2}\right)
}(x^{2})-2(n+1)L_{n+1}^{\left( -\frac{1}{2}\right) }(x^{2})\right) ,
\end{equation}%
where $L_{n}^{(\alpha )}(x)$ are the generalized Laguerre polynomials.

\noindent \textbf{1D-Schr\"{o}dinger equation with potential $V^{(3)}(x)$ (%
\ref{eq:V3}):}
\begin{equation}
E_{0}=-\frac{5}{6},\quad E_{n\geq 1}=n+\frac{7}{6},
\end{equation}%
\begin{equation}
f_{0}(x)=\frac{\sqrt{2}}{\pi ^{\frac{1}{4}}}\frac{e^{-\frac{x^{2}}{2}}}{%
1+2x^{2}},
\end{equation}%
\begin{equation}
f_{n\geq 1}(x)=\frac{1}{\pi ^{1/4}\sqrt{2^{n}(n+2)(n-1)!}}e^{-x^{2}/2}
\end{equation}%
\begin{equation}
\times \left( \frac{4x}{(1+2x^{2})}H_{n-1}(x)+H_{n}(x)\right) .
\end{equation}%
where $H_{n}(x)$ are the Hermite polynomials.

\noindent \textbf{1D-Schr\"{o}dinger equation with potential $V^{(4)}(x)$ (%
\ref{eq:V4}):}
\begin{equation}
E_{0}=-\frac{3}{2},\quad E_{n\geq 1}=n+\frac{5}{2},
\end{equation}%
\begin{equation}
f_{0}(x)=\frac{2\sqrt{6}}{\pi ^{\frac{1}{4}}}\frac{e^{-\frac{x^{2}}{2}}}{%
3+12x^{2}+4x^{4}},
\end{equation}%
\begin{equation}
f_{n\geq 1}(x)=\frac{1}{\pi ^{1/4}\sqrt{2^{n}(n+4)(n-1)!}}e^{-\frac{x^{2}}{2}%
}
\end{equation}%
\begin{equation}
\times \left( \frac{8x(3+2x^{2})}{(3+12x^{2}+4x^{4})}H_{n-1}(x)+H_{n}(x)%
\right) ,
\end{equation}%
where $H_{n}(x)$ are the Hermite polynomials.

\noindent \textbf{Two-component 1D Schr\"{o}dinger equation (\ref%
{eq:Two-Component_NLS}):} Using transformation $\psi _{+}=u+v$ and $\psi
_{-}=u-v$, the equation produces two ``towers" of eigenvalues,
\begin{equation}
E_{n}^{(\pm )}=n+\frac{1}{2}\pm c,
\end{equation}%
and the same eigenfunctions as above:
\begin{equation}
f_{n}^{(\pm )}(x)=\frac{1}{\pi ^{1/4}\sqrt{2^{n}n!}}H_{n}(x)e^{-\frac{x^{2}}{%
2}}.
\end{equation}

\noindent \textbf{A wave equation in the anti-de Sitter space (\ref{eq:AdS}):%
}
\begin{equation}
E_{n}=2n+3,
\end{equation}%
\begin{equation}
f_{n}(x)=\frac{2\sqrt{n!(n+2)!}}{\Gamma \left( n+\frac{3}{2}\right) }\cos
^{3}(x),P^{\left( \frac{1}{2},\frac{3}{2}\right) }(\cos 2x),
\end{equation}%
where $P^{\left( \frac{1}{2},\frac{3}{2}\right) }$ are Jacobi polynomials.


\end{document}